\documentclass[iop]{emulateapj}
\usepackage{color}
\usepackage{lscape}
\usepackage{subfigure}
\usepackage{natbib}
\usepackage{amsmath}
\slugcomment{Submitted to ApJS}
\shorttitle{HPS I}
\shortauthors{Adams \it{et al.}}

\begin{document}
\title{HETDEX pilot survey for emission-line galaxies - I. Survey design, performance, and catalog\altaffilmark{*}}
\author{Joshua J. Adams\altaffilmark{1}, Guillermo A. Blanc\altaffilmark{1}, 
Gary J. Hill\altaffilmark{2,3}, Karl Gebhardt\altaffilmark{1,3}, 
Niv Drory\altaffilmark{4}, Lei Hao\altaffilmark{1,5},
Ralf Bender\altaffilmark{4,6}, Joyce Byun\altaffilmark{1,7}, 
Robin Ciardullo\altaffilmark{8},
Mark E. Cornell\altaffilmark{2},
Steven L. Finkelstein\altaffilmark{9}, Alex Fry\altaffilmark{1,10},
Eric Gawiser\altaffilmark{11}, Caryl Gronwall\altaffilmark{8},
Ulrich Hopp\altaffilmark{4,6}, Donghui Jeong\altaffilmark{1,3,12}, 
Andreas Kelz\altaffilmark{13}, 
Ralf Kelzenberg\altaffilmark{4}, Eiichiro Komatsu\altaffilmark{1,3}, 
Phillip J. MacQueen\altaffilmark{2},
Jeremy Murphy\altaffilmark{1}, 
P. Samuel Odoms\altaffilmark{2}, 
Martin Roth\altaffilmark{13}, Donald P. Schneider\altaffilmark{8},
Joseph R. Tufts\altaffilmark{1,2,14}, Christopher P. Wilkinson\altaffilmark{2}}
\altaffiltext{*}{This paper includes data taken at The McDonald Observatory of The University of Texas at Austin.}
\altaffiltext{1}{Department of Astronomy, University of Texas at Austin, 1 University Station C1400, Austin, TX 78712, USA}
\altaffiltext{2}{McDonald Observatory, University of Texas at Austin, 1 University Station C1402, Austin, TX 78712, USA}
\altaffiltext{3}{Texas Cosmology Center, University of Texas at Austin, 1 University Station C1400, Austin, TX 78712, USA}
\altaffiltext{4}{Max-Planck Instit\"{u}t fur extraterrestrische Physik, Giessenbachstra\ss e, 85741 Garching bei M\"{u}nchen, Germany}
\altaffiltext{5}{Current address: Key Laboratory for Research in Galaxies and Cosmology, Shanghai Astronomical Observatory, Chinese 
Academy of Sciences, 80 Nandan Road, Shanghai 200030, China}
\altaffiltext{6}{Universitt\"{a}s-Sternwarte der Ludwig-Maximilians-Universit\"{a}t, Scheinerstra\ss e 1, 81679 M\"{u}nchen, Germany}
\altaffiltext{7}{Current address: Department of Astronomy, Cornell University, 610 Space Sciences Building, Ithaca, NY 14853, USA}
\altaffiltext{8}{Department of Astronomy and Astrophysics, The Pennsylvania State University, 525 Davey Lab, University Park, PA 16802, USA}
\altaffiltext{9}{George P. and Cynthia Woods Mitchell Institute for Fundamental Physics and Astronomy, Department of Physics and Astronomy, Texas A\&M University, 4242 TAMU, College Station, TX 77843, USA}
\altaffiltext{10}{Current address: Department of Astronomy, University of Washington, Box 351580, U.W., Seattle, WA 98195, USA}
\altaffiltext{11}{Department of Physics and Astronomy, Rutgers, The State University of New Jersey, 136 Frelinghuysen Rd, Piscataway, NJ 08854, USA}
\altaffiltext{12}{Current address: California Institute of Technology, Mail Code 350-17, Cahill Center 326, Pasadena, CA, 91125, USA}
\altaffiltext{13}{Astrophysikalisches Institut Potsdam, An der Sternwarte 16, D-14482 Potsdam, Germany}
\altaffiltext{14}{Current address: Las Cumbres Observatory Global Telescope Network, Inc., 6740 Cortona Dr. Suite 102, Santa Barbara, CA 93117, USA}
\begin{abstract}
\par We present a catalog of emission-line galaxies selected solely by their 
emission-line fluxes using a wide-field integral field spectrograph. This work is 
partially motivated as a pilot survey for the upcoming 
Hobby-Eberly Telescope Dark Energy Experiment (HETDEX). We describe the 
observations, reductions, detections, redshift classifications, line fluxes, and counterpart information 
for 397 emission-line galaxies detected over 169 $\sq\arcmin$ with a 
3500-5800\AA\ bandpass under 5\AA\ full-width-half-maximum (FWHM) spectral resolution. The survey's best sensitivity 
for unresolved objects under photometric conditions is between 
$4-20\times10^{-17}$ erg s$^{-1}$ cm$^{-2}$ depending on the wavelength, and Ly$\alpha$ luminosities between 
$3-6\times10^{42}$ erg s$^{-1}$ are detectable. 
This survey method complements narrowband and color-selection techniques in the search for high redshift 
galaxies with its different selection properties and large volume probed. 
The four survey fields within the COSMOS, GOODS-N, MUNICS, and XMM-LSS areas are rich with existing, complementary data. 
We find 104 galaxies via their high redshift Ly$\alpha$ emission at $1.9<z<3.8$, and the majority of the remainder 
objects are low redshift [OII]3727 emitters at $z<0.56$. The classification between low and high redshift objects 
depends on rest frame equivalent width, as well as other 
indicators, where available. Based on matches to X-ray catalogs, the active galactic 
nuclei (AGN) fraction amongst the Ly$\alpha$ emitters (LAEs) is 6\%. We also analyze the survey's completeness and 
contamination properties through simulations. We find five high-$z$, highly-significant, resolved 
objects with full-width-half-maximum sizes 
$>44$ $\sq\arcsec$ which appear to be extended Ly$\alpha$ nebulae. 
We also find three high-$z$ objects with rest frame 
Ly$\alpha$ equivalent widths above the level believed to be achievable with normal star formation, EW$_{0}>240$\AA. 
Future papers will investigate the physical properties of
this sample.  
\end{abstract}

\keywords{galaxies: formation --- galaxies: evolution ---galaxies: high-redshift --- cosmology: observations} 


\section{Introduction}
\par The Hobby-Eberly Telescope (HET) Dark Energy Experiment (HETDEX) \citep{Hil04,Hil08} will 
survey 60 \sq \arcdeg\ spread throughout 420 \sq\arcdeg\ to discover 0.8 million new Lyman-$\alpha$ 
emitting galaxies (LAEs) over $1.9<z<3.5$ and use them to 
map the expansion history of the universe. A further $\sim$1 million 
low-$z$ galaxies will have their redshifts determined, primarily in the 
[OII]3727 transition, over $0<z<0.47$.  The primary HETDEX science goal is to 
measure the dark energy equation of state at high redshift by using the three-dimensional 
power spectrum of 
LAE positions and redshifts \citep{Jeo06,Koe07,Jeo09,Sho09}. 
An important secondary goal of HETDEX is to investigate the 
physical properties of star forming galaxies, through Ly$\alpha$ and [OII] emission, 
using vastly greater statistics and volumes than currently available. The survey will 
use an array of 150 integral field spectrographs on the upgraded 10~m HET \citep{Ram98,Sav10} called the 
Visible Integral field Replicable Unit Spectrograph \cite[VIRUS;][]{Hil10}. 
\par The HETDEX Pilot Survey (HPS) is the pathfinder to the full HETDEX survey. This 
pilot survey provides a direct test of equipment, data reduction, target properties, observing procedures, and 
ancillary data requirements to HETDEX by using one integral field spectrograph, named the 
VIRUS prototype \cite[VIRUS-P;][]{Hil08b},  on the 
2.7~m Harlan J.~Smith telescope at the McDonald Observatory over 111 nights. 
To do this, the pilot survey uses 
the novel technique of blind, field-of-view, wide-field contiguous spectroscopy 
to find emission line objects over a broad redshift range. While large numbers of narrowband-selected 
LAEs have been assembled by previous 
surveys \citep[e.g.][]{Hu96,Cow98,Rho00,Steid00,Ouc03,Hu04,Hay04,San04,Pal04,Ven05,Gaw06,Gro07,Nil07,Ouc08,Nil09,Gua10,Til10}, 
these surveys are heterogeneous in nature, with different depths and equivalent width (EW) limits. The HPS 
is designed to produce a homogeneous sample of LAEs over an extremely large volume, 
1.03$\times$10$^{6}$ Mpc$^3$h$_{70}^{-3}$, that is nearly an order of magnitude larger than 
the largest existing blind spectroscopic survey, 2.5$\times$10$^{5}$ Mpc$^3$h$_{70}^{-3}$ \citep{Cas10}, and 
vastly larger than other blind surveys \citep{Pir04,van05,Xu07,Saw08,Mar08}. This allows us to evaluate potential redshift evolution 
of LAE properties and to make comparisons to color-selected high redshift galaxy populations
\citep[e.g.][]{Ste96,Ste99,Dad04,Kor10}. The HPS also enables us to find a 
large sample of lower redshift galaxies selected through, 
primarily, their [OII]3727, H$\beta$, and [OIII] emission and study their properties over a 
lower redshift ranges (up to $z=$0.56, 0.19, 0.17, and 0.16 for [OII], H$\beta$, [OIII]4959, and 
[OIII]5007 respectively).
\par The paper is organized as follows. In \S \ref{sec_setup} we describe the instrumental 
capabilities of VIRUS-P, the type and quality of data taken, the necessary calibrations, 
and the imaging compiled to 
aid source classification. We detail the data reduction steps, with special care given 
toward tracking systematic errors 
in \S \ref{sec_reduc}. In \S \ref{sec_cuts}, we describe the 
methods used to recover objects to the survey's statistical limits 
and analyze the effect of noise contamination and the emission-line flux measurements. 
In \S \ref{sec_class}, we 
present our classification methods, 
relying primarily on imaging counterpart likelihoods and 
equivalent width measurements. The 
contamination of the high redshift LAE sample by active galactic 
nuclei (AGN) is presented as well as example classifications. The 
final emission-line catalog and its summary properties are given 
in \S \ref{sec_cat}. Finally, in \S \ref{sec_con}, we review the analysis and 
describe its place in future projects.  
\par In this work, we adopt a standard $\Lambda$CDM cosmology with H$_0$=70 
km s$^{-1}$ Mpc$^{-1}$, $\Omega_{M}$=0.3, and $\Omega_{\Lambda}$=0.7. All 
magnitudes are quoted in the AB system \citep{Oke83}. All wavelengths are corrected to 
vacuum conditions in the heliocentric frame with an assumed 
wavelength-independent index of refraction for air at the observatory's altitude of n~$=1.00022$. 
\section{Observations}
\subsection{Instrumental configuration}
\label{sec_setup}
\par The Visible Integral-field Replicable Unit Spectrograph Prototype (VIRUS-P) 
was designed for this pilot survey and is described in 
\citet{Hil08b} and references therein. The instrument is a fiber-based 
Integral Field Spectrograph (IFS) fed at f/3.65 on the McDonald Observatory's 
2.7m Harlan J. Smith telescope. A small focal reducer sits just prior to the 
Integral Field Unit (IFU) input in the lightpath of the telescope's f/8.8 focus. 
Originally, VIRUS-P operation used a focal reducer 
labeled FR1, but all data taken after September of 2008 used a 
second focal reducer labeled FR2, which has significantly improved 
efficiency below 4000\AA\ compared to FR1 (see \S \ref{sec_fluxcal}). 
Auto-guiding and sky transparency measurements were performed with an 
off-the-shelf Apogee Alta camera installed into a 
field position $\sim9$\arcmin\ north 
of the IFS field of view (FOV). The guider has a square 20.25$\sq\arcmin$ 
FOV and uses a B+V filter with a mean wavelength of 5000\AA\ at a platescale 
of 0\farcs53 per pixel. 
\par
\begin{figure}
\centering
\epsscale{1.0}
\includegraphics[scale=0.7,angle=0]{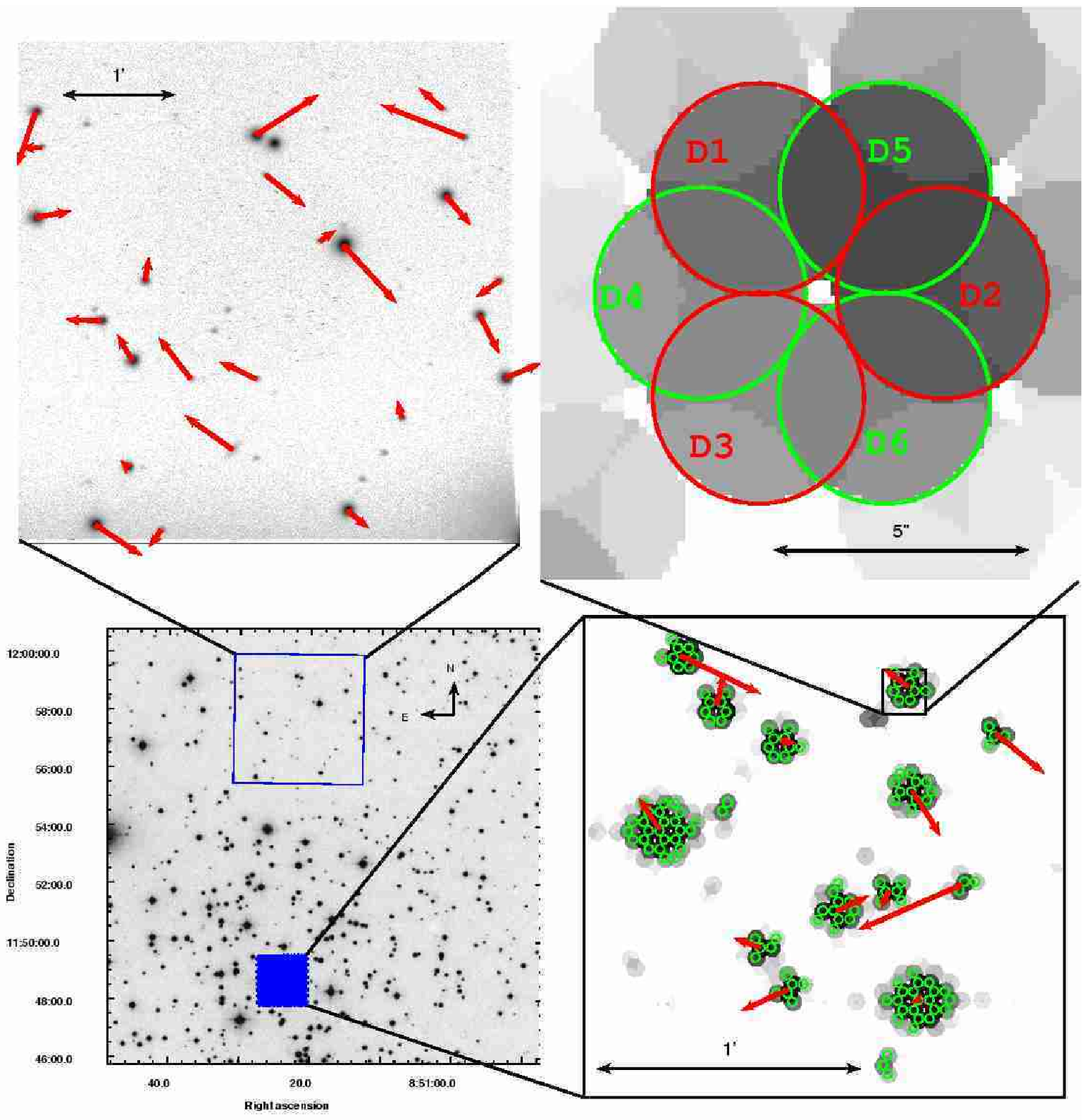}
\caption{The layout of a VIRUS-P observation and the quality of a guider-based 
astrometric solution. \textit{Bottom left}: The footprints 
of the spectroscopic science FOV and the northernly offset guider FOV overlayed on a 
Digital Sky Survey (DSS) image of the open cluster M67. This type of field is used to 
calibrate the astrometry of the guider and the fibers as discussed in 
\S \ref{sec_astrom}. \textit{Top left}: An expanded view of the 
VIRUS-P guider field with residuals from the astrometric model. The residuals are shown as
red vectors scaled by 60$\times$. The rms is 0\farcs42. \textit{Bottom right}: An expanded view of the science FOV. The 
continuum map is generated from the IFS data summed over 
4100\AA$<\lambda<$5700\AA. Fibers that have significant flux and 
border other significant fibers are highlighted with green circles and bunched as 
point source detections for the astrometric fit. The residuals are shown as 
red vectors scaled by 60$\times$. The rms is 0\farcs21. The residuals in the IFU 
are less than the residuals in the guider as both fields have a similar number 
of degrees of freedom, but 
the guider has more datapoints. \textit{Top right}: The 
expanded view of one fiber moving through the six dither 
positions. The pattern, marked with D1-D6, gives very nearly an oversampling of two. The small offsets 
necessary to complete the dither patterns are controlled by sending offsets to the guider.  
} 
\label{fig_astrom_resid}
\end{figure}

 Two different IFUs have been 
used over the course of this pilot survey. Fiber bundle IFU-1, used prior to 
March 2008, spans $1\farcm70\times
1\farcm77$ with 244 functional and 3 broken 200 $\mu$m core diameter (4\farcs235 on-sky) fibers. 
IFU-2 spans $1\farcm61\times1\farcm65$ with 246 functional and 0 broken fibers of the same core size. 
There is no significant difference in throughput between the bundles. 
Both IFUs are of the densepak type \citep{Bar98} with a filling factor near 1/3, requiring at least three dithered positions to 
fully sample the FOV. This survey utilizes a six position dithering pattern as 
illustrated in Figure \ref{fig_astrom_resid}. The 
nearly $\times$2 oversampling delivered by this dithering pattern provides improved spatial 
registration between detected spectral objects and imaging-based continuum counterparts. The 
wavelength range on VIRUS-P is adjustable from 3400-6800\AA\, and a set of volume phase 
holographic gratings delivering various spectral resolutions are available. For this survey the instrument 
was set to cover 3500-5800\AA\ at resolutions that range from 4.5-5.5\AA\ full width half maximum 
(FWHM) over the whole dataset through a 831 lines mm$^{-1}$ grating that 
delivers a dispersion of 1.1\AA\ pixel$^{-1}$ in the unbinned charge-coupled device (CCD) mode. 
The spectral resolution over that range weakly and gradually varies with wavelength 
and between different fibers due to CCD surface shape 
deviations from planarity, camera design limits, and the residual camera alignment errors. 
The data are recorded on a 2k$\times$2k CCD with 15~$\mu$m pixels in 
a custom built, LN$_2$ cooled, vacuum-sealed camera \citep{Tuf08} with electronics 
that deliver between 3.6-4.2 e$^-$ read noise, making the sky 
background the dominant source of noise at all wavelengths in our 20 minute exposures. The 
data have been taken with $2\times1$ binning along the dispersion direction to minimize read noise 
and still maintain a Nyquist sampling of the instrumental line profile.
\par Several instrumental properties determine the survey's calibration needs. The instrument's 
scattered light properties have been discussed in \citet{Ada08}. A weak 
in-focus ghost of atmospheric OH lines redder than the targeted wavelength range 
was found to exist at discrete wavelengths. These lines are easily distinguished by their deviations 
from calibrated wavelength solutions and fiber trace positions. The strength of the scattered light varied 
over time as alignments changed and baffling was implemented, but the ghost's strength was at maximum $3\times$ the 
resolution element noise, and more characteristically below 
the noise in any one fiber. The scattered light affected one resolution element per fiber.
Extra masking installed around the grating 
solved this issue for all data taken after September 2008. All emission-line sources discussed in this paper 
from observations prior to the installation of the grating mask have been visually inspected to not 
lie in the affected regions. 
\par The lab testing and characterization of the VIRUS-P fibers, with 
particular attention to transmission and focal ratio degradation, has been 
investigated in \citet{Mur08}. A high stability in each fiber's 
throughput over a night, at minimum, is crucial toward the survey's goals. 
IFU mounting practices have been established from these 
tests to yield fiber stability sufficient for our purposes. To facilitate mounting on the HET as well as the 
Smith telescope, the IFU was made longer than otherwise necessary. Since the IFU demonstrated inferior 
performance when coiled, the fibers were left uncoiled for most of this pilot survey. 
When the 
IFU bundle is properly uncoiled, it is measured on-telescope to be stable over nightly operating conditions 
to 1\% root-mean-squared (rms) for the most affected fibers 
and 0.3\% rms for the median fiber. We will explore the effect of this 
potential systematic on the data in \S \ref{sec_systemat}. There, we will show that the 
VIRUS-P fiber stability is not an important issue for 
emission-line detections, but can dominate the uncertainty in continuum estimates. 
\par The mechanical design of VIRUS-P has been presented in 
\citet{Smi08}. The instrument's mechanical structures are all made from aluminium to 
achieve a uniform coefficient of thermal expansion between components and to maintain the optical alignment. 
The gimbal mount connecting VIRUS-P to the telescope allows VIRUS-P to swing into a horizontal position 
for any pointing of the equatorially-mounted telescope. 
This ensures that the trace patterns of fibers on the CCD remains constant to high precision 
over a night. Although a $<0.05$ pixel trace shift per night is desired, this could 
not always be accomplished. A trace could shift by up to 0.3 pixels with
temperature under some operating conditions. 
Consequently, data reduction steps were developed to identify and compensate for 
this subtle systematic; these are described in \S \ref{sec_reduc}. 
There is not an atmospheric differential corrector 
installed on the telescope. We discuss the atmopheric effects on emission-line 
source astrometry in Appendix \ref{ap_adr} and the absolute flux calibration of the data 
in \S \ref{sec_fluxcal}. All 
observations were taken with airmasses below two.

\subsection{Data collection}
\par We obtained regular fall/winter/spring dark time observations from September 2007 
to February 2010 on the McDonald 2.7m Harlan J. Smith telescope. These observing runs 
are summarized in Table \ref{tab_obs_run}. In total, out of our allocation of 113 nights, 
61 were useful for this project. 
We constructed datacube mosaics in four science fields: the Cosmological Evolution 
Survey \cite[COSMOS;][]{Sco07}, the 
Hubble Deep Field 
North \cite[HDFN;][]{Wil96} and the surrounding Great Observatories Origins 
Deep Survey North \cite[GOODS-N][]{Dic03}, 
the Munich Near-IR Cluster Survey \cite[MUNICS;][]{Dro01}, and the 
XMM Large Scale Structure field \cite[XMM-LSS;][]{Pie04}. We 
completed 27, 13, 16, and 4 field pointings, respectively in these fields, by taking three 20-minute 
exposures at each of the 6 dither positions. Our effective observation area, 
accounting for mosaic overlap, 
is 169.23 $\sq\arcmin$ over the wavelengths $\sim$3500-5800\AA\ with a spectral resolution of 
$\sim$5\AA. This corresponds to survey volumes of 
1.03$\times10^{6}$ Mpc$^3$h$_{70}^{-3}$ for LAEs and 4.24$\times10^{4}$ Mpc$^3$h$_{70}^{-3}$ for [OII] sources. As 
described in \S \ref{sec_fluxcal} and shown in Figure \ref{fig_flux_lim}, we give the 
survey's flux and luminosity limits as a function of wavelength under photometric 
conditions for the case of a spectrally unresolved, point source emission-line object well centered on a fiber. 
\par In addition to the science data, the following calibration data were obtained one or twice each night. Spectrophotometric 
standard stars from \citet{Mas88} were observed. Flats near zenith of 
the dawn and dusk sky were taken. Calibration with dome lamps was 
explored but abandoned when none were found with sufficient blue-to-red flux balance. 
Sets of bias frames were taken and used to construct a master bias for each run. 
HgCd arc lamps were used to illuminate a dome screen for wavelength calibration. Custom line lists for the HgCd lamps 
were made by observing the lamps with the 2.7m's Tull Coud\'{e} Spectrograph \citep{Tul95} at 
R=60k. The Coud\'{e} wavelength calibration was made from ThAr lines. For most of the observing runs, guider frames 
were saved at intervals of 2-10 seconds, depending on the guider star brightness and transparency. 
The collection of guider frames was 
prevented 13\% of the time due to human error and guider equipment failure. For those observations, 
the flux calibration was done assuming the median of the observed atmospheric transmission (\S \ref{sec_fluxcal})
from the dataset's remaining observations.
\subsection{Astrometry}
\label{sec_astrom}
\par The position of a faint source is not well determined by 
the IFS data alone since most pointings lack sufficiently bright stars to establish an 
astrometric solution for the frame. Instead, the 
positions of stars in the offset guider camera were used to determine the 
fiber positions; this required precise calibration of the relative 
astrometry between the fiber array and the offset guider. The relative fiber-to-fiber positions 
of both IFUs were measured in the laboratory 
and verified to be very regular due to the precise machining. Illumination 
and direct imaging in the lab showed that IFU-2 has 
exceptional uniformity in its fiber matrix, and no deviations 
from the designed pitch of 340~$\mu$m could be measured to 
an accuracy of 1~$\mu$m. IFU-1 is somewhat less 
uniformity in its fiber matrix than IFU-2. We have 
mapped the centroid of each fiber to within 0.3$\mu$m, or 0\farcs007, at the 
nominal plate scale.
\par The transformation from guider field position to science field position was 
calibrated by on-sky measurements. Whenever the guide camera was replaced, we 
obtained data under a six dither pattern on open 
clusters at low airmass. In total, seven astrometric solutions were derived, 
each yielding the plate scales, offsets, and rotations of two image planes 
under a standard tangent projection \citep{Gre93}. We found a adequate fits 
with constant plate scales determined for each IFU axis yielding twelve 
degrees of freedom in a non-linear transformation from guider and IFS pixel 
positions to celestial coordinates. We first determined 
guider positions by using SExtractor \citep{Ber96} to measure the 
positions of stars and match to coordinates from the United States Naval 
Observatory's (USNO) Nomad 
catalog \citep{Zac05}. Similarly, the continuum intensities of USNO stars in the 
fibers were measured by summing flux over the wavelength 
range 4100\AA$<\lambda<$5700\AA; this region was 
chosen to mimic the guider wavelength response and minimize 
atmospheric refraction differences. 
Fibers containing signal significantly above the noise were matched with 
significant detections in adjacent fibers. Centroids were calculated 
for each source and again matched to the Nomad catalog. A simplex method \citep{Pre92} was then 
used to find the least squares minimum robustly in the presence of the many local minima. We show 
in Figure \ref{fig_astrom_resid} the fit quality in a representative solution. The range of systematic 
uncertainty in our seven eras of astrometric solutions 
was 0\farcs17-0\farcs51 with a median of 0\farcs31. 
\par We further measured the stability of the astrometry over 
many months from flux standard stars. We anticipated any drift to be negligible due to the 
design of plastic pins which located the IFU head against the telescope 
mounting surface. However, we found substantial month-to-month systematic variations of 
order 1\farcs8 rms. The only clear 
dependence was a declination term with temperature, which we attribute to a thermal 
expansion of the guider camera mount. However, this expansion cannot explain 
the bulk of the astrometric scatter. Since we fin much smaller astrometric 
scatter in any one month, the monthly removal and
remounting of the IFU
input head from the telescope between observing runs is the plausible 
source of drift. So, we have chosen to estimate 
an empirical month-by-month offset in the astrometric zeropoint which lowers the median 
monthly rms to 0\farcs6 and ranges from 0\farcs0-1\farcs0.
\par Coarse positional sampling by the large fibers and low S/N limitations  
forms the final component of the astrometric error budget. In order to quantify 
this uncertainty, we have simulated the positional recovery for a range of emission 
line sources. We describe those simulations in \S \ref{sec_comp}. The result is a fit 
to the random astrometric uncertainty with a functional form of 
$\sigma_{r,random}=0\farcs348+2\farcs04/(S/N)$. 
\par We can assess the completeness of our error budget by measuring the 
observed positional offsets of emission-line objects found with 
high confidence counterparts. 
As explained in \S \ref{sec_class}, a comparison of our fiber detections with 
broadband imaging shows that 55\% of our emission-line detections have an isolated 
counterpart detected with $\ge90$\% confidence. Through a comparison, we find a 
mean offset of $\Delta\alpha=-0\farcs53\pm0\farcs05$ and $\Delta\delta=0\farcs39\pm0\farcs05$ between 
the fiber-based emission-line source positions and the broadband photometric centers. The source 
of this offset is not certain, but we apply it to all our reported 
emission-line positions. After correcting for this offset, the counterpart 
associations were iterated to produce our final emission-line positions. In Figure \ref{fig_astrom_hist} we present the 
distribution of the data offsets to test the 
error budget. This error budget serves as an important input in the method (\S \ref{sec_class}) for 
assigning broadband counterparts in crowded fields to the emission-line sources. 
\subsection{Flux calibration and transparency}
\label{sec_fluxcal}
\par The majority of the observations were not taken under photometric 
conditions, hence a proper flux calibration requires a realtime measurement 
of the atmospheric transparency. Unlike some modern wide-field imagers, 
the VIRUS-P field of view is not large enough to 
contain photometrically calibrated stars in the majority of its arbitrary 
pointings. However, the 
offset guider with a larger field of view has a size sufficient 
for this continuous calibration 
purpose. We recorded all guide camera 
exposures sampled at 2-10 seconds that were contemporary with the 
IFS science exposures. The guider exposure times varied depending on the guide star brightness. 
Basic bias-subtraction 
and flat-fielding reductions were implemented on the guider frames. We performed aperture photometry on all 
stars detected. When 
available, we used 
Sloan Digital Sky Survey (SDSS) measurements \citep{SDS08} for our calibrations; 
otherwise we used the USNO-B1.0 survey \citep{Mon03}. The SDSS 
photometric precision is quoted at below 1\% for guide stars used, 
typically V$<19$. The 
USNO-B1.0 photometric precision is typically much worse, $\sim$0.25 magnitudes, 
and this directly leads to an important uncertainty in line fluxes for objects 
in the MUNICS and XMM-LSS fields. Accordingly, we have added in quadrature a 15\% 
error, assuming the median of three guide stars per field, to the flux and 
equivalent width (EW) measurements for the MUNICS and XMM-LSS sources. 
We treat these errors as random, since multiple and independent sets of 
stars were used in different mosaic pointings and multiple spectrophotometric 
standards were observed. A color term was fit from the 
guider data considering its non-standard, wide-bandpass filter, a new 
zeropoint was calculated each month to correct for periodic equipment changes and mirror cleanings, 
and non-photometric extinctions were found for each frame after removing a standard airmass 
term of 0.186 mag AM$^{-1}$. Typically, we had two to five stars per field that were 
bright enough for this purpose. The resultant distribution of zeropoint offsets due to transparency, 
$\Delta\mbox{zp}$, is given in Figure \ref{fig_extinc}. By measuring the 
scatter in the zeropoint offset from all the stars available in each frame, we find 
a mean uncertainty of 6\% in the guider-based photometric correction.
\par The flux calibration of IFS data was done in a manner similar to that for 
longslit spectroscopy, but 
with some additional steps to compensate for fiber sampling patterns. We used the spectrophotometric 
stars and calibrations of \citet{Mas88} observed under a six-dither pattern. Airmass 
extinction coefficients for photometric conditions with a curve specifically modeled 
for McDonald Observatory are applied. This extinction curve is similar to the Kitt Peak curve 
supplied with IRAF. The bright standards allowed us to 
determine both the source position relative to the fiber grid and the 
seeing Point Spread Function (PSF), which in turn yields the exact fiber sampling. 
In contrast, fainter emission-line sources require 
statistical sampling corrections that are discussed in \S \ref{sec_fl_meas}. In order 
to determine the percentage of incident flux captured over the six dither 
positions, we employed the following analysis. We began by considering 
the spectra for all fibers positioned within a large radial aperture (operationally, 
8\arcsec) from the stellar centroid. and adopting a seeing model with a 
2D circular, Gaussian PSF. The broadband flux of each fiber was measured by summing over a large 
wavelength range (operationally, 4000\AA$<\lambda<$5500\AA). The PSF and 
Gaussian normalization were determined through a nonlinear least squares minimization by 
assuming the spatial response of each fiber was tophat. The sampling correction was then 
formed from the ratio of the 
Gaussian normalization to the sum of the broadband flux measurements. Then, the spectral count 
rates of the relevant fibers were resampled to a common wavelength scale, co-added, and 
normalized using the 
sampling correction. By using such a broad, circular aperture, we ensured 
that the effects of atmospheric 
differential refraction on the 
co-added spectrum were negligible. The final spectral flux calibration curve was then formed from 
the ratio of the published, absolute flux density to the sampling corrected data count rate. 
Spectrophotometric standards were taken under a range of conditions, so their 
comparison required a further correction for transparency as estimated 
from the guider measurements. Once done, 
we find an rms between all flux calibration curves of 9.3\% and 8.5\% for FR1 and FR2. We find 
no trend with wavelength in this scatter and so validate the assumed gray zeropoint correction 
for all guider transparencies at these levels of uncertainty. The final catalog will list 
the random line flux errors, but the whole sample may be considered to also 
be subject to the $\sim$10\% flux calibration systematic uncertainty just discussed. We do not fold the 
systematic into the tabulated values as relative comparisons within the sample 
should not be subject to it. 
\par Several statistics from this flux calibration analysis summarize the survey's performance. 
First, the range of atmospheric transparencies for recorded data is shown in Figure \ref{fig_extinc}. 
These statistics are biased against periods of weather too poor to attempt observation and 
represent only the best 60\% by time. The median nonphotometric transparency penalty to this survey in 
the observable periods is 0.28 magnitudes. The total system throughput is shown in Figure \ref{fig_thru} as 
the fraction of light recorded after passing through one photometric airmass (zenith), 
the telescope, the focal reducer, and the VIRUS-P instrument. The curves for the two 
focal reducers show a dramatic difference: FR2 performs better than FR1 at all wavelengths, but 
particularly in the blue where FR1 has only half the throughput of FR2. 
\par
\begin{figure}
\centering
\subfigure{\includegraphics [scale=0.3,angle=-90]{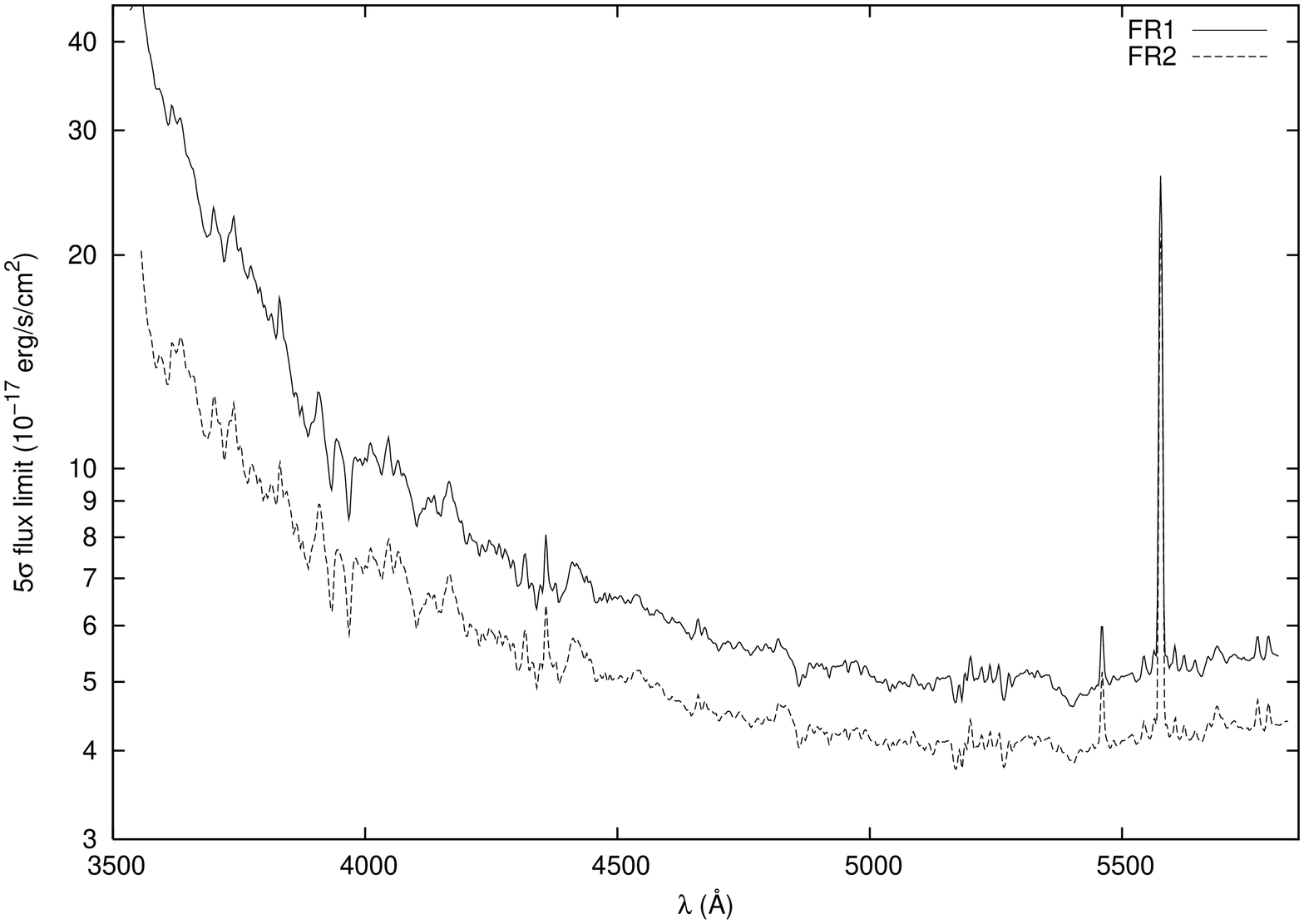}}
\subfigure{\includegraphics [scale=0.3,angle=-90]{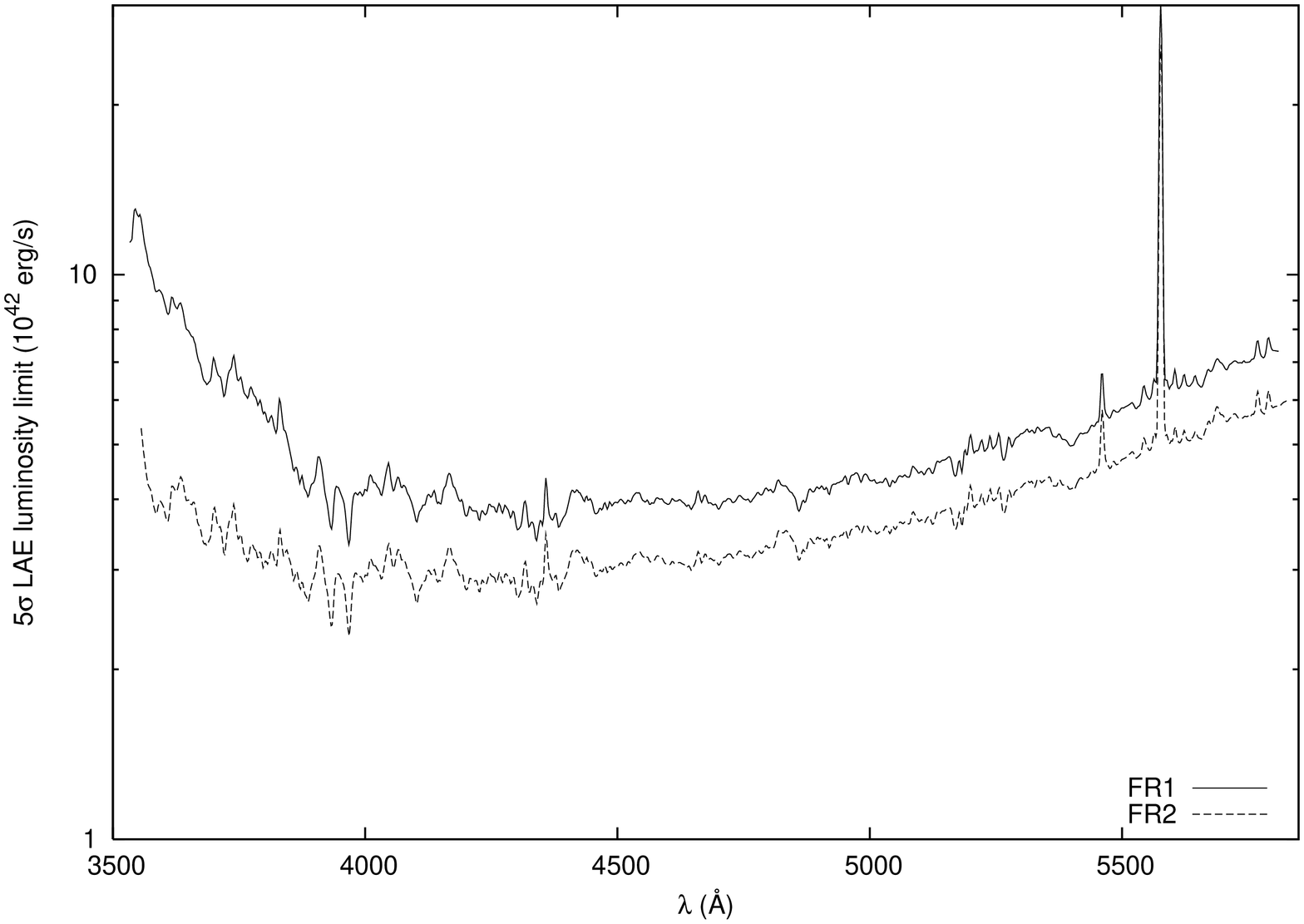}}
\caption{\textit{Left} The 5$\sigma$ detection limit under photometric conditions for 
an emission-line object perfectly centered in a fiber in three 
20 minute exposures. Different source positions 
can improve or decrease this limit by $\sim$15\% which is 
captured in our completeness calculation. In both figures, curves are given 
for the two focal reducers, FR1 and FR2. 
\textit{Right} The 5$\sigma$ luminosity limit under photometric conditions 
for objects detected in the Ly$\alpha$ line.}
\label{fig_flux_lim}
\end{figure}

\begin{figure}
\centering
\subfigure{\includegraphics[bb=51 51 554 554,clip=true,scale=0.3,angle=-90] {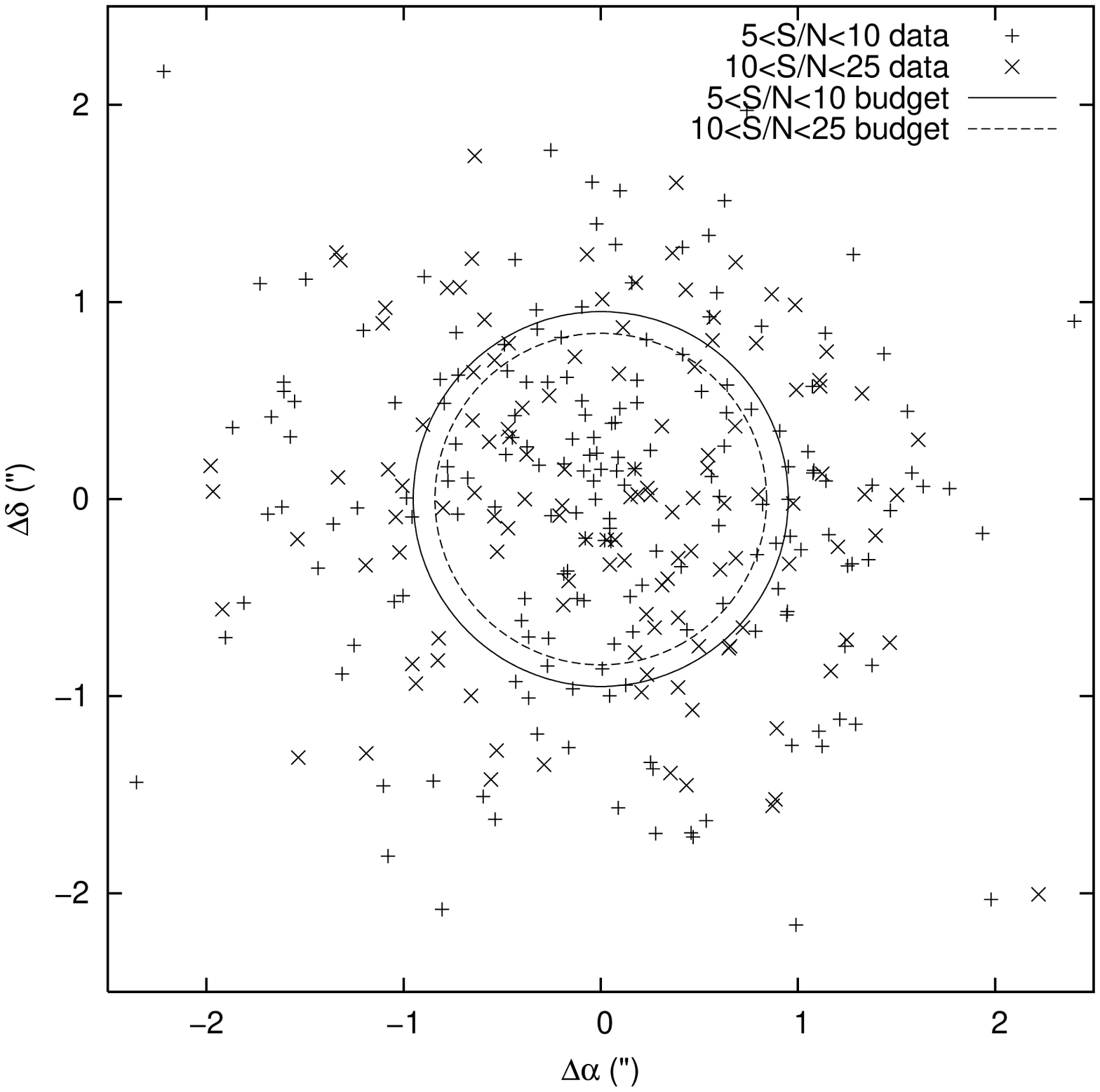}}
\subfigure{\includegraphics[bb=51 51 554 568,clip=true,scale=0.3,angle=-90] {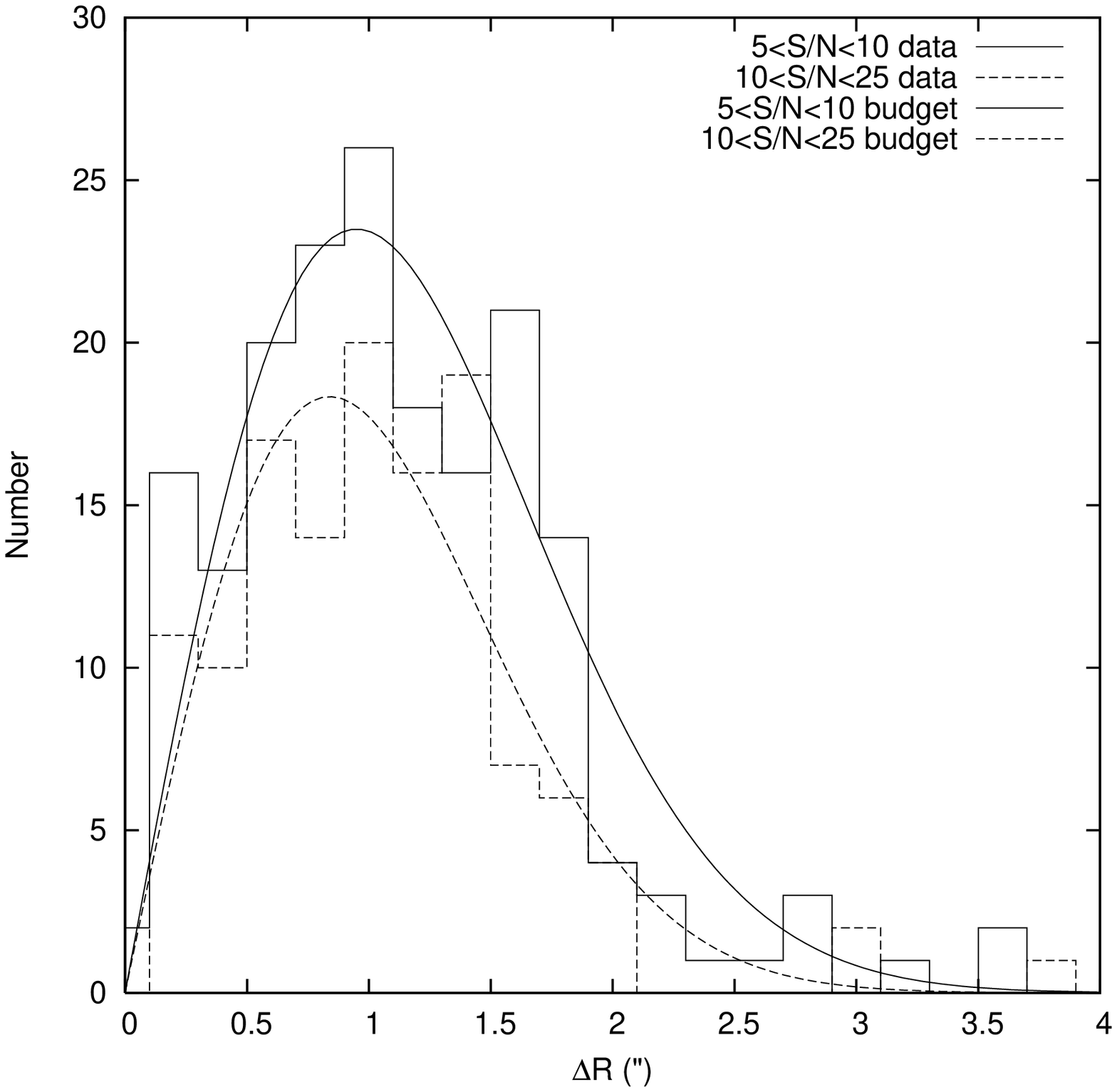}}
\caption{\textit{Left}: The positional offsets from catalog emission-line 
detection positions and broadband counterpart image 
positions divided into two S/N bins. The 
residual correction discussed in \S \ref{sec_astrom} has been applied. 
The 1$\sigma$ astrometric 
error budget for each bin is also drawn with 
radii 0\farcs95 for $5<S/N<10$ and 0\farcs84 
for $10<S/N<25$. \textit{Right}: A 
histogram of the same data shown with a Rayleigh distribution. 
The same dispersions are used to demonstrate the appropriate characterization 
of the astrometric error as a two-dimensional Gaussian function.}
\label{fig_astrom_hist}
\end{figure}

\begin{figure}
\centering
\includegraphics [scale=0.4,angle=-90]{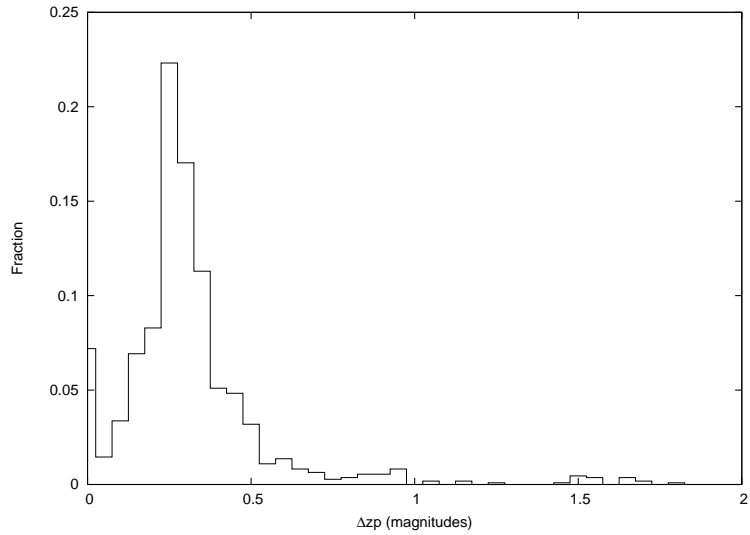}
\caption{The distribution of zeropoint offsets due to 
non-photometric transparency as measured with the guider camera. This 
distribution represents the best 60$^{\mbox{th}}$ percentile 
of the observing allocation with the remaining 40\% being 
too poor to guide or requiring dome closure.}
\label{fig_extinc}
\end{figure}

\begin{figure}
\centering
\includegraphics [scale=0.4,angle=-90]{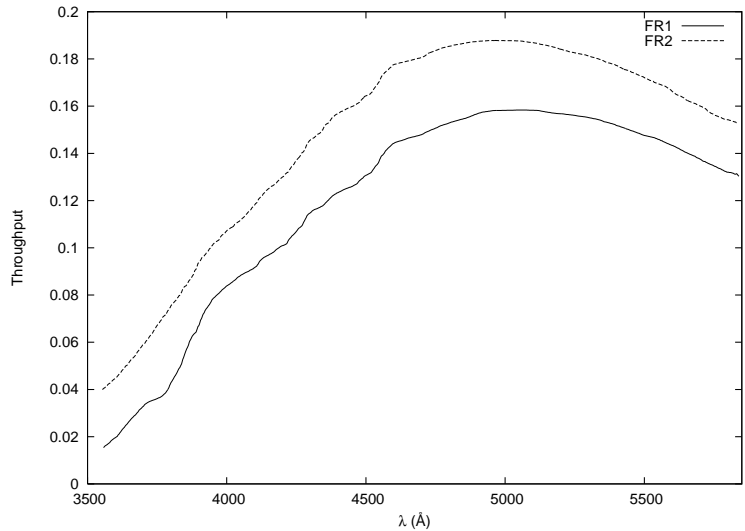}
\caption{The total system throughput of VIRUS-P, the 2.7m telescope, and 
the atmosphere at an airmass of one. Curves are given for the two eras of focal 
reducers, FR1 and FR2.}
\label{fig_thru}
\end{figure}

\begin{figure}
\centering
\epsscale{1.0}
\includegraphics [scale=0.7,angle=0]{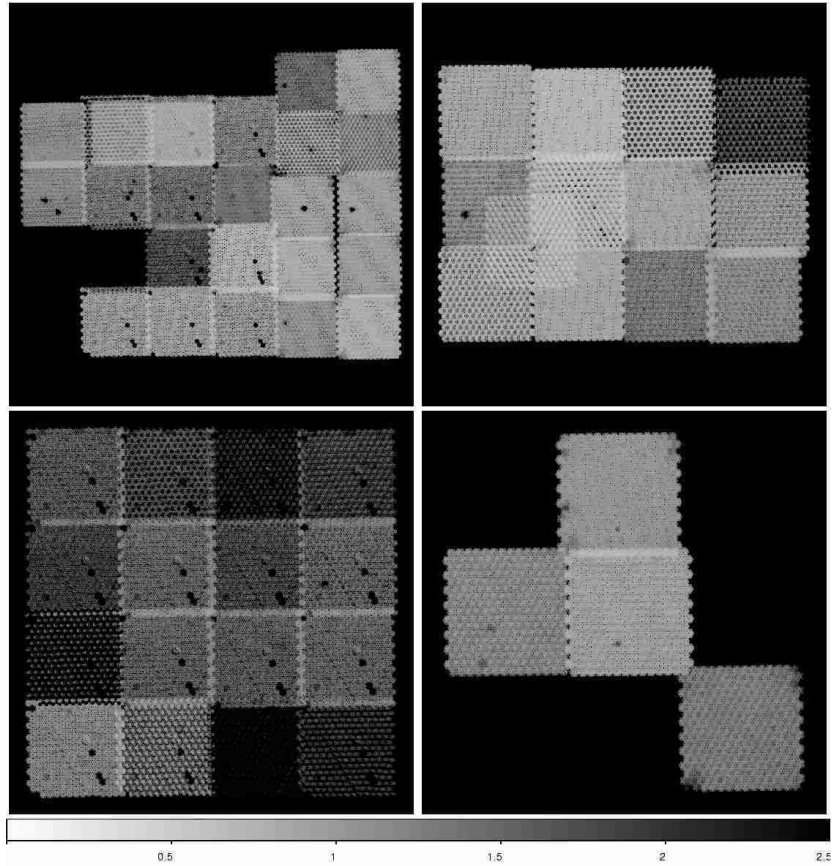}
\caption{Sensitivity maps (1$\sigma$ per detection element) 
at 4500\AA\ in 10$^{-17}$ erg s$^{-1}$ cm$^{-2}$. The three 
broken fibers in fiber bundle IFU-1 are evident. \textit{Top left}: COSMOS, 
\textit{Top right}: GOODS-N, \textit{Bottom left}: MUNICS, \textit{Bottom right}: XMM-LSS.} 
\label{fig_sens}
\end{figure}

 The combination of read noise, system throughput, and sky brightness determine the detection limit 
for an unresolved emission-line source. Figure \ref{fig_flux_lim} shows the 
5$\sigma$ limit in a detection element (defined as $\pm$2$\times$ the instrumental dispersion or 
$\pm$1.9$\times$ binned pixels), which is nominally the survey's photometric limit with some 
modulation for sources sampled under different fiber positions. The luminosity limit 
for LAEs is also shown in Figure \ref{fig_flux_lim}. The exact limits will 
be further explored in \S \ref{sec_cuts} and compensated for with the completeness limit derived in 
\S \ref{sec_comp}. Finally, in Figure \ref{fig_sens} we give the sensitivity maps
at 4500\AA\ for spectrally unresolved point sources, taking into account mosaic 
overlap, bright objects, dead fibers in IFU-1, guider measured extinctions, and the range in
airmass over the dataset. 
Small gaps in the map are due to the slightly different sizes of IFU-1 and IFU-2, and the 
failure to complete the desired six dither pattern in one COSMOS pointing by only 
completing a three dither pattern. 
Finally, five fields were chosen to overlap with previous fields 
for cases where transparency in the first pass yielded poor depth. 
\subsection{Ancillary imaging}
\label{sec_ancil}
\par This survey discovers and spectroscopically measures LAEs 
in one pass, as opposed to narrowband surveys that often require spectroscopic 
confirmation on a subsample. The depth and bandpass restrictions of VIRUS-P, however, still make discrimination 
between LAEs and low-$z$ contaminants challenging. For both LAEs and 
[OII] emitters at many redshifts, we expect to have only one strong emission line in the VIRUS-P bandpass. 
Respectively, [OIII]$\lambda$5007, [OIII]$\lambda$4959, and H$\beta$ will be lost at 
$z>0.158$, $z>0.170$, and $z>0.193$, and the survey's spectral resolution does not resolve the [OII] doublet. Furthermore, 
the variation observed in local galaxies for strong line ratios \citep{Ken92} never guarantees that two 
statistically significant lines will be detected. By necessity, we resort to an EW cut, as used extensively in
LAE narrowband surveys, to classify single emission-line detections. We discuss the EW cut further in \S \ref{sec_class}. 
However, the VIRUS-P spectra are not sufficiently sensitive for continuum 
detections for the majority of the emission-line detections. To reach the necessary limits, we must supplement the spectra 
with deep imaging. 
\par This dataset's fields are located in regions of the sky with existing deep images and catalogs 
\citep{Dro01,Fer99,Cap04,Cap07,Ilb09}. The XMM-LSS field does not have a published catalog but is covered by the 
Canada-France-Hawaii Telescope Legacy Survey\footnote{Based on observations obtained with MegaPrime/MegaCam, a 
joint project of CFHT and CEA/DAPNIA, at the Canada-France-Hawaii Telescope (CFHT) which is operated by the 
National Research Council (NRC) of Canada, the Institut National des Science de l'Univers of the Centre National 
de la Recherche Scientifique (CNRS) of France, and the University of Hawaii. This work is based in part on data 
products produced at TERAPIX and the Canadian Astronomy Data Centre as part of the Canada-France-Hawaii Telescope 
Legacy Survey, a collaborative project of NRC and CNRS.} (CFHTLS) wide field W1. The deep MUNICS images, which were not 
part of the original publications, consist of B$_J$, g', i', and z' data taken with the Large 
Area Imager for Calar Alto (LAICA) on the Calar Alto Observatory 3.5m, with zeropoints made 
by matching stellar photometry to the published catalog. 
Instead of using the literature catalogs, we have chosen to produce our own SExtractor 
catalogs on the images and error maps; this ensured a consistent analysis for the fields and 
pushed the S/N to a lower threshold for a more complete emission-line association. 
We list select properties of the relevant broadband data in Table \ref{tab_Schlegel}. The 
table also gives the Galactic extinction values \citep{Sch98} we 
applied to the continuum and emission-line fluxes under the extinction curve 
fit of \citet{Odo94}. 
\par Care was taken in the photometry to 
ensure our photometric colors were robust. Two measures of seeing FWHM are relevant: the 
one for the particular band where a Kron \citep{Kro80} aperture is 
measured (FWHM$_{Kron}$) and another larger value to which the other photometric bands will 
be matched (FWHM$_{match}$). 
For each field, we 
formed a detection image by stacking the deeper 
available bands without matching each band's seeing (see Table \ref{tab_Schlegel}). The 
detection parameters of SExtractor were then set to find a minimum of three neighboring pixels 
detected with 1$\sigma$ significance over sky without filtering. Since we will only be using 
sources with 3$\sigma$ significance in their photometry, the exact detection weights and filters have 
little importance. Also, the return of spurious continuum sources from the low significance 
thresholds is acceptable for our application. A chosen band with good depth for each field, labeled here as $i$, was 
compared to the detection image using SExtractor dual image mode, in order 
to measure flux densites in a blending corrected Kron 
aperture, $\hat{f}_{\nu,i,Kron}$. The Kron ellipse dimensions $a$ and $b$ were also measured. 
Blending correction was crudely accomplished with the SExtractor {\tt AUTO} flux measurements and the flag
{\tt MASK\_TYPE} set to {\tt CORRECT}. Under this setting, SExtractor sums the flux from the 
opposite side of the Kron aperture whenever it encounters pixels covered by multiple 
Kron apertures. In the remaining 
bands, labeled here as $j$, each frame was matched in seeing to FWHM$_{match}$ and run in dual detection mode to 
measure the flux density in a circular aperture of diameter
1.4$\times$FWHM$_{match}$, $\hat{f}_{\nu,j,circ}$. The term 
$f_{corr}=(1-e^{-0.5ab/\sigma_{Kron}^2})$ then forms a correction factor for the fraction of flux lost to the Kron aperture 
from a point source under seeing with dispersion $\sigma_{Kron}$. The final aperture-corrected flux density 
in each band $j$ was then estimated from Equation \ref{eq_bb_imag}. Standard error 
propagation was applied. 
\begin{equation}
\label{eq_bb_imag}
\hat{f}_{\nu,j}=\frac{\hat{f}_{\nu,i,Kron}\times \hat{f}_{\nu,j,circ}}
{\hat{f}_{\nu,i,circ}\times f_{corr}}
\end{equation}
\par This resultant source catalog was used only in cross-correlation 
with our VIRUS-P emission-line catalog to identify object counterparts. The 
method of assigning counterparts is described in \S \ref{sec_class}. The emission line fluxes 
are subtracted off from the broadband measurements according 
to the filter transmission curves as supplied by \citet{Bra08} once 
counterparts are assigned. 

\section{Data reduction}
\label{sec_reduc}
\par The science goals of this survey required the development of a custom reduction pipeline. Several 
IFS reduction pipelines already exist \citep[e.g.][]{Val92,Zan05,Tur06,San06,San10} and are well suited to 
many applications. In particular, we first tried using a predecessor of {\tt p3d} \citep{San10,Bec02}. 
The crucial limitation of the {\tt p3d} package and all other IFS pipelines at the time, is that they 
resample the spectrum of each fiber onto a common wavelength scale early in the processing. This 
step correlates errors and complicates the detection statistics. In fact, we found by 
running simulated, source-less VIRUS-P data through {\tt p3d} that many more resolution elements 
were flagged to have 5$\sigma$ significance than was possible from the input Poisson 
statistics. The use of {\tt p3d} would have either 
produced too high a contamination fraction or required higher S/N cuts and 
survey flux limits. This consideration led us to 
develop a set of scripts and FORTRAN routines collectively called Vaccine. 
Many of the pipeline steps are standard to all spectroscopic reductions. 
However, the primary Vaccine requirement to avoid data resampling is done in a manner 
similar to the \citet{Kel03} pipeline developed for longslit spectroscopy and affects the flat fielding and 
sky subtraction steps.
\subsection{Preliminaries}
\par The first operation done to each VIRUS-P frame is to 
measure a single bias value from the overscan regions, subtract it 
from the frame's data section, and trim the overscan. A 
master bias then is created from all the overscan-subtracted 
biases taken during an observing run (typically 100 to 200 frames). 
Overall, the noise statistics in bias frames were remarkably stable and 
indistinguishable over weeks. Next, we cleaned the images with a bad pixel mask 
made by exposing the camera to scattered white light and 
finding the pixels with relative quantum efficiency outside 
10\% of the CCD's median. The VIRUS-P CCD has very clean cosmetics: 
besides the two rows nearest the readout register, this bad pixel mask 
only contained thirteen total pixels in three patches. Data combination 
for all co-additions of frames is accomplished using the biweight estimator \citep{Bee90}; 
this algorithm was chosen for its robust performance regarding outliers such 
as cosmic rays. The master bias and individual overscans 
are subtracted from all calibration, science, and 
flux standard frames. Calibration frames, consisting of arc frames and twilight flats, 
are taken at the beginning and end of each observing night. The dawn arcs and flats 
were preferentially used over those frames taken in dusk, as they were a better match 
to the temperature of the night-time conditions. 
\par As is common to both IFS and slitlet multi-object spectroscopy, the 
traces of all fibers are not strictly parallel to the CCD pixels or to each other. 
The fiber profiles, taken from a flat field calibration, must be traced to 
define an extraction aperture of each fiber. 
Moreover, the dispersion axis is not necessarily parallel to each fiber's trace. 
However, with the camera alignment in VIRUS-P, we found the maximum deviation 
of this misalignment is 0.2 resolution elements, so we ignored this 
distinction and defined the dispersion axis along the fiber trace to be 
perpendicular to the cross-dispersion direction. This assumption effectively 
broadens, slightly, the resolution in some fibers. The tracing is then 
made by fitting Gaussian functions to cuts along the 
cross-dispersion axis at a series of wavelengths for each fiber. The 
Gaussian centers are fit by a fourth order polynomial 
across the CCD. This fit was tested against 
repeated flats and shown to be precise to $<0.1$ pixels across the CCD. Trace 
information is displayed for the user, who can iterate 
the fit tolerances if required. All further operations 
are done in the traced coordinates with cross-dispersion 
apertures of five pixels. Vaccine propagates errors 
for all operations starting with the read noise and 
keeps track of the Poisson noise from sources and the background sky. 
\subsection{Wavelength calibration}
\par An automated peak finding algorithm is run on the 
arc lamp frames, and line identifications are made from a user entered 
initial wavelength solution. Typically, seven unblended HgCd lines are found 
with their central pixel locations determined by a Gaussian fit to the line profile. 
The pixel-to-wavelength mapping is then fit with a fourth order polynomial in the 
dispersion direction. The first order term of that 
polynomial is found to vary smoothly for all fibers as a function 
of the cross-dispersion direction. Hence, or increased accuracy, this first order term is refit 
as a function of the cross-dispersion distance from the 
camera optical axis using a fourth order polynomial. Finally, 
the wavelength polynomial as a function of dispersion direction pixel is 
refit, this time 
with the constrained first order term. 
The residuals of this procedure are typically one hundredth 
the size of a resolution element and the solutions are stable to a tenth 
of a resolution element over several weeks. 
\par The heliocentric correction is found for each frame by using 
a FORTRAN implementation (written by G.~Torres\footnote{http://tdc-www.harvard.edu/iraf/rvsao/bcvcorr/bcv.f}) 
of the IRAF task {\tt bcvcorr} in the {\tt rvsao} package \citep{Kur98}. The small, $<$1 km s$^{-1}$ differences 
in heliocentric velocities for exposures at the same dither position but taken over different nights 
are ignored and only the mean heliocentric correction between them is applied. All reported 
wavelengths are in the heliocentric frame. A correction to vacuum conditions is made 
assuming an index of refraction for air of n=$1.0002$ for all observed wavelengths. 
\subsection{Flat fielding}
\par Typically fifteen twilight flats were taken each night and combined using the 
biweight estimator. To 
ensure high S/N in the twilight flats, each frame was exposed to 
near but below the CCD's 1\% nonlinearity specification 
which occurs at 50\% of full well. Four 
signals are present in the twilight flats, 1) the solar spectrum, 
2) the relative throughputs between fibers, 3) the fiber 
profile in the cross-dispersion direction, and 4) the 
relative pixel-to-pixel responses. To remove the 
first of these we 
employed a bspline fit \citep{Die93} constrained by input from 
large subsets of fibers. Such a fit is 
robust against outlier datapoints (i.e. our cosmic rays or faint sources that 
fill a subset of the data) and fits curvature that a linear interpolation would miss. The 
advantage of the bspline fit is best leveraged when a spectrum 
is highly supersampled, and the camera's optical distortions naturally 
deliver this quality in different fibers, predominantly as a smooth 
function of cross-dispersion direction. However, the slight (10\%) spectral resolution 
variation across the CCD disfavors a single fit for all the 
fibers' data. As a compromise, we consider each fiber with its 
twenty nearest fibers in CCD coordinates. Within 
these sets the spectral resolution variations at any wavelength 
are less than 2\%. We do not make more complicated 
corrections for the spectral resolution variation beyond this. 
The bspline fit for each fiber, serving as a model of the solar spectrum, 
is then divided into the original flat field data, resulting in a precision between 
different sets of frames to $<$1\% rms. 
\subsection{Background subtraction}
\par The majority of VIRUS-P fibers and resolution elements in this blind survey 
record blank sky. This enables the noise if our sky model to be driven 
down by stacking measurements over many fibers, so long as the noise is 
statistical. By using the 50 nearest fibers in the 
cross-dispersion direction, the 
statistical noise in the sky can be reduced to only 14\% of a single fiber's 
noise. In this way, the uncertainty in the post-sky-subtracted data 
can be made very close to that of the pre-sky-subtracted data (as long as 
the flat-fielding systematics are understood). 
Our sky background models were formed identically to the flat field models. 
\par We note, however, that this semi-local sky estimation method 
is only robust for sources that fill a small fraction of the 
combination window, which on-sky is approximately 
$\Delta\alpha$=100\arcsec\ by $\Delta\delta$=20\arcsec. No bright, broadband 
sources have such sizes in the survey fields. Moreover, in order to further 
avoid oversubtracting bright sources, we constructed an object mask prior to the bspline 
fit. Any fibers that yield $>$2$\sigma$ significance in the continuum, as 
estimated by combining the data and errors across all VIRUS-P wavelengths, 
were placed in the object mask. 
\subsection{Data combination}
\label{sec_comb}
\par The count rates in the three frames taken at each dither position were first corrected 
by the airmass-based photometric extinctions and the guider-based 
transparency measurements and then combined. The three frames and the 5 pixel 
cross-dispersion aperture delivered fifteen input values to the biweight 
estimator at each wavelength. The VIRUS-P flux standard frames are passed through 
Vaccine exactly as the primary science data. Finally, the science spectra 
(and errors) are 
scaled by the flux calibration (\S \ref{sec_fluxcal}) to form a set of calibrated, 
one-dimensional spectra at each fiber and dither position. 
\subsection{Systematic errors}
\label{sec_systemat}
\par We identify three potential sources of systematic error in VIRUS-P data, 
one unimportant, and two that require monitoring. 
First, we discuss why crosstalk between fibers 
is not important in VIRUS-P data. Next, we identify the effects of 
throughput variations and the accuracy of flat field cross-dispersion 
profiles on the error budget as the most prominent systematics. 
Finally, we describe an empirical, frame-specific estimate of the systematics 
that must be added to the random errors. 
\par IFS crosstalk occurs when the profile of a fiber in the cross-dispersion 
direction significantly overlaps that of other fibers projected nearby on the CCD. 
We make no crosstalk correction in Vaccine for two reasons. First, the fibers are 
measured to have cross-dispersion profiles of 4 pixels FWHM size. 
This is a factor of 2 smaller than the center-to-center fiber 
spacing on the CCD and larger than that found in many IFS instruments. 
As a result of our 5 pixel extraction aperture, sources of 
equal strength in neighboring fibers imply only a $<0.5$\% contamination. 
Second, the blind field selection of this survey 
leaves most fibers seeing only uniform sky background and leaves 
little risk from cross-talk contamination. A fiber 
aligned on a source will usually be isolated and trade an equal 
flux from the background sky with its crosstalk neighbors. The flux calibration 
(\S \ref{sec_fluxcal}) steps use the same cross-dispersion 
aperture, and therefore correct for the source flux lost by crosstalk.
\par The stability in the throughput of fibers can cause significant 
systematic errors in some measurements. As discussed earlier, our fiber 
throughput is very stable, with 1\% rms 
variation at worst and 0.3\% median variation over a night. 
However, our background sky is 25-40$\times$ stronger 
than the statistical noise limits in each resolution element. As a result, 
the systematics can overwhelm the statistical errors in spectral apertures of
six resolution elements or more during the worst stability conditions. Continuum 
estimates using large wavelength ranges may thus be severely affected 
in our survey, and we make no claims on such properties. However, the 
situation for emission lines is far better. First, the systematics 
in a detection element (approximately two resolution elements) 
are at worst 56\% of the statistical error and at 
median are 13\% before background subtraction. 
Second, most of the throughput variation is captured in the 
background subtraction step. As described in \S \ref{sec_cuts}, before we detect emission lines 
we subtract off a locally 
estimated continuum value using roughly ninety independent spectral pixels. 
Since fiber throughput variations manifest 
uniformly across wavelength, the spurious signal is a small multiple 
of the sky spectrum and relatively featureless over our bandpass 
(exempt for the bright [OI] 5577\AA\ sky line which we mask 
prior to all detections). The systematic error in a 
post-background-subtraction detection element therefore drops to 5.9\% of 
the statistical error in the extreme case and 1.4\% of the statistical 
error in the median case. We include this systematic uncertainty in both the 
detection and flux calibration error budgets via the 
empirical correction described below. 
\par The final known source of systematic error is occasional variability 
in the cross-dispersion profile that occurs with time and temperature 
for different fibers. These profile changes can appear as both a trace position shift and 
a width change, and while small, are important. Between 
twilight flats spaced eight hours apart and through maximum dome temperature 
changes of ten Celsius degrees, we have measured trace shifts of 
up to 0.3 pixels and profile FWHM changes of 0.3\AA. 
Our goal was to limit this systematic to 10\% for any pixel in the flat. The FWHM 
variation already meets this criterion, but the maximum 
trace shift is too large by a 
factor of six. Moreover, although the trace shift also appears to be coherent between 
adjacent fibers on the CCD, it sometimes goes in opposite directions at the 
opposite ends of the fiber bundle, as if the traces are subject to a ``breathing mode.'' We have 
developed a heuristic solution that mitigates this problem. The core idea is to measure the offset 
over subsets of fibers, alter the flat fields to maintain the 
fiber-to-fiber and pixel-to-pixel patterns but resample the 
fiber profile to produce a shifted flat tailored to each exposure. 


\par For each pre-sky-subtracted 
data frame, the fiber centroids at each wavelength 
along the cross-dispersion direction are calculated with 
respect to the corresponding flat. These 
trace shift estimates are then median smoothed with their 
twelve nearest fibers on the CCD. Rather than presume a cross-dispersion 
profile shape, which displays non-Gaussian features, 
we use sinc interpolation to resample the profile. Linear interpolation 
fails to recover the strong curvature in this profile. In each fiber and each 
wavelength, the flat field is resampled at the fiber-specific estimated 
offset relative to the polynomial trace peak. However, additional smoothing is 
still required to leave pixel-to-pixel features 
unaltered. To do this, we run a boxcar smoother 
of eighty-one pixels along the dispersion direction 
for both the original flat field and the sinc resampled flat field. The 
biweight of each forms a pure profile model in the original and resampled frames, 
and the pixel-to-pixel variations are isolated in a separate image. 
The total, shifted flat is then formed by multiplying the pure, shifted 
profile model by the isolated pixel-to-pixel estimate. A final scaling is then applied to maintain the 
fiber-to-fiber throughputs and total flat normalization, as sinc interpolation does not 
automatically conserve flux. The use of these shifted flats rather than the original flats 
results in lower systematic errors and meets the goal of $<10$\% flat field profile error. 
\par
\begin{figure}
\centering
\includegraphics [scale=0.4,angle=-90]{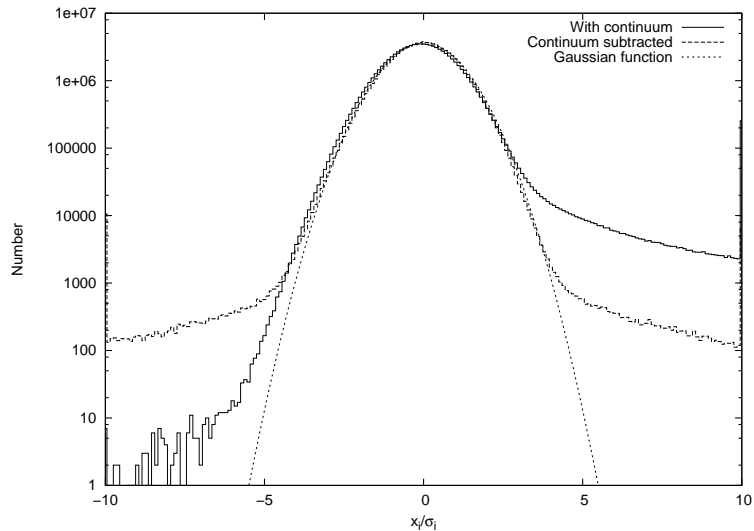}
\caption{A histogram of the ratio between the reduced data and error 
for all 87.8M indepedent spectral elements in this survey. Most 
elements only see the sky background with residuals 
consistent to a normally distribution noise model. 
A small but obvious fraction of the elements also see bright, positive 
signal from continuum sources. The normalized Gaussian 
function shows the expected distribution in the absence of 
signal and systematics. The influence of the 
fiber throughput systematic (\S \ref{sec_systemat}) likely 
broadens the distribution. This is most evident 
on the negative side which becomes nearer the normal 
distribution after background subtraction. The 
distribution after continuum subtraction appears 
much more symmetric and with a better matching 
width. The data wings at high and low ends represent the 
fiber positions with strong continuum. The 
boxcar-based continuum fitting is a rather 
crude tool that does not characterize all the 
continuum signal, but the emission 
line catalog is uncompromised by its use.}
\label{fig_dat_hist}
\end{figure}

 To capture any remaining systematics we have made a second, independent 
estimate of the error using the rms of the fifteen measurements that 
go into the final data combination (\S \ref{sec_comb}). This error 
estimate is itself noisy, but the ratio between this 
empirical error and our formal error over all pixels is 
useful as a diagnostic. We find the median of this ratio per frame is 
0-20\% above the random noise alone, and the median over all data 
is 5\%. Therefore, we increase the errors by this amount prior to the 
detection steps. Figure \ref{fig_dat_hist} shows the 
distribution of all 87.9 million independent datapoints divided 
by the error estimates of this dataset. Versions prior to 
and after continuum subtraction are shown. If 
the dataset were entirely without signal, if all the 
systematics were understood, and if all the noise 
were uncorrelated, the distribution should match 
the given Gaussian function with a dispersion of unity. 
Clearly  the distribution is asymmetric, distorted on the positive end by 
signal and the negative end presumably by the 
previously discussed fiber throughput variations. However, the 
continuum subtracted data with the fiber throughput 
variation removed are much 
more symmetric and show a distribution 
that is a much better match to the Gaussian width. Emission line 
objects are detected in the continuum-subtracted data, and the 
noise model is validated. 

\section{Emission line source selection}
\par The controlled selection of emission-line objects is the 
next step in producing this survey's catalog. The 
primary task of the detection process is to optimally use the 
source signal that has been distributed into, potentially, 
several fibers. The challenge is to push to a 
high completeness level at low S/N under a contamination 
constraint. The approach we adopt is to define emission-line detection 
seed apertures at a low S/N significance, test the 
combination of the seed apertures and all nearby fibers 
on sky, and allow the aperture to grow if the 
significance of the encompassed signal increases. 
The growth process is iterated. 
To understand the completeness and contamination rates 
of this method, we also present simulations with mock data. In 
similar datasets such as blind longslit spectroscopy \citep{Gil10} and 
grism spectroscopy \citep{Meu07}, detection algorithms based on 
data convolution have been used. We have 
tested this approach on our dataset, but found it 
inferior in completeness to our adopted technique (see \S \ref{sec_comp}). 
\subsection{Detection method}
\label{sec_cuts}
\par Several terms require definition before we describe the 
detection method. A fiber position carries a set of 
neighboring fibers, defined as all other
fibers offset by $\leq3$\arcsec\ in their center-to-center 
coordinates. The detection 
aperture starts with one fiber and, by iteration, 
is allowed to grow by accepting neighboring fibers. A 
detection aperture may be composed of multiple fibers and has its 
own set of neighbors, defined as the union of all 
neighbors to the current member fibers. The S/N of 
a potential emission line is calculated in a specific 
spectral window around the fit central wavelength. We 
define this detection window as spanning $\pm2\sigma_{res}$ 
where $\sigma_{res}$ is the dispersion of the 
VIRUS-P resolution element (2.2\AA). Within this window, 
data are summed and errors added in quadrature. Pixels that straddle the 
window are included by their fractional overlap. 
\par We begin with the fully calibrated spectra, errors, and fiber 
sky coordinates. First, a local continuum for each fiber is estimated 
and removed through a 200\AA\ wide biweight boxcar. Second, seed 
apertures are defined as all pixels that have 1$\sigma$ 
positive significance under a 6\AA\ wide boxcar smoothing. Seeds 
are merged when found in the same fiber and at contiguous wavelengths. 
Third, a Gaussian model is fit to each seed with 
variable width, wavelength, and 
intensity using a data window of 30\AA. We anticipate 
emission-line widths for LAEs to lie below the VIRUS-P spectral 
resolution, but the detection method is designed to be general 
to all line widths. We experimented with 
basing the detection aperture on the 
Gaussian function's fit width instead of the 
instrument's resolution, but simulations showed that the broad 
fits produced an unacceptable level of contamination. Fourth, fits with the seed apertures and 
each of the neighboring fibers are made. When making fits using 
multiple fibers, each fiber's emission-line intensity is allowed to vary, but constrained 
to a common wavelength and width. Fifth, if the inclusion of any prospective neighboring fiber 
increases the 
total S/N over a particular threshold, the fiber with the greatest increase is added to the 
detection aperture. Operationally, we use a threshold of $\Delta S/N=0.3$. Sixth, these steps are iterated until the 
apertures no longer grow or the aperture size reaches six fibers. The cut at six fibers is chosen 
because in the dither pattern, a 
point source can be equidistant from at most six fibers. Seventh, a final significance cut is made on the potential 
detections. If the detections had 
only been made using single, independent apertures, simple counting 
statistics could be used to meet the $<10$\% contamination goal. 
For example, when applied to the luminosity function of \citet{Gro07}, 
our S/N$\ge$5 cut and no galactic extinction implies that we should 
see 2.4 LAEs per VIRUS-P pointing under photometric conditions. 
Similarly, a VIRUS-P pointing (over six dithers) 
has 756k independent resolution 
elements, so a S/N$\ge$5 cut would deliver 8\% contamination. Unfortunately, the 
more complicated detection algorithm used here is not so straightforward to assess. 
While the growth steps will recover some sources that would otherwise be missed, they can also 
bundle noise from neighboring fibers. We therefore have made simulations of mock noise frames in order to 
optimize our selection thresholds. 
\subsection{False source tests}
\label{sec_false_det}
\par
\begin{figure}
\centering
\includegraphics [scale=0.4,angle=-90]{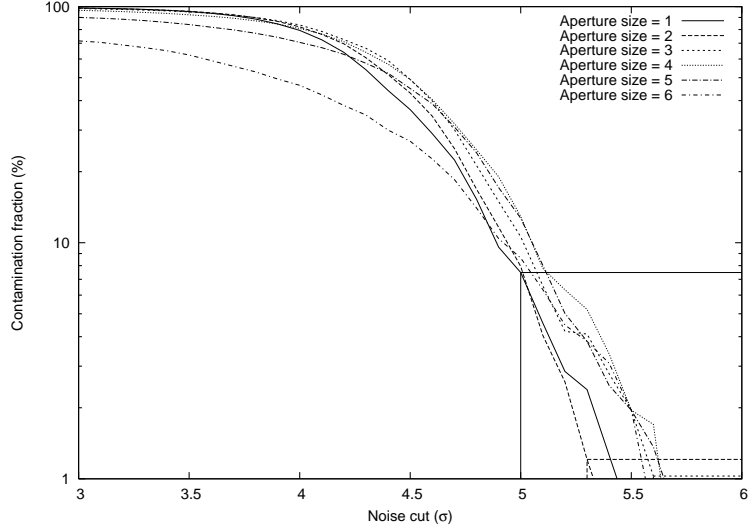}
\caption{The fractional contamination to the 
LAE sample by S/N cut. The detection method used is 
described in detail in \S \ref{sec_cuts}. The 
LAE predictions are the same as in \S \ref{sec_fluxcal}. 
Curves are given for differently sized collections of neighboring 
fibers. The growth of apertures was allowed 
whenever the S/N was increased by $>0.3$ with the 
inclusion of another fiber. 
At a high, constant S/N, the greatest contamination 
is produced by large apertures. The optimum sized aperture 
for a point source under all dither-source 
geometries is two at median. To optimize our selection, 
we make a staggered 
series of S/N cuts based on the number of fibers 
used, $N$, 
as S/N$>5.0+0.3\times(N-1)$. The horizontal and vertical 
lines show the evaluation points of this cut 
to the simulation curves. This procedure predicts 10\%$\pm$1.6\% 
noise contamination.}
\label{fig_false_curves}
\end{figure}

 To test for false sources, we began by simulating full, two-dimensional spectral data for twenty-five 
VIRUS-P fields using the observed 
median sky brightness. The mock data were made with noise realizations 
from the actual sky background and CCD read noise but were otherwise without sources. The 
fields were then analyzed for emission-line sources exactly as in \S \ref{sec_cuts} for 
all detections that reached S/N~$\ge3$. 
The number of spurious sources were then compared to the expected number of true LAEs 
\citep{Gro07} as a function of S/N cut, aperture, size, and survey depth. Evidence 
indicates that the LAE luminosity function does not evolve strongly at $z=3$ and higher 
redshifts \citep{Ouc08}, but there is less certainty about the rate of evolution over the lower 
redshifts that we also probe \citep{Nil09,Cas10}. The results of this 
analysis are shown in Figure 
\ref{fig_false_curves}. Interestingly, at higher S/N the larger apertures 
begin to contribute the most contamination. Under the typical survey observing 
conditions and the majority ($\sim$80\%) of 
source-fiber geometries, the optimal number of fibers to include in a 
simultaneous detection is two. Point source emission objects, which we anticipate 
most LAEs to be \citep{Bond10}, rarely ($<$~5\%) benefit from fiber apertures of four or more. 
Conversely, the S/N for extended low-$z$ objects is often improved by including more 
fibers, so we should not avoid large apertures altogether. 
Finally, it is clear that a common cut of S/N~$\ge$~5 
would deliver an unacceptably high rate of (60\%) contamination. The situation can be 
improved by varying the S/N limit as a function aperture size. The choice we adopt is for an 
aperture of $N$ fibers to have a S/N cut of $S/N\ge5+0.3\times(N-1)$. Under the 
assumption of a non-evolving LAE luminosity function, we predict a 10\%$\pm$1.6\% contamination 
of spurious sources to the LAE sample. We project there are 
17$\pm$3 spurious sources in the data catalog. A sample essentially free of 
contamination can be produced by using this catalog with a S/N$>6$ cut, which 
by the limited number statistics of these simulations may contain 0$^{+5}_{-0}$ spurious sources.
\par In addition we have also performed an empirical test for spurious sources by 
analyzing the inverse of the survey data frames. All sources with a detected 
continuum were masked (so that we would not find the inverse of absorption features as spurious sources), 
and our detection algorithm was re-run. This analysis found 7 spurious sources 
in 28 fields; a rate that is significantly lower 
than that estimated from the simulations. This suggests our estimate of the systematic 
error is conservative and the true contamination fraction likely lies somewhere between 4-10\%.

\subsection{Completeness tests}
\label{sec_comp}
\par Not every source at the flux limit of Figure \ref{fig_flux_lim} will be 
recovered by the detection scheme. Beyond the usual statistical 
fluctuations introduced by noise, different source 
positions and seeing variations will cause the signal to be distributed over 
a different numbers of fibers and cause varying fractions of light to be lost to 
the gaps between fibers. While this partial image sampling is 
an undesirable feature, IFS mitigates these uncertainties 
compared to serendipitous longslit observations \citep{Rau08,Lem09,Cas10}, where the slit 
losses can range (nearly uniformly) from 0-100\%.
\par
\begin{figure}
\centering
\includegraphics [scale=0.4,angle=-90]{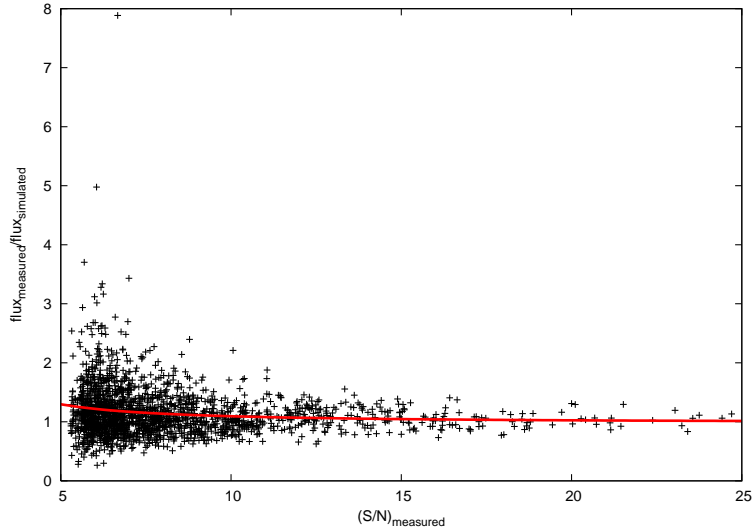}
\caption{The ratio of measured and input fluxes from 
point source simulation and a least-squares fit (\S \ref{sec_comp}). The 
curve-of-growth method is used here to measure emission line flux. The 
upturn at low S/N is expected from the Eddington 
bias. This ratio can be applied to the LAE flux measurements 
of Table \ref{tab_line_list1} for some applications. The average correction is fit as 
flux$_{\mbox{measured}}$/flux$_{\mbox{true}}=
$0.98+0.74/(S/N)$_{\mbox{measured}}$+4.27/(S/N)$_{\mbox{measured}}^2$}
\label{fig_frat_sim}
\end{figure}

\begin{figure}
\centering
\includegraphics [scale=0.4,angle=-90]{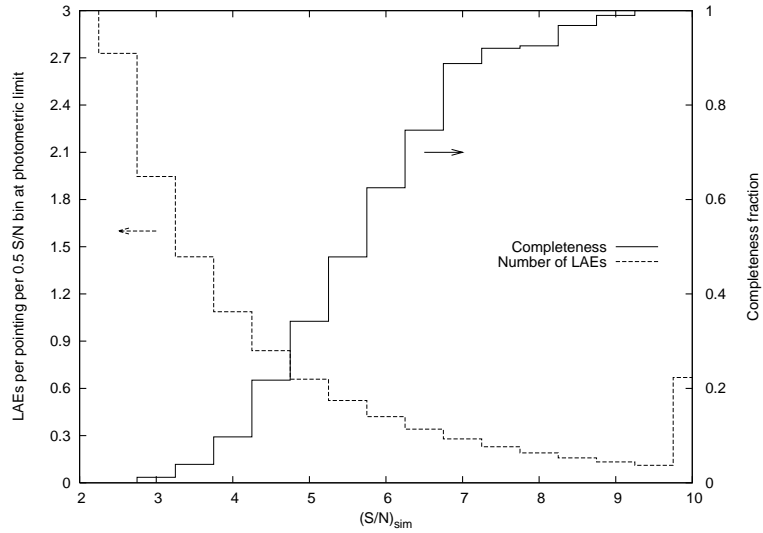}
\caption{The survey's catalog completeness function for 
multiples of the flux limit. When combined, this function, 
the survey's photometric wavelength-based flux limit 
in Figure \ref{fig_flux_lim}, and the actual mosaic 
pattern with photometric calibration in Figure \ref{fig_sens} 
define the limits necessary for volume and luminosity 
function calculation \citep{Bla10}.}
\label{fig_compl}
\end{figure}

 We have simulated our completeness limit using 25 mock 
fields of full, two-dimensional data with noise generated from the 
mean McDonald sky spectrum and the CCD read noise. Each simulated 
image contained 3000 emission-line sources randomly chosen in position 
and wavelength, but constrained to avoid object blending and 
spaced by the seeing from the IFU edges. We used the same detection routines 
as for the real data. For all these simulations, the seeing was held 
constant at the survey's 1\farcs5 FWHM median. 
These mock sources were modelled as spectrally unresolved point sources with 
fluxes randomly drawn from an unevolving \citet{Gro07} LAE luminosity 
function over the luminosity range 
$41.5<\log L(\mbox{erg s$^{-1}$})<44.5$ where the lower bound was chosen to 
yield S/N=0.5 over most of the wavelength range. Figure 
\ref{fig_frat_sim} compares our simulated emission-line fluxes to the 
fluxes that were measured. As the S/N decreases, the error in our measurements 
increases. Moreover, at the faintest limits, there is a slight 
systematic trend, with the measured fluxes being over-estimated. This is the 
well-known Eddington \citep{Edd13,Edd40} correction which, if ignored, 
can lead to an under-estimate of a luminosity function's slope. 
The least-squares fit shown in the figure 
will be used to statistically correct all our LAE fluxes prior to luminosity function 
computation. The completeness results are shown in Figure \ref{fig_compl}. 
We reach 50\% and 95\% corrected completeness at 5.6$\sigma$ and 8.3$\sigma$ respectively. Compared to 
a step function completeness limit at $S/N>5$ at the photometric limit of this survey which we 
consider the ideal goal, the 
number of detected LAEs is degraded by 13\%. The long, low S/N tail helps mitigate the loss of 
objects to the non-ideal completeness.
\par
\begin{figure}
\centering
\includegraphics [scale=0.4,angle=-90]{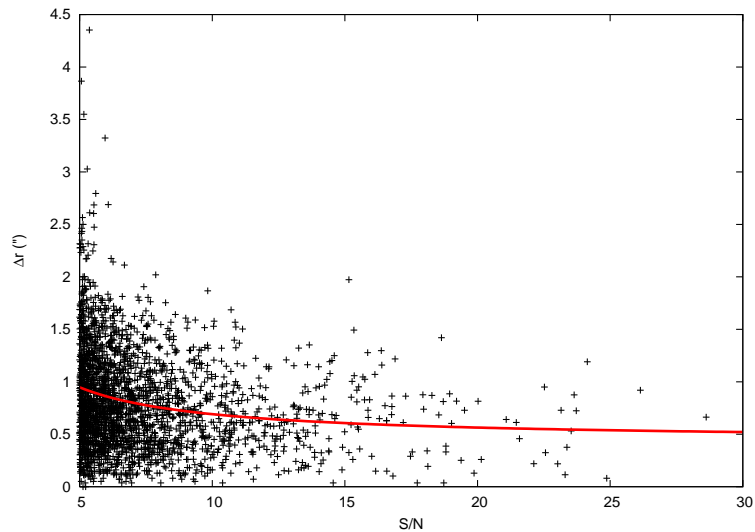}
\caption{The distribution of simulated source 
positional errors under VIRUS-P 
sampling patterns with a fit to the dispersion 
as a function of measured S/N. The maximum liklihood 
fit to the peak of a Rayleigh distribution 
gives $\sigma=0\farcs348+2\farcs04/(S/N)$.}
\label{fig_sim_astrom}
\end{figure}

 Our source simulations also allow us to quantify the statistical astrometric 
error as a function of S/N. This is an important ingredient to our 
algorithm for associating VIRUS-P emission-line objects with sources 
found in broadband imaging (see \S \ref{sec_class}). 
If we adopt a Rayleigh dsibtribution for the form of the radial errors, i.e. 
$\sigma=a+b/(S/N)$, then a maximum likelihood fit for the coefficients yields 
$a=0\farcs348$ and $b=2\farcs04$. 
Figure \ref{fig_sim_astrom} shows this relation, with the individual measurements overplotted.
\par
\begin{figure}
\centering
\includegraphics [scale=0.4,angle=-90]{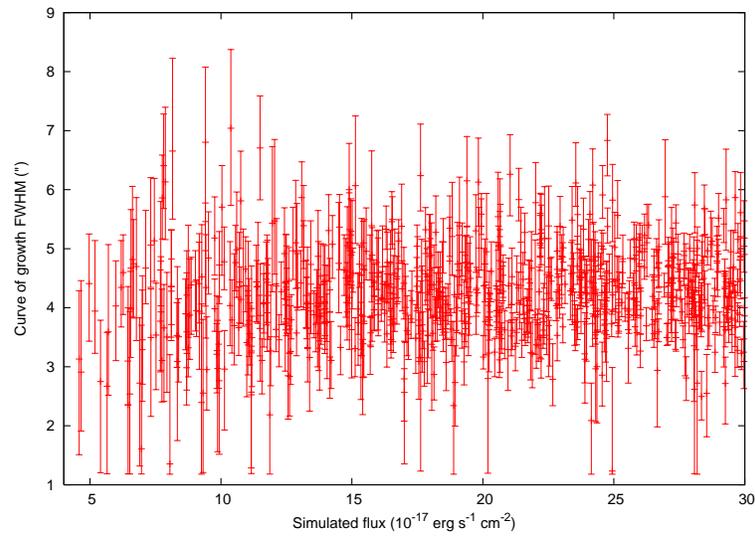}
\caption{The distribution of measured source 
sizes in simulated data. The simulation input 
source sizes were drawn from the survey 
seeing distribution. The large fiber sizes 
set a large resolution threshold. Based on the 
99.7\% confidence interval, we only claim significantly 
resolved measurements for curve-of-growth 
FWHM sizes $>6.81$\arcsec.}

\label{fig_sim_sz}
\end{figure}

 The large VIRUS-P fibers lead to poor
spatial resolution. Nevertheless, we have also simulated one mock 
field of 3000 point sources at and above the survey's 
flux limit and seeing distribution in an effort to quantify the 
minimum resolvable source size. To do this, we modeled the seeing FWHM distribution as a 
Gaussian function centered on 1\farcs5 with a 
dispersion of 1\arcsec\ but truncated below 1\farcs2. With the 
oversampled pattern of dithers, we expect the Nyquist limit to be 
near the diameter size of a fiber. The same curve-of-growth 
photometry routines as described in \S \ref{sec_fl_meas} 
were used to measure the sizes of simulated point sources. Figure \ref{fig_sim_sz} shows 
the distribution of emission line flux and measured size. The distribution 
is mostly flat with either flux or source S/N. Based on the simulation, we 
label a threshold of 7\farcs5 as the resolution limit of our survey. This can be 
compared to the usual definition for Ly$\alpha$ blobs, i.e. emission over an isophotal area of $>16$\sq\arcsec\ 
at a certain surface brightness threshold. The Ly$\alpha$ blob surveys of \citet{Mat04} 
and \citet{Yan10} used thresholds of $2.2\times10^{-18}$ erg s$^{-1}$ cm$^{-2}$ arcsec$^{-2}$ and 
$5\times10^{-18}$ erg s$^{-1}$ cm$^{-2}$ arcsec$^{-2}$, respectively. Our HETDEX pilot survey should 
detect many Ly$\alpha$ blobs based on this flux limit, but will only be able to resolve the 
very largest objects. The full HETDEX survey will have $\sim$3$\times$ better 
spatial resolution. 
\subsection{Line flux measurement}
\label{sec_fl_meas}
\par A source's detection aperture 
described in \S \ref{sec_cuts} does not contain the total source flux. 
The imposed S/N cut omits some fraction of the flux in the detection 
aperture; this fraction is a function of source strength and orientation to the 
fiber dither pattern. In order to determine an unbiased emission 
line flux in the presence of these complications, we describe here a 
curve-of-growth procedure used to measure a source's total line flux after detection. 
While other total flux estimators are possible, we advocate this 
method as generally robust against the range of sizes and morphologies 
encountered in the survey and the rather 
large astrometric errors and seeing variations inherent in this dataset. 
The algorithm is similar to curve of growth (CoG) \citep{Ste90} fits previously 
developed for CCD imaging photometry, but is new to spectrophotometry.
\par
\begin{figure}
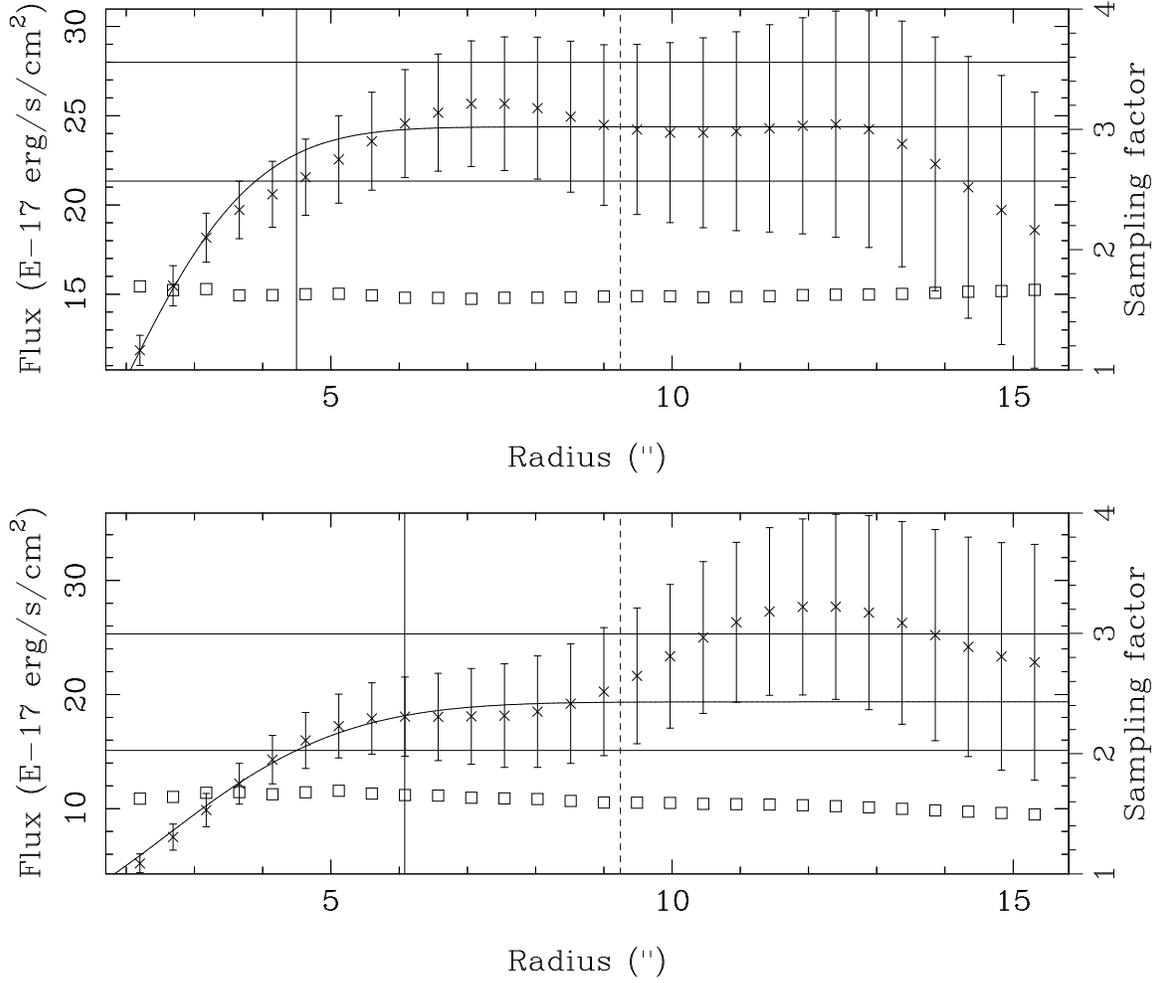

\centering
\subfigure{\includegraphics[scale=0.7,angle=-90] {cog_LAE.eps}}
\\
\subfigure{\includegraphics[scale=0.7,angle=-90] {cog_lowz.eps}}
\caption{The data, fits, and final evaluation for the 
curve of growth line flux measurements of a LAE (\textit{top}) and 
an [OII] emitter (\textit{bottom}). In both, the open square symbols 
display the cumulative sampling factor on the right hand scale. 
The sampling factor is written in Equation \ref{eq_fluxobs} as 
$N(\Delta r<r_{fib},r,\theta)$ and is the average number of 
fibers overlaying the surface enclosed in radius $r$. 
The points with errors show the estimated cumulative flux on the left hand 
scale. The vertical dotted lines mark where the fit is truncated. This 
threshold has been selected with consideration towards being significantly 
larger than the widest objects found and small enough to limit 
unnecessary noise. The 
vertical solid line marks the radius to the fit where 90\% of the flux in 
enclosed. The horizontal solid lines show the 
1$\sigma$ confidence of the fit's normalization. The top fit 
returns a total flux of $24^{+3.6}_{-3.0}\times10^{-17}$ 
erg s$^{-1}$ cm$^{-2}$. The bottom fit returns a total flux of 
$19^{+6.0}_{-4.3}\times10^{-17}$ erg s$^{-1}$ cm$^{-2}$. The 
errors are correlated on the displayed scale, but the Monte Carlo 
fit varies the data from each fiber independently to 
generate proper errors.}
\label{fig_cog}
\end{figure}

 We begin a flux measurement by considering the positions, central 
wavelengths, and line widths ($\sigma_{det}$) obtained from the emission-line detection 
algorithm described in \S \ref{sec_cuts}. A circular aperture 
is formed around the centroid emission-line position of variable radius. Fibers 
overlapping this aperture are given fractional weights determined by their 
enclosed areas. Specifically, we form fifteen apertures linearly spaced between 
radii 2\farcs2 and 9\farcs0. In each aperture, the enclosed fibers 
have their continuum-subtracted data summed and errors summed in quadrature 
for wavelengths within $\pm2\sigma_{res}$ of the 
detection wavelength. A spectral correction factor is 
defined as the flux fraction of a Gaussian line profile that 
falls within the fixed, spectral window defined by Equation \ref{eq_fluxcorr}.  
\begin{equation}
\label{eq_fluxcorr}
f_{spec,corr}=\mbox{erf}(\sqrt{2}\sigma_{res}/\sqrt{\sigma_{res}^2+\sigma_{det}^2})
\end{equation}
Note that the fluxes returned by directly summing all fibers in a circular aperture 
of radius $r$, $\hat{f}(r)_{raw}$, may oversample or undersample 
the source flux depending on the data completeness and 
overlap regions of mosaic. For example, the ideal six dither pattern 
produces an oversampling of very near two. Let the number of 
fibers at a particular position lying within one 
fiber radius, $r_{fib}$, be $N(\Delta r<r_{fib},r,\theta)$ in polar coordinates. 
Equation \ref{eq_fluxobs} gives the raw flux measured for 
arbitrary sampling of a source with total flux $f_{total}$ and normalized profile $P(r,\theta)$; 
$f(r)_{samp}\equiv\int^{r}_{0} f_{total} P(r,\theta) r dr d\theta$ 
is an estimate of the cumulative flux corrected for sampling. This approximation 
is correct when $N(\Delta r<r_{fib},r,\theta)$ does not systematically 
depend on $r$, which is nominally true for the randomly positioned 
observations presented here. The approximation is necessary to cleanly estimate an 
unbiased flux without knowing the exact profile. 

\begin{equation}
\label{eq_fluxobs}
\begin{split}
\hat{f}(r)_{raw}=\int^{r}_{0} f_{total} N(\Delta r<r_{fib},r,\theta) 
\times P(r,\theta) r dr d\theta \approx \\
\hat{f}(r)_{samp} \times \frac{\int^{r}_{0} N(\Delta r<r_{fib},r,\theta) r dr d\theta}{\pi r^2}
\end{split}
\end{equation}
We fit, by nonlinear least squares minimization, a cumulative two-dimensional 
Gaussian function, $A_{CoG}\times(1-e^{-0.5r^2/\sigma_{CoG}^2})$, to the 
highly correlated distribution $\hat{f}(r)_{samp}$, 
where we enforce the limits $1\arcsec<\sigma_{CoG}<10\arcsec$. In addition, we 
create Monte Carlo realizations by varying each fiber's intensity 
from the best-fit model. The 
CoG datapoints are highly correlated, so we took care to estimate the errors 
from the uncorrelated data of each fiber. The final, total 
flux estimate is given by Equation \ref{eq_fluxfin}, with errors similarly propagated from the raw data and the 
uncertainty in $\sigma_{CoG}$. Figure \ref{fig_cog} gives curve of growth examples for an [OII] emitter and a LAE. 
\begin{equation}
\label{eq_fluxfin}
\hat{f}_{total}=A_{CoG}/f_{spec,corr}
\end{equation}
\par
\begin{figure}
\centering
\includegraphics [scale=0.4,angle=-90]{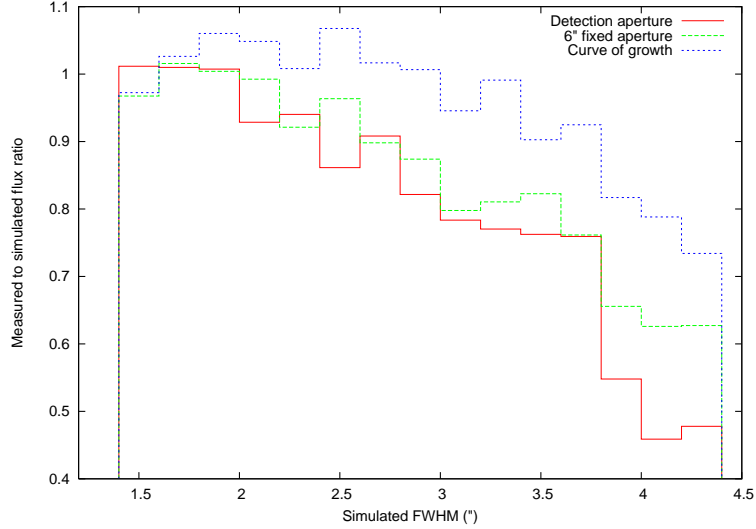}
\caption{Binned ratios of measured and input 
fluxes for simulated data under a range of source sizes. 
The curve-of-growth flux estimator is preferred as the least biased 
for extended sources compared to either the 
detection set of fibers or a fixed radius set of 
fibers as the photometry aperture.}
\label{fig_test_CoG}
\end{figure}

 We tested the reliability of the curve-of-growth flux measurement, particularly for 
correlated errors with the source size, by using the simulated data discussed in 
\S \ref{sec_comp}. We first measured the 
flux from the fibers chosen as the detection aperture (\S \ref{sec_cuts}), and 
compared this to the simulated flux. The mean and dispersion of the 
measured-to-simulated ratio are 
0.93 and 0.31; unsurprisingly, the fluxes are systematically underestimated. 
Next, the set of all fibers within 6\arcsec\ of the detected position 
was used as the flux aperture. This reduced the scatter found by the fixed aperture 
method, but a systematic error still remained with a mean of 0.94 and dispersion of 
0.20. Finally, the 
curve-of-growth flux measurement was considered. Under this procedure, the bulk 
systematic flux measurement error vanished, giving a mean of 1.00 while still 
maintaining a low dispersion of 0.23. 
All three flux estimation methods are shown in Figure \ref{fig_test_CoG} 
against the simulated source size. A systematic offset with input 
source size can be seen for all cases, but the curve-of-growth 
photometry is preferred as the least biased method investigated. 

\section{Source classification}
\label{sec_class}
\par An emission-line galaxy catalog is of limited value without secure redshift identifications. 
Unfortunately, the uncertainty in identifying single emission lines is a common hindrance to 
high-redshift galaxy surveys \citep[e.g.,][]{Ste00}. We here describe the 
two steps necessary to robustly assign 
redshifts to the emission-line catalog. Tables \ref{tab_line_list1} and \ref{tab_line_list2} present the 
catalogs. We give the detailed description of these tables in \S \ref{sec_samp}. We further summarize the 
statistics of commonly found objects and compare the sample to other works 
where available.
\subsection{Spectral classification}
\label{sec_spec_clas}
\par As mentioned in \S \ref{sec_ancil}, the presence of multiple, strong emission-lines 
can be used to identify some low-$z$ objects, but the absence of such lines is not 
sufficient evidence to classify a source as an LAE. We begin all source classifications 
by cross-correlating the primary emission line at various assumed redshifts to 
other bright, expected emission lines. We automatically search all the detection spectra for 
MgII2798, [OII]3727, H$\gamma$4341, H$\beta$4861, [OIII]4959, and [OIII]5007 
assuming the detected line to be, variously, [OII]3727, H$\beta$4861, 
[OIII]4959, and [OIII]5007. At high redshift we test Ly$\alpha$ for the 
presence of CIV1549. We have manually tried using the other, 
commonly weaker lines as confirmation of the primary detection, but have found only 
two cases of interest. For emission line index 4 of Tables \ref{tab_line_list1} and \ref{tab_line_list2},
the CIII]1909 line is detected with the also-significantly detected
[OII]3727 line of index 5. For emission line index 85 of the same tables, 
the broad MgII2798 line is brighter than the also-significantly detected 
[OII]3727 line. We have also mis-identified index 400 as an [OII] emitter; it is known 
to be an [OIII]5007 emitter from the literature \citep{Bar08}, but we find no other 
detections at other wavelengths.
\par
\begin{figure}
\centering
\includegraphics [scale=0.6,angle=-90]{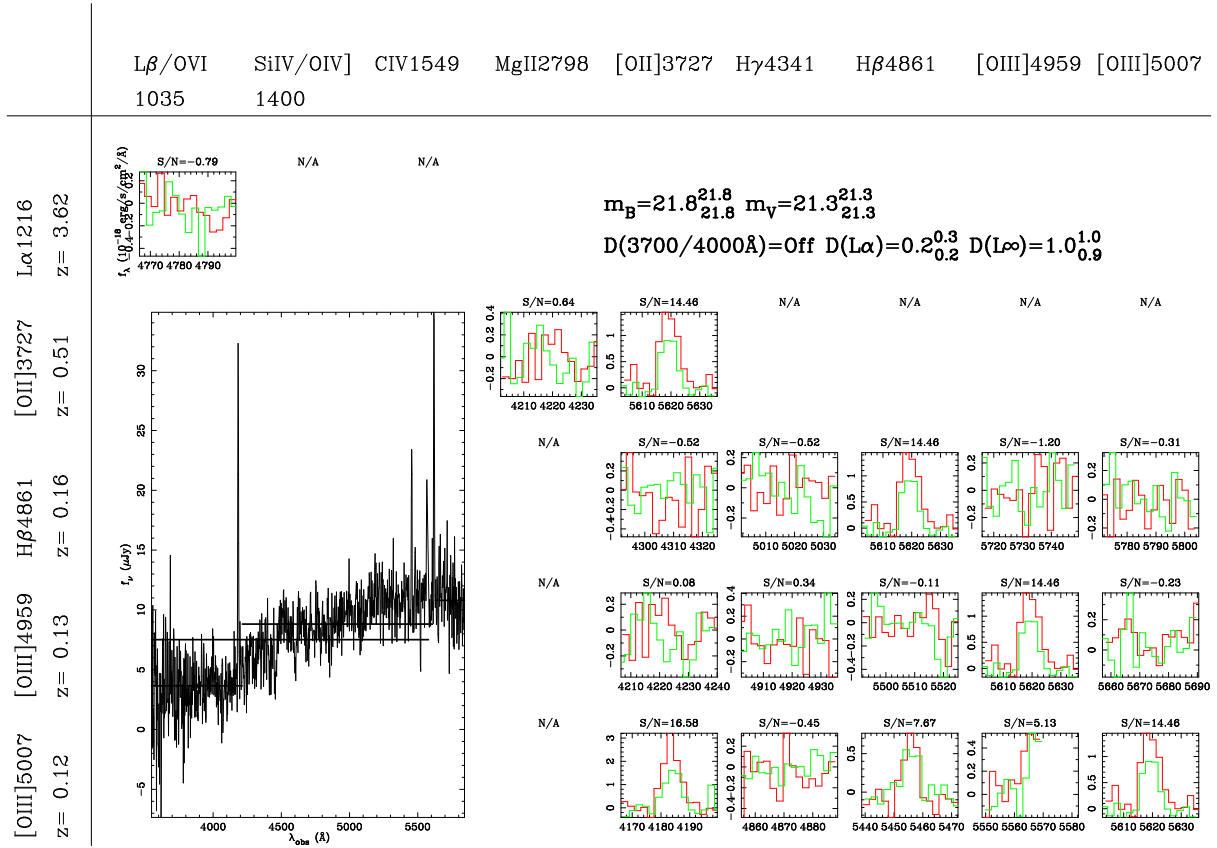}
\caption{The attempts to find matching lines to an [OIII]5007 detection 
in index 323 of Table \ref{tab_line_list1}. The detection is 
formed from two fibers represented independently with the 
red and green lines. 
The leftmost column shows the prospective 
identifications of our originally detected line at 5619\AA. For 
each prospective identification, we attempt fits to the emission-line 
possibilities in the top row. In this case, matches to [OII] and 
H$\beta$ both give a clear identification. The [OIII]4959 is 
detected but overlapping with the mask around 5577\AA. This 
technique only rarely gives positive evidence for Ly$\alpha$ 
classification by ruling out low$-z$ emission line 
combinations because our wavelength bandpass is not much larger than common 
bright optical emission-line spacings. However, is often useful in 
classifying transitions between low redshift options as in this case. The 
full spectrum is shown in the large window to the left. Various 
continuum regions are evaluated by assuming the primary emission line to be 
Ly$\alpha$ and [OII] shown by the horizontal lines. The continuum 
fits are used to look for various breaks as calculated in the 
upper right. This galaxy is also identified by significant 
detections of H$\beta$, [OIII]4959, and [OII] in entries 
325, 326, and 327 of Table \ref{tab_line_list1}.}
\label{fig_redet}
\end{figure}

\begin{figure}
\centering
\includegraphics [scale=0.6,angle=-90]{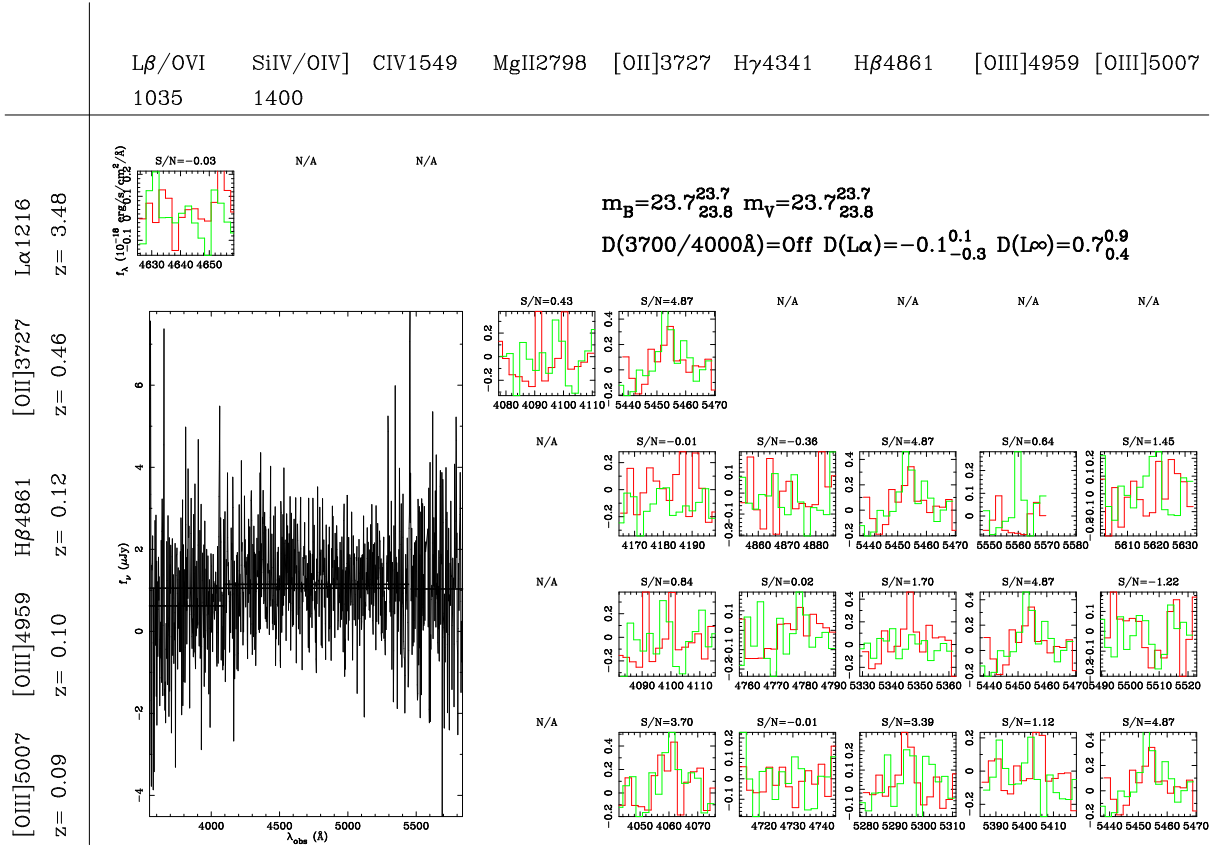}
\caption{The automated search for multiple emission line in 
index 430. The format is the same as in Figure \ref{fig_redet}. 
This is one of the few cases where the consideration of marginally 
significant counterpart lines aids the classification. The 
primary detection is revealed to be [OIII]5007, with 
marginal detections in [OII] and H$\beta$ that did not 
make the primary emission line catalog.}
\label{fig_redet2}
\end{figure}

 A demonstration of this 
cross-correlation process for a multiple-emission-line source is shown in Figure \ref{fig_redet}. 
We find only two cases where the correlation against data below the catalog 
signal-to-noise cut aids classification as shown in Figure \ref{fig_redet2}. The first is 
emission line index 234, which is 
formally a single emission line detection. However, we find that an identification 
of the primary line with [OIII]5007 leads to a S/N=3.2 
detection at the wavelength of [OII], a S/N=5.1 detection at H$\beta$. and 
a S/N=3.6 detection at [OIII]4959. The second case is 
emission line index 430 which is
also a single emission line detection. We again find 
that an identification
of the primary line with [OIII]5007 leads to a S/N=3.7
detection at the wavelength of [OII] and a S/N=2.9 detection at H$\beta$.
In practice, the primary utility in the emisison line cross-correlation is to 
discriminate between various low-$z$ possibilities 
with high S/N detections. 
\subsection{EW-based classification}
\label{sec_EWrule}
\par In any LAE survey at sufficient redshift, the most likely contaminants 
are [OIII]5007 and [OII]3727. Many of the former objects can be identified by the 
presence of [OIII]4959 or H$\beta$. The latter may be identified by either splitting 
the [OII] doublet, or by using line equivalent width as a discriminant \citep{Cow98}. 
Since we lack the resolution to split [OII]3727, we follow \citet{Gro07} 
and require LAEs to have EW$_{\mbox{rest}}>20$\AA\. A number of different EW 
estimators are possible with measurements in many filters. 
We look at two ways to estimate the EW using broadband data, and conclude that the cleanest 
selection of LAEs is obtained when the $R$-band data is used alone. 
\par
\begin{figure}
\centering
\subfigure{\includegraphics[scale=0.3,angle=-90] {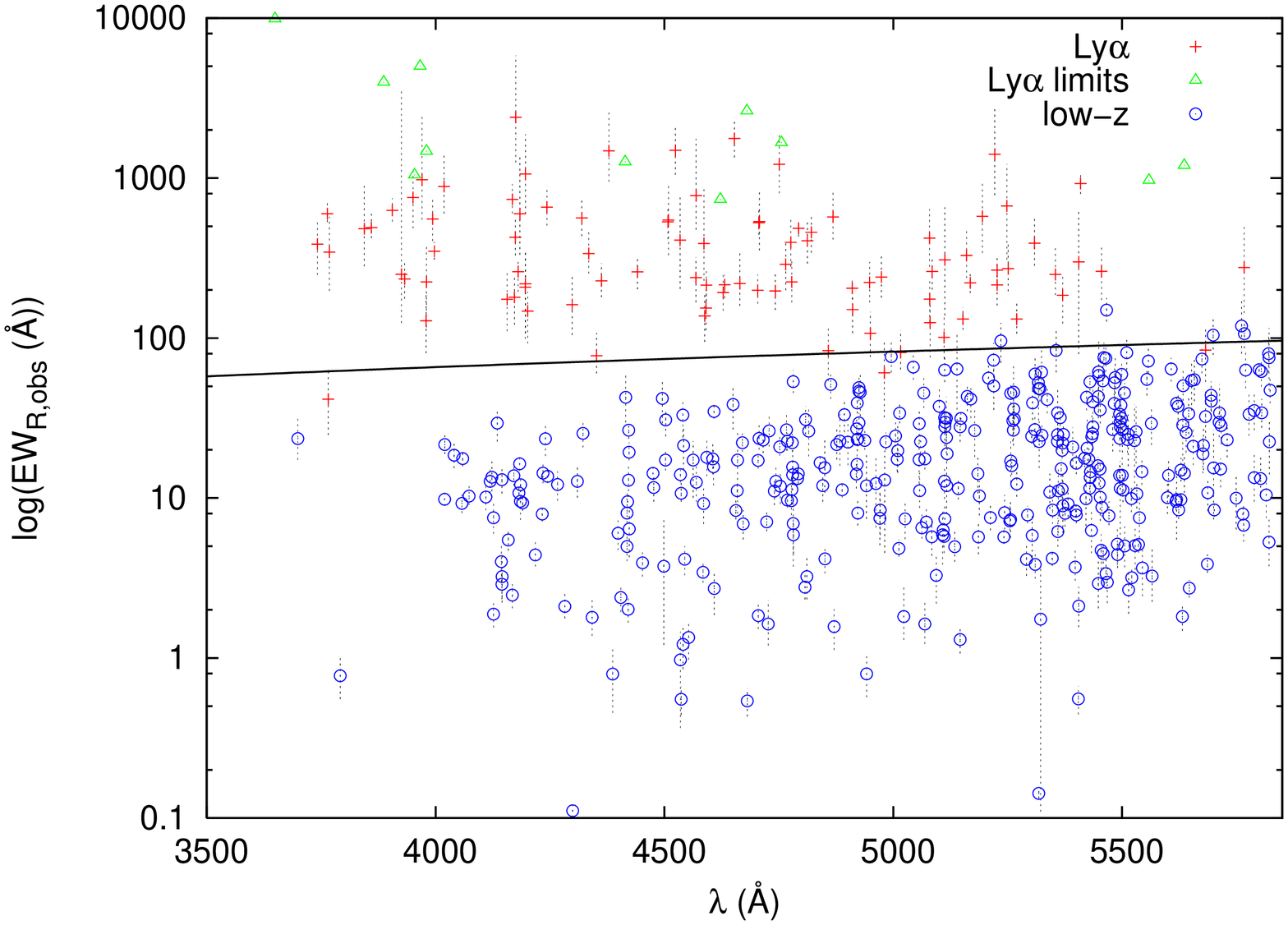}}
\subfigure{\includegraphics[scale=0.3,angle=-90] {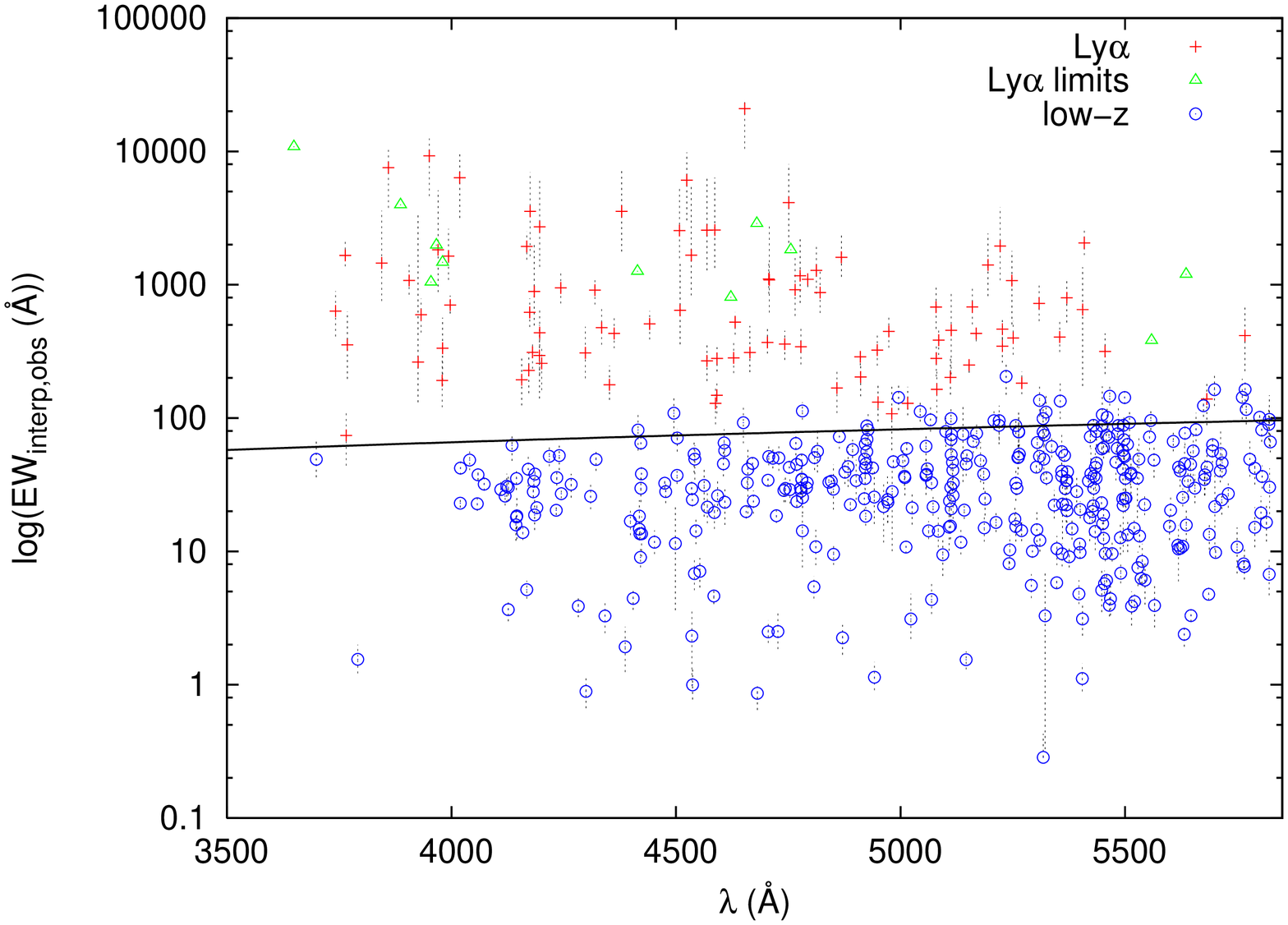}}
\caption{The distribution of observed wavelength 
and observed EW. The line marks the usual 
EW cut used in narrowband imaging and adopted here. 
Exceptions to the EW selection are discussed in \S \ref{sec_EWrule}. 
\textit{Left} Continuum estimated only from the $R$-band photometry 
(or the i'-band in MUNICS). \textit{Right} Continuum estimated from 
interpolation with the two nearest filters bounding each emission line's 
wavelength.
}
\label{fig_lamb_EW}
\end{figure}

\begin{figure}
\centering
\subfigure{\includegraphics[scale=0.3,angle=-90] {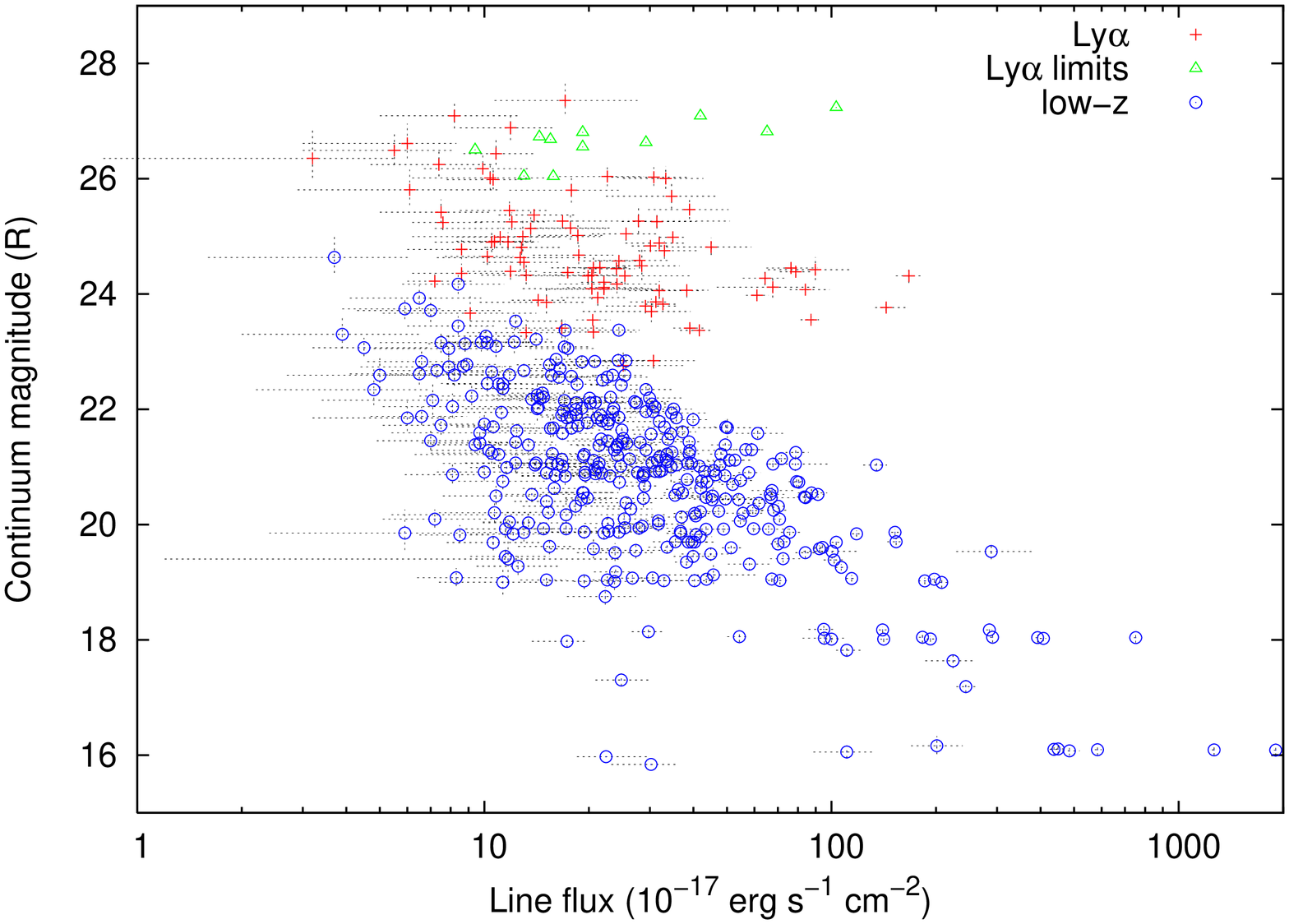}}
\subfigure{\includegraphics[scale=0.3,angle=-90] {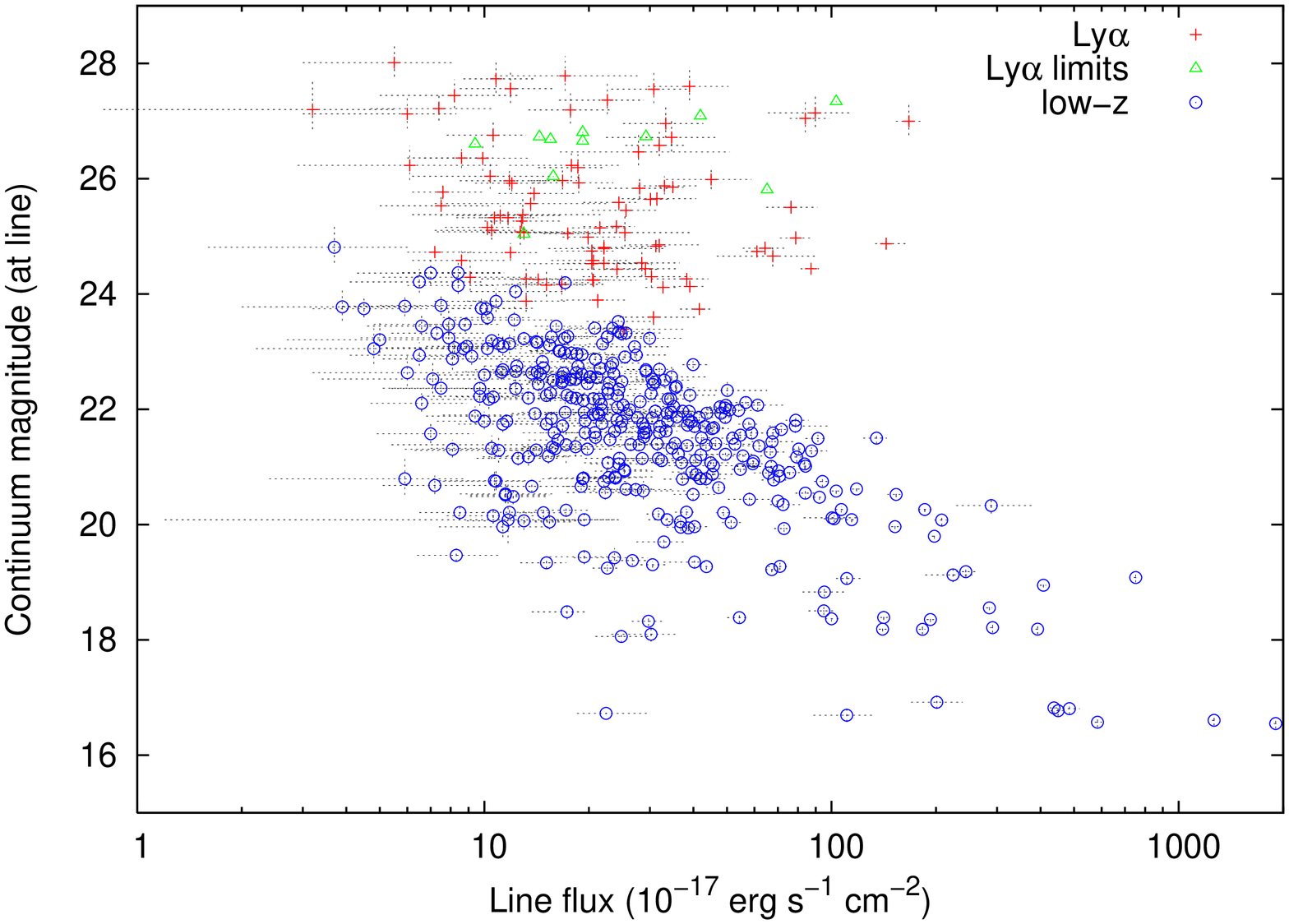}}
\caption{The distribution of emission line flux and broadband magnitude. 
\textit{Left} Continuum estimated only from the $R$-band photometry 
(or the i'-band in MUNICS). \textit{Right} Continuum estimated from 
interpolation with the two nearest filters bounding each emission line's 
wavelength.
}
\label{fig_flux_EW}
\end{figure}

 The observed wavelengths and EWs are shown in Figure \ref{fig_lamb_EW}. Emission lines 
without counterparts are shown as limits. We calculate the EW first by using the 
nearest-available filter that lies redward of the entire sample. For XMM-LSS, GOODS-N, 
and COSMOS, this is the $R$-band. MUNICS 
lacks an $R$-band image, so we used i'. 
The redward choice is important to avoid attentuation by the intergalactic medium (IGM) 
for these data and the Lyman break. Although there may be some diversity 
in LAE dust content \citep{Fin09}, it appears most LAEs at our redshifts of interest have 
only small amounts of dust and exhibit flat continua \citep{Gaw07,Gua10,Bla10}. Of course, 
low redshift, star-forming galaxies may also exhibit flat continua or, more likely, some level of a 
Balmer/4000\AA\ break, but by extrapolating the continua from the $R$-band, the 
low redshift EWs will be somewhat underestimated while the LAE EWs should remain 
unaffected. Still, while such a property is beneficial to the classification process, 
an unbiased EW is also desireable for physical studies. So, we next calculate the 
EWs in the right panel of Figure \ref{fig_lamb_EW} by interpolating each emission-line 
with the two nearest, bounding broadband filters. Clearly, the 
high and low EW populations have more overlap in the interpolated EW measurements. 
For this reason, we adopt the $R$-band EW in our classification scheme. 
Figure \ref{fig_flux_EW} shows the emission 
line flux against continuum magnitude for each emission line. We have also checked the 
GALEX \citep{Mar05} GR4/GR5 database for all objects. None of the LAE classified objects 
are GALEX sources, while most of the low-$z$ classified objects do have 
counterpart GALEX detections. 
\par There are nine objects for which we make exceptions: four low EW 
objects we identify as LAEs and five high EW sources we believe are low redshift 
interlopers. 
\par 1) The lowest wavelength exception is observed at $\lambda$=3765.6\AA\ with 
EW$_{\mbox{obs}}=41_{-17}^{+21}$\AA\ as 
index 313. If this were [OII], the galaxy would be extremely nearby (45 Mpc) 
away and have M$_{R}=$-10.5. The photometric redshift 
of \citet{Ilb09} suggests the line to be Ly$\alpha$ and excludes all the low-$z$ options 
with 95\% confidence. 
\par 2) The next low EW object is in the MUNICS 
field as index 51 and has m$_{i'}$=23.7. The detected wavelength is 
4981.6\AA\ with EW$_{\mbox{obs}}=61_{-29}^{+38}$\AA. 
The case for this object is not terribly strong, but the dim continuum and 
lack of a GALEX detection suggest this to be an LAE. 
\par 3) The third low EW object 
is in the GOODS-N field as index 447 at wavelength 5017.2\AA\ with 
EW$_{\mbox{obs}}=81_{-18}^{+31}$\AA. It was originally listed as an 
Lyman Break Galaxy (LBG) in \citet{Ste03}, but no redshift measurement exists in the 
literature. The counterpart has m$_{R}$=24.2. 
\par 4) The final low EW object is 
in the MUNICS field as index 92 at wavelength 5683.3\AA\ with
EW$_{\mbox{obs}}=84_{-31}^{+34}$\AA. The counterpart has 
m$_{i'}$=23.3, but no GALEX detection. Again, this is a borderline classification. 
\par Next, we consider the five high EW objects reclassified 
as being at low redshifts. 
\par 5) The first high EW low$-z$ object is 
in COSMOS as index 289 at wavelength 5235.9\AA\ with
EW$_{\mbox{obs}}=96_{-26}^{+28}$\AA. It does not have a GALEX detection, but 
the counterpart has m$_{R}$=23.4. The photometric redshift
of \citep{Ilb09} suggests the line to be [OII] and excludes all other 
reasonable options with 95\% confidence. 
\par 6) This COSMOS object is index 234 with m$_{R}$=23.4, 
wavelength 5466.7\AA\, and EW$_{\mbox{obs}}=149_{-23}^{+28}$\AA. 
As discussed in \S \ref{sec_class}, the source 
shows low significance emission lines such that the primary detection is likely 
[OIII]5007. Such an identification is possible since unlike [OII], 
[OIII]5007 can have extremely high EWs \citep{Hu09}. The source also has a GALEX detection. 
\par 7) This GOODS-N object is index 356 at m$_{R}$=22.8, wavelength 5700.5\AA\, and 
EW$_{\mbox{obs}}=104_{-26}^{+32}$\AA. It has a GALEX detection and has a 
measured redshift in \citet{Bar08} as being from [OII] emission. 
\par 8) This is index 439 from the GOODS-N field with m$_{R}$=24.2, 
wavelength 5762.4\AA\, and EW$_{\mbox{obs}}=119_{-39}^{+52}$\AA. The 
object is detected with GALEX. 
\par 9) This is index 94 from the MUNICS field with 
m$_{i'}$=21.0, wavelength 5768.4\AA\, and EW$_{\mbox{obs}}=107_{-20}^{+22}$\AA. 
The object is detected with GALEX.
\par We next review the likely levels of 
contamination in the LAE sample from low redshift objects based on previous studies. 
The frequency of EW in bright, rest-frame-optical 
lines at low redshift has been studied in \citet{Ham97}, \citet{Hog98}, 
\citet{Tre98}, \citet{Sul00}, \citet{Gal02}, and \citet{Tep03}. 
By combining the observation that $\sim$2\% of [OII] emitters have EW$_{\mbox{rest}}>60$\AA\ 
\citep{Hog98} with the $0<z<0.4$ [OII] luminosity function of \citet{Sul00} and assuming no 
redshift evolution of either the [OII] or LAE EW distributions \citep{Gro07}, we estimate 
our that sample may contain 1.6 high EW [OII] interlopers. Similarly, if we 
use the local luminosity function of \citet{Gal02}, we predict zero high EW interloping [OII] emitters. 
As a second comparison, \citet{Kak07} 
presents narrowband imaging and limited spectroscopic follow-up in their 
search for low metallicity galaxies. They find [OIII]5007 at $z=0.63$ 
and $z=0.83$, and [OII] at $z=1.19$ and $z=1.45$ at high enough 
EW values to contaminate our sample. By comparing 
their high EW [OII] number density to the Schechter \citep{Sch76} 
function fits of \citet{Ly07} at $z=1.18$ and $z=1.47$ without extinction 
corrections, we find the high EW [OII] fraction should only be 3\%. In 
contrast, the same analysis suggests the 
high EW [OIII]5007 fraction is much higher (33\%). However, there is no evidence 
for such a large fraction of high EW [OIII]5007 over 
our redshifts of interest, and the VIRUS-P bandpass will always enclose 
[OII] and [OIII]4959 when [OIII]5007 is observed. Thus, neither high EW [OII] 
nor [OIII]5007 emitters should form important catalog contaminants. The wavelength 
spacing between AGN lines is smaller than for [OII] and the other optical lines, so 
AGN lines should be identifiable with multiple detections. For example, 
index 4 is our only CIII]1909 detection, identified with a co-detection in index 5 
as CIV1549. The best available contamination estimate is that this LAE sample contains 0-2 contaminants 
from mis-identified redshifts if only EW information is used.
A complementary question is how our rest-frame EW affects the selection of high-redshift 
galaxies. Assuming the LAE distribution in \citet{Gro07}, the answer is that $\sim$21-26\% of 
potential detections are lost by the EW cut. 
\par We briefly state how we propogate errors to the EW estimation. In cases where the flux density measurement 
is very noisy, the usual first-order error propagation breaks down. Importantly, the error on equivalent 
width becomes asymmetric in the case of a low S/N continuum even if the original errors 
on flux and flux density are symmetric. One simple solution is to treat the 
maximum liklihood distributions in flux and flux density as Gaussian functions, transform 
the flux into EW and flux density, and define the EW errors using the extrema of the 
68\% confidence interval. 
Similarly, in the case of asymmetric errors for the line fluxes, we use the same 
equation evaluated with each one-sided error to arrive at final EW limits. When 
we find no upper limit, we list the upper uncertainty as 1000\AA.

\subsection{Counterpart association}
\label{sec_cntr}
\par The coarse spatial resolution of our VIRUS-P survey often prevents us 
from associating with certainty a given emission line to a unique broadband counterpart. However, 
stringent redshift probabilities can often be made by marginalizing over 
all possible counterparts and their implied rest-frame emission-line equivalent widths. 
We quantify this association probability by using the astrometric error, discussed in 
\S \ref{sec_astrom}, and the differential number counts for the $R$-band images. Since 
MUNICS lacks $R$-band data, we use the i'-band there. The exact band choice 
for this step is not critical, so long as the filter samples a fairly 
flat spectral region for both low-$z$ and high-$z$ objects. We describe 
the method for $R$-band continuum association as it applies to 
equivalent-width-based redshift discrimination. We use the 
same formalism for AGN association through 
X-ray data in \S \ref{sec_AGN}.  
\par The problem of assigning counterpart probabilities to detections 
in multiple bandpasses has been explored by Bayesian methods 
in \citet{Sut92} and is commonly implemented in X-ray surveys 
\cite[e.g.][]{Luo10}. We choose not to use the Bayesian 
technique here since it requires assuming a prior on the continuum 
counterparts to the emission-line detections. We instead make 
a simpler, frequentist estimate that still uses the 
information from multiple candidates. The only assumed 
inputs are the astrometric error and the number counts of 
background and foreground objects. 
\par The probability of an emission line being associated with any one 
image-based counterpart can be constructed as the joint probability of 
all the remaining imaging detections being unassociated and 
drawn from established number counts and the preferred counterpart 
having the observed offset evaluated against the astrometric error budget. 
For simplicity, we treat all the individual probabilities as independent; this 
simplication is justified since the range of distances in our redshift 
range is much larger than cross-correlation scales between galaxies. 
We begin by identifying all the significant imaging detections within some large 
area of the detected emission line. We then define: $i$ 
as the set of all imaging detections 
in the survey field, $\Delta r_i$ as the angular offset 
between the position of the emission line and the 
centroid of counterpart $i$, $S_i$ as the 
flux density of counterpart $i$ (or X-ray flux in a defined 
bandpass), $\sigma_i$ as the astrometric 
error for the emission line under consideration and counterpart $i$, and 
$n(S)$ as the differential number count of galaxies in the observed bandpass. 
We begin by assembling the set of imaging detections with cardinality $j$ as 
$C_j=(j \in i : \Delta r_j < 10\arcsec)$. The exact value of the angular limit is not 
important so long as it is several times the astrometric error. The chance of a 
superposition by one or more imaging detections without them being actual 
counterparts is then $P_{nc,j}=1-f(0,\lambda)$ 
where $f(n,\lambda)=\frac{\lambda^n e^{-\lambda}}{n!}$ is the Poisson 
probability distribution and $\lambda$ is the expectation value for the 
number of galaxies brighter than $S_j$ within $\Delta r_j$, so 
$\lambda=\pi \Delta r_j^2 \int_{S_j}^{\infty}n(S)dS$. Alternatively, the 
detection $j$ may be the true counterpart. If we model the astrometric 
error as a two-dimensional 
Gaussian distribution in the astrometry error and take its 
cumulative evaluation from infinity, the chance of measuring the true counterpart 
at $\Delta r_j$ or further is $P_{c,j}=\mathrm{exp}({\frac{-\Delta r_j^2}{2\sigma^2}})$. It may 
also be that we have not measured the true imaging counterpart, 
either due to imaging depth or the emission-line detection being spurious, in 
which case all imaging detections must be explained as 
superpositions.
\par We give in Equation \ref{eq_spat_prob} the full joint probabilities 
assembled from the individual probabilities just described, under the assumption that 
either one or none of the imaging detections is the true counterpart to the 
emission-line detection. A similar calculation is done to evaluate the 
significance of X-ray counterparts in \S \ref{sec_AGN}. 
\begin{equation}
\label{eq_spat_prob}
P=\left\{
\begin{array}{l}
\frac{\displaystyle\prod_{1\leq k\leq j}P_{nc,k}}
{\displaystyle\prod_{1\leq k\leq j}P_{nc,k}+\displaystyle\sum_{1\leq\ k\leq j}
\left( P_{c,k}\times\displaystyle\prod_{1\leq m\leq j,m\neq k}P_{nc,m}\right)}\\
: \mbox{no counterpart}\vspace{12pt}\\
\frac{\displaystyle P_{c,k}\times\displaystyle\prod_{1\leq m\leq j,m\neq k}P_{nc,m}}{
\displaystyle\prod_{1\leq k\leq j}P_{nc,k}+\displaystyle\sum_{1\leq\ k\leq j}
\left( P_{c,k}\times\displaystyle\prod_{1\leq m\leq j,m\neq k}P_{nc,m}\right)}\\
: \mbox{counterpart\ } k
\end{array}
\right.
\end{equation}
In the case of imaging, the astrometric error is dominated by the 
positional uncertainty of the emission-lines, but in the case of X-ray data 
the positional uncertainty of both the emission-line and X-ray detections are comparable 
and important. 
The normalization is 
simply chosen to make the probabilities sum to unity.  
\par In order to match $R$-band objects, we performed a least squares minimization fit to the R-band differential number counts 
of \citet{Fur08} with a double power law function to estimate $n(S)$ as given 
in Equation \ref{eq_r_pl}. 
\begin{equation}
\label{eq_r_pl}
n(f_{\nu}) [\mbox{per\ } \sq\arcdeg]=\left\{
\begin{array}{l}
5142\times (f_{\nu}/10^{-28})^{-1.996}\\
: f_{\nu}>7.81\times 10^{-30} \mbox{erg/s/cm$^2$/Hz}\\
\\
10882.6\times (f_{\nu}/10^{-28})^{-1.702}\\
: f_{\nu}\leq7.81\times 10^{-30} \mbox{erg/s/cm$^2$/Hz}
\end{array}
\right.
\end{equation}
\par In practice, we consider a threshold distance of 1$\arcsec$ when calculating 
the expected number counts of sources based on common seeing conditions to avoid 
the claim of total counterpart certainty at $\Delta r_{j}=$0 regardless of other counterpart 
options, although the 
exact threshold makes little difference. Finally, in defining $\Delta r_j$, 
we take the radial offset from the emission-line centroid to the 
nearest position contained in the imaging's Kron aperture instead of the 
Kron aperture center. This is motivated by the 
fact that [OII] emission in nearby galaxies may be from HII regions 
located at large galactocentric radii.
\par Object classification starts by identifying all imaging catalog counterparts within 
10\arcsec\ from the emission-line centroid. The association probability for each possible 
counterpart is calculated using Equation \ref{eq_spat_prob}. The emission line is 
classified under the EW rule of 
\S \ref{sec_EWrule} to be at either low (most likely as [OII]) or high (most 
likely as Ly$\alpha$) redshift. In 74\% of the cases, 
the best counterpart probability exceeds 90\%; 
we refer to these objects as the isolated sample. 
In another 3\% of the cases, our analysis is most consistent with there being 
no broadband counterpart; we classify these sources as 
LAEs since all the image depths imply EW$_{\mbox{Ly$\alpha$,rest}}>20$\AA. 
For the remaining cases, our classification is less certain due to there being multiple 
likely counterparts; 
nevertheless the most probable association is always presented. We 
illustrate step-by-step two representative classification 
cases in \S \ref{sec_LAE} and \ref{sec_lowz}, and a case with less certainty in 
\S \ref{sec_uncert}.
\par We confirm the proper classification of many low redshift objects by 
observing multiple emission lines. In total, there are 118 emission-line sources with one or more
associated emission lines in combinations of [OII], H$\delta$, 
H$\gamma$, H$\beta$, [OIII]4959, and [OIII]5007. Of these, all are classified automatically 
by our EW cut as being at low redshift. 
\subsection{AGN contamination}
\label{sec_AGN}
\par We attempt to identify the Ly$\alpha$ sources that are AGN through existing X-ray data. 
All our survey fields have either Chandra or XMM/Newton coverage, although to quite non-uniform depths. 
We use the point-source catalogs 
of \citet{Elv09,Cap09} in COSMOS, \citet{Ale03} in GOODS-N, \citet{Pie04} in XMM-LSS, and \citet{Wat09} 
in MUNICS. The data covering MUNICS is described in \citet{Sev05} but not cataloged. 
The same methodology for determining broadband imaging counterparts 
is applied to the X-ray data. The cataloged X-ray spatial uncertainty is 
added in quadrature to the emission-line spatial uncertainty, and the fit 
of \citet{Capp07} from 2-10 keV is used for the differential number count. 
Unlike with the imaging counterparts, the 
association of an X-ray source with a VIRUS-P emission line is nearly binary in nature: 
there is either a single counterpart or no counterpart. Table \ref{tab_AGN} summarizes our 
results by listing the fraction of X-ray sources in the low and high-$z$ objects. We find 
6-8\% contamination of LAEs by AGN over all the fields with the range depending on what 
fraction of the line detections we attribute to noise. AGN contamination 
is likely a strong function of flux limit, but we compare briefly to other, deeper surveys. 
The sample of \citet{Gaw07,Gro07} at $z=3.1$ contains 1\% AGN contamination, the 
sample of \citet{Nil09,Nil10} at $z=2.3$ contains 6-15\% AGN contamination, and the 
sample of \citet{Gua10} at $z=2.1$ contains 5\% AGN contamination.  
These numbers, all utilizing X-ray detections of AGN, are consistent with the 
value we find. However, other work with mid-IR and far-IR AGN identification has 
potentially shown a much higher AGN fraction of 75\% at $z=2.2$ \citep{Bon10}. We do 
not perform any mid-IR or far-IR AGN analysis here. There is no 
significant variation between AGN fractions of GOODS-N with the deeper 
X-ray data and the COSMOS field. The small number statistics and shallower 
X-ray data in MUNICS and XMM-LSS explain the lack of AGN detections in those 
fields. 
\par
\begin{figure}
\centering
\includegraphics [scale=0.4,angle=-90]{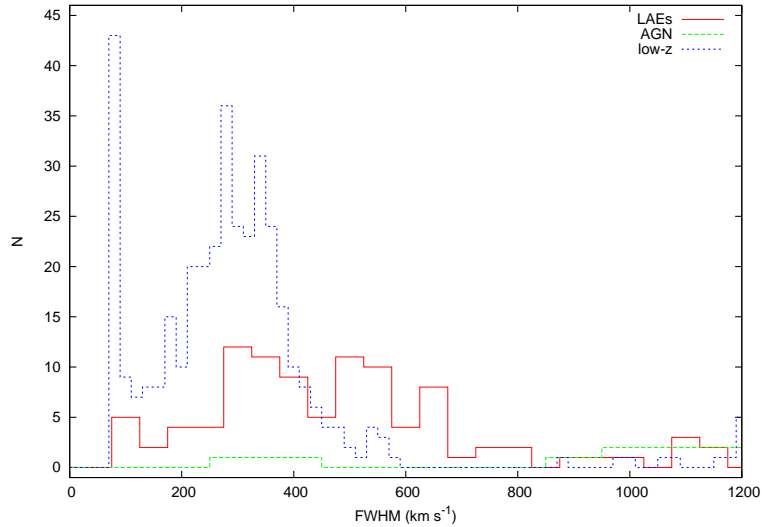}
\caption{The line width distributions for all survey objects. We have subtracted 
an instrumental resolution of 300 km s$^{-1}$ in quadrature. There is significant overlap between 
all populations making width-based classification impossible. No attempts have been 
made here to deconvolve the blended 
[OII] doublet. The low-$z$ objects are generally contained to low widths, but 
the LAE distribution overlaps heavily with both the AGN and low-$z$ distributions.}
\label{fig_width}
\end{figure}

 There are two other potential indicators of AGN activity: broad Ly$\alpha$ emission-line widths 
(which may also be seen in Lyman-$\alpha$ ``blobs'' where AGN activity is not evident 
\citep{Fra01,Bow04,Mat06,Smi07,Sai08}), and 
the presence of CIV1549 over a fraction of the redshift range. 
The distributions of line widths, without any deblending of the [OII] doublet, 
is given in Figure \ref{fig_width}. From the distributions, it is clear that line width 
information does not aid object classification. We find two cases where broad line 
objects (FWHM line widths $>500$ km s$^{-1}$) have been classifed as Ly$\alpha$ without any X-ray 
detections, but none at $>1000$ km s$^{-1}$. However, only one Ly$\alpha$ and CIV1549 
source (indices 461 and 462) was not detected in X-ray.
\subsection{Example sources}
\par The rules to classify the emission-line objects have been described, but 
the display of the steps on actual VIRUS-P data is useful to establish 
confidence and the range of objects encountered. We will walk through 
the evidence for one emission source of each type and then give a 
summary display of representative subsamples. 
\subsubsection{LAEs}
\label{sec_LAE}
\par
\begin{figure}
\centering
\includegraphics [scale=0.6,angle=-90]{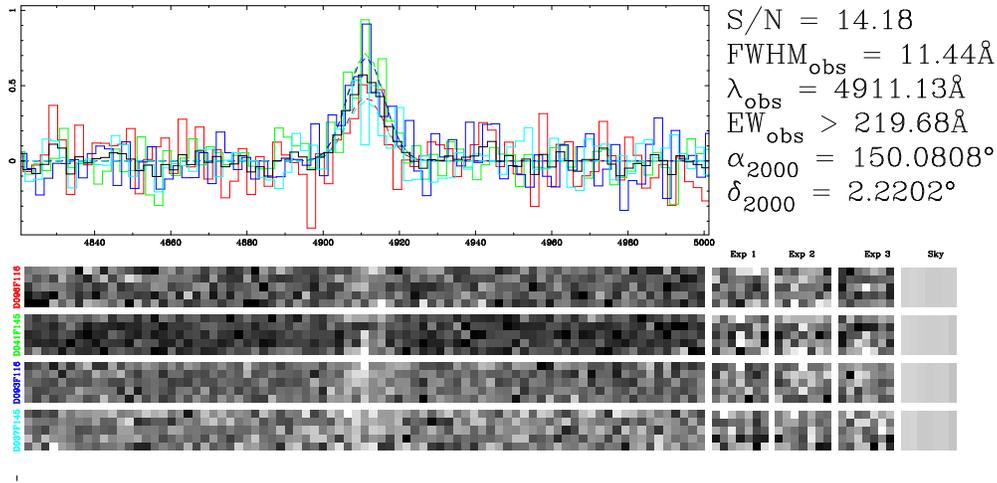}
\caption{The VIRUS-P emission-line detection image for index 229 
which lies in the COSMOS field. This object is undoubtably a real detection 
with one of the largest S/N ratios we 
find for any high-$z$ object. In this case, the aperture is grown to include 
four fibers. The four rows in the Figure's bottom half show the spectra 
from the four detection fibers. The right side, square cut-outs show the 
spectra from individual 20-minute exposures and the sky model. The 
three exposures in each fiber are then biweight combined into the 
two-dimensional, bottom-left spectra and the one-dimensional spectra 
in the upper-left line plot. The 
collapsed, one-dimensional spectra are color-coded by fiber number. 
The Gaussian fits to each fiber are given by dotted curves. 
For visual clarity, the spectra are resampled and stacked into the black 
histogram. Continuum is not detected within a 200\AA\ boxcar around the line, 
and the high level of flux permits the Ly$\alpha$ classification from the 
spectrum alone in this rare case. The tabulated EW instead is based 
on the flux density of the imaging counterpart. The quoted central 
wavelength in this figure has not yet had the heliocentric 
and vacuum corrections applied as is done with the tabulated values.}
\label{fig_highz_det}
\end{figure}

\begin{figure}
\centering
\includegraphics [scale=0.4,angle=-90]{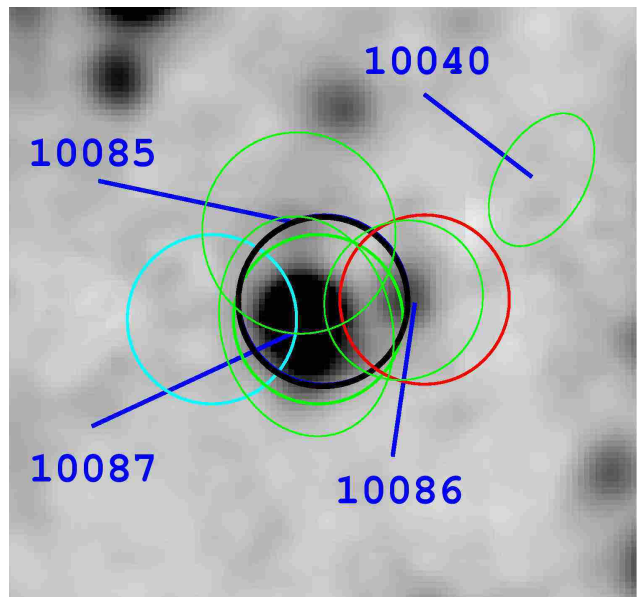}
\caption{The ground-based, $V_j$ imaging cutout for 
the Table \ref{tab_line_list1} index 229. 
The four color circles represent fiber positions and are 
color coded in accordance with the spectra of Figure \ref{fig_highz_det}. 
The black circle indicates the emission line centroid. 
The Kron apertures from the imaging 
catalog are drawn as green, numbered ellipses. The best centered and brightest source is \#10087 with an 
association likelihood of 87\% and r$^+$=23.4. The next two most likely 
counterparts are \#10085 at 6\% and r$^+$=25.1 and \#10087 at 
5\% and r$^+$=24.9. Either of these would also make the LAE EW cut, and it is 
possible that one or more LAEs at similar redshifts are jointly contributing 
to the emission-line flux, but this is unlikely.
}
\label{fig_highz_ima}
\end{figure}

\begin{figure}
\centering
\subfigure{\includegraphics[bb=41 46 372 666,clip=true,scale=0.33,angle=-90] {quick_highz1.eps}}
\subfigure{\includegraphics[scale=0.2,angle=-90] {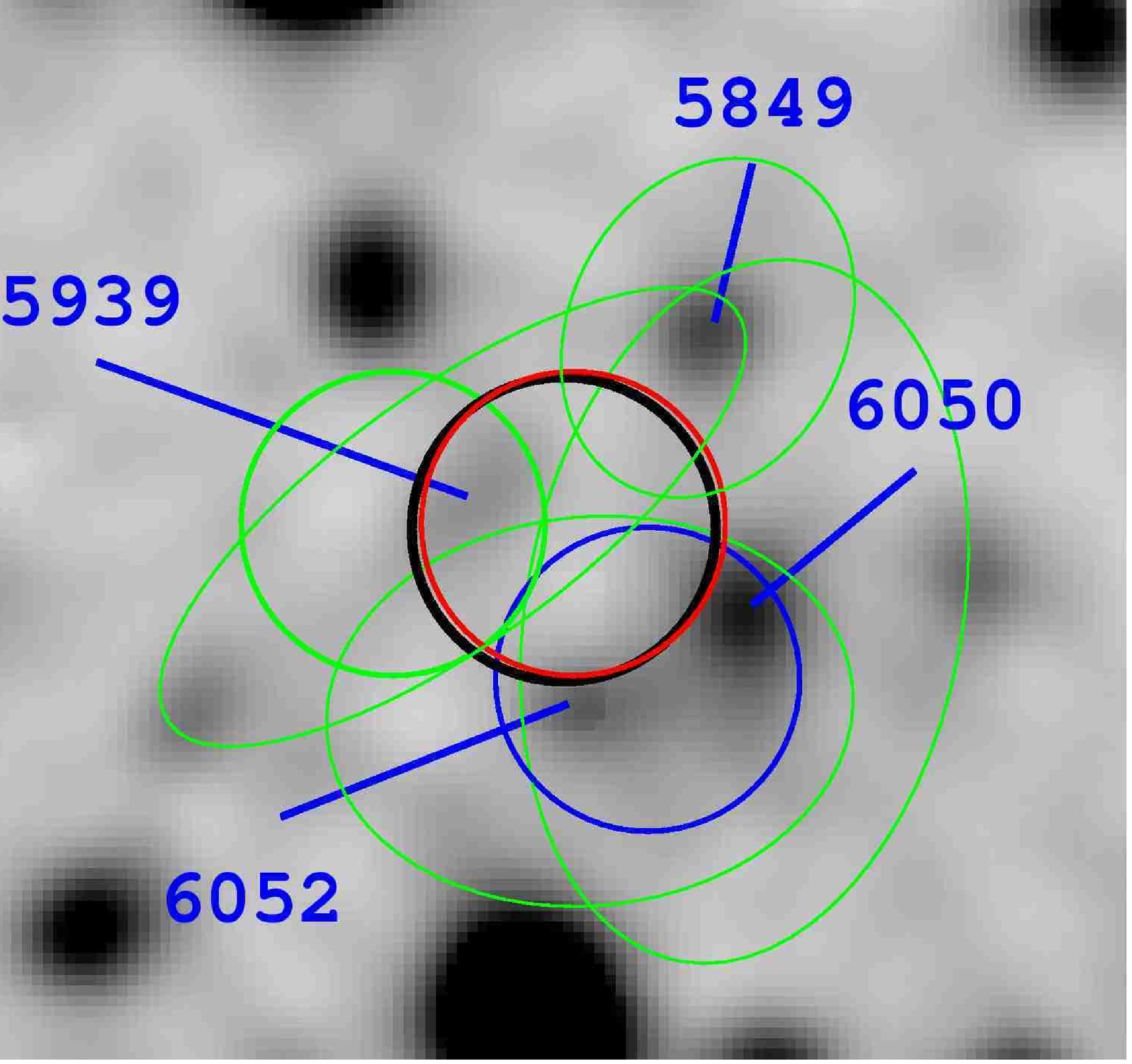}}
\vspace{-22pt}
\\
\subfigure{\includegraphics[bb=41 46 372 666,clip=true,scale=0.33,angle=-90] {quick_highz2.eps}}
\subfigure{\includegraphics[scale=0.2,angle=-90] {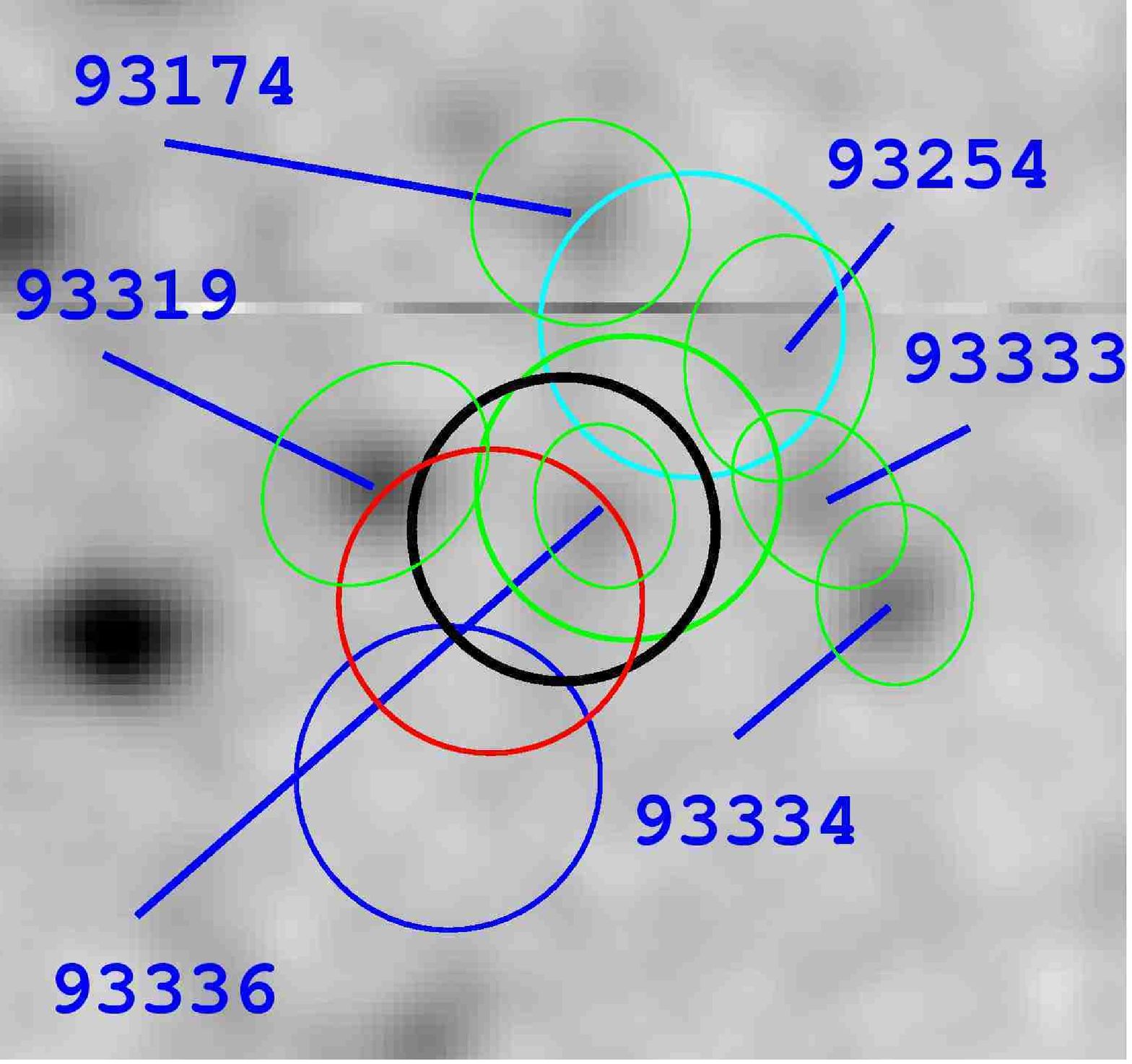}}
\vspace{-22pt}
\\
\subfigure{\includegraphics[bb=41 46 372 666,clip=true,scale=0.33,angle=-90] {quick_highz3.eps}}
\subfigure{\includegraphics[scale=0.2,angle=-90] {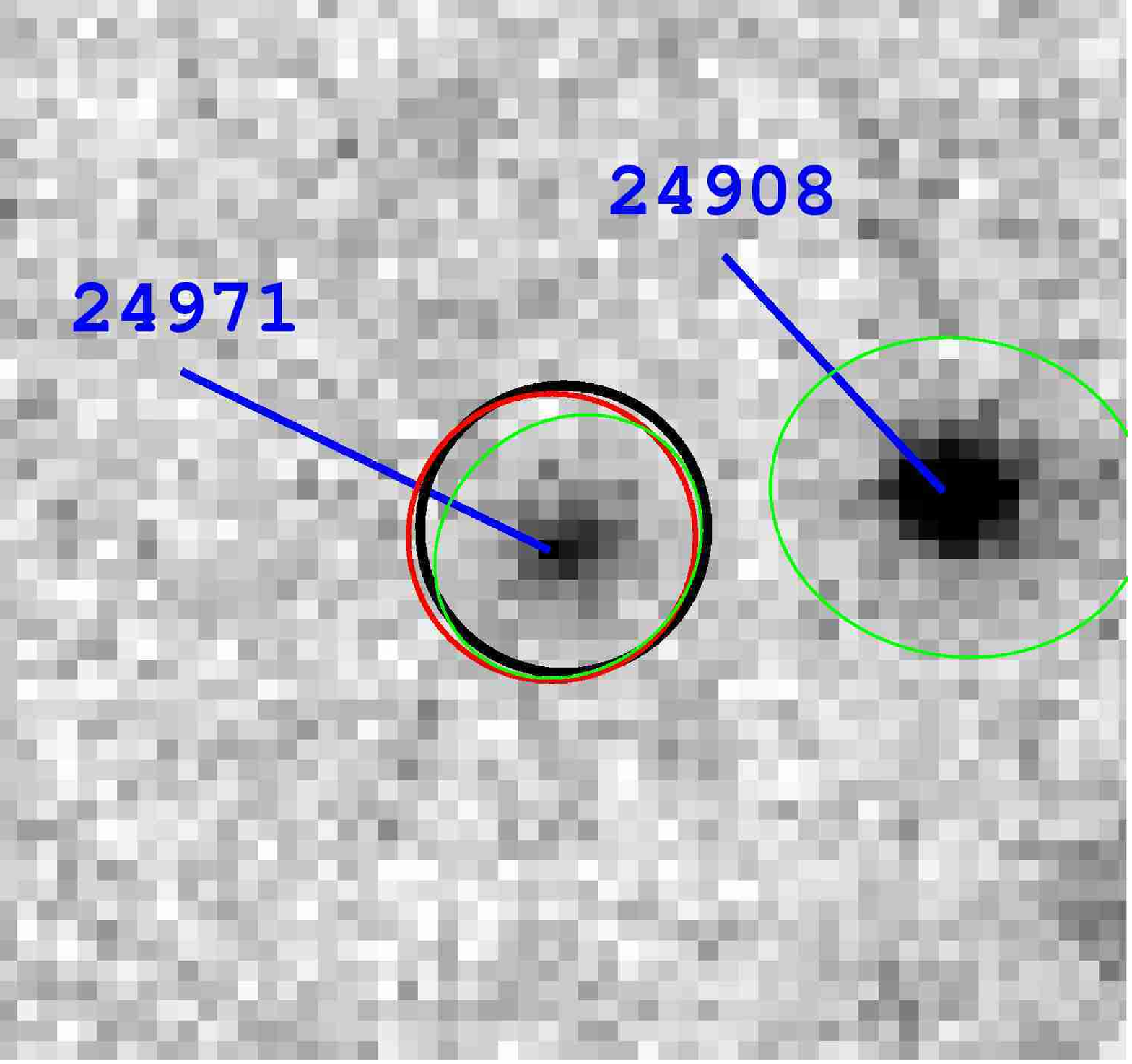}}
\vspace{-22pt}
\\
\subfigure{\includegraphics[bb=41 46 372 666,clip=true,scale=0.33,angle=-90] {quick_highz4.eps}}
\subfigure{\includegraphics[scale=0.2,angle=-90] {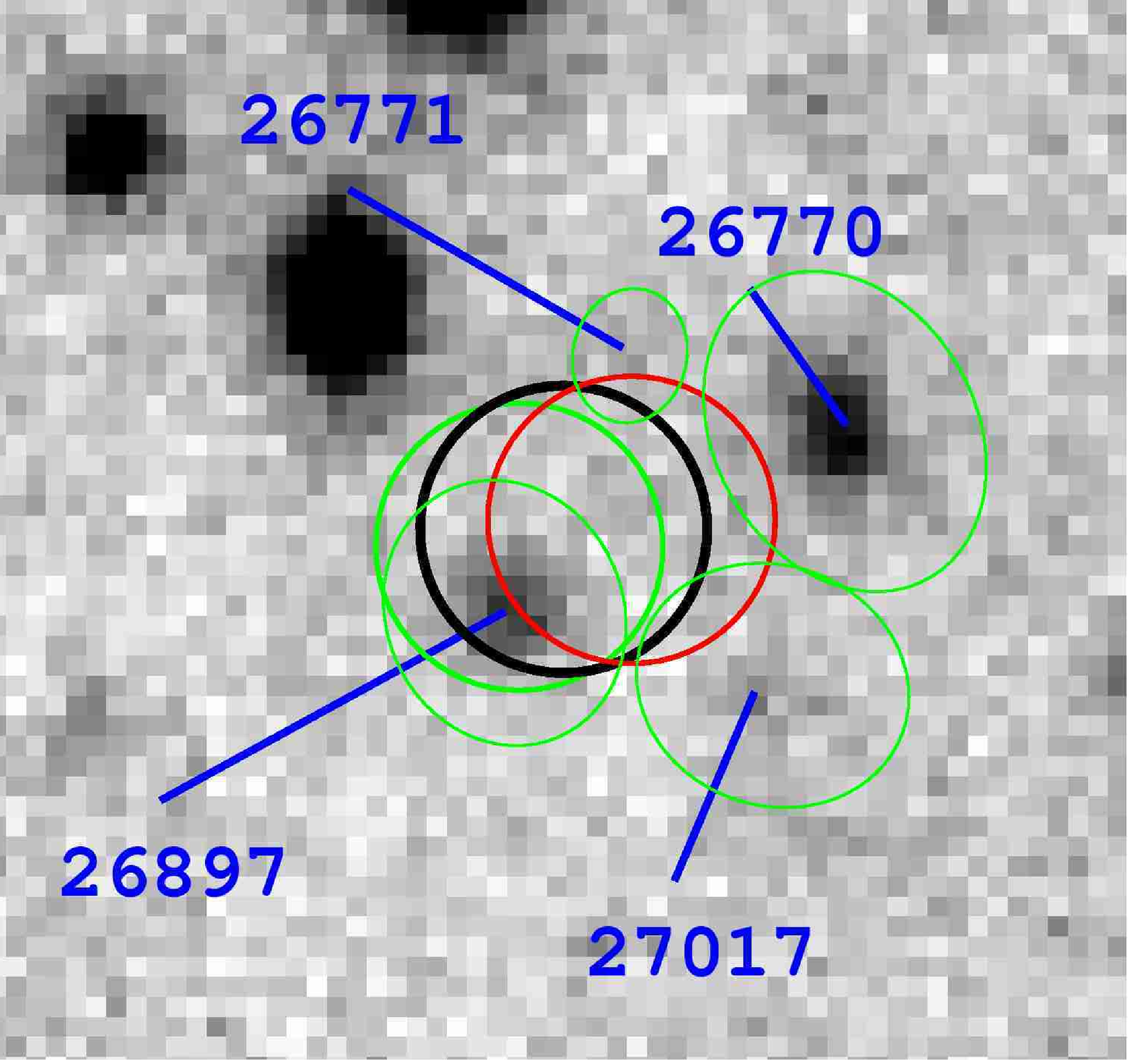}}
\vspace{-22pt}
\\
\subfigure{\includegraphics[bb=41 46 372 666,clip=true,scale=0.33,angle=-90] {quick_highz5.eps}}
\subfigure{\includegraphics[scale=0.2,angle=-90] {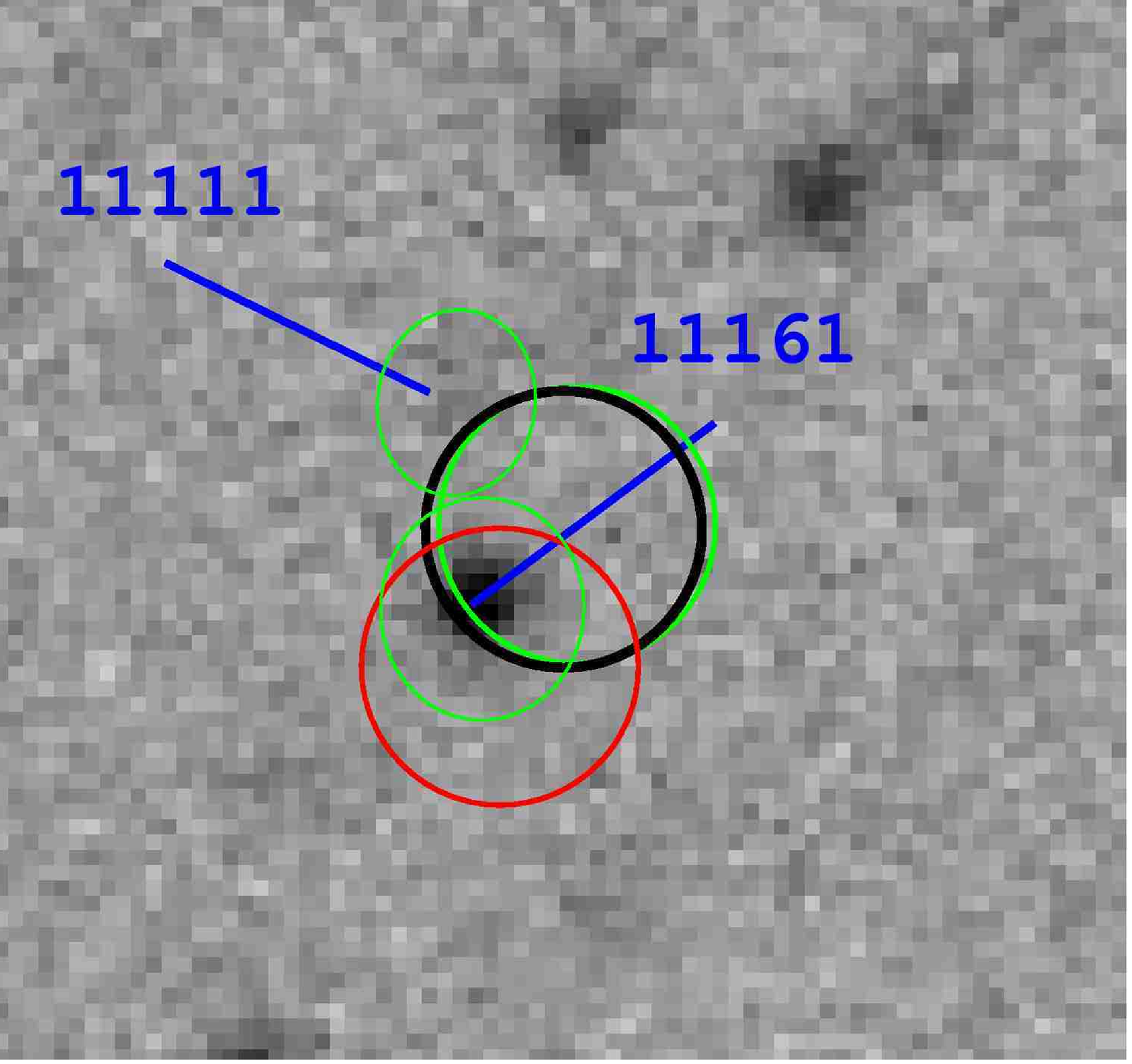}}
\vspace{0pt}
\caption{Five example detections of LAEs with the same formats as Figures 
\ref{fig_highz_det} and \ref{fig_highz_ima}. The first two lie in the COSMOS field, the 
next two lie in the GOODS-N field with the first redshift 
previously measured and the second new, and the final one lies in the MUNICS field. The entries 
from Table \ref{tab_line_list1} for these five are indicies 223, 160, 341, 402, and 62. The 
best continuum counterpart matches are to \#'s 5939, 93336, 24971, 26897, and 11161.}
\label{fig_high_mos}
\end{figure}

 The detection image of Figure \ref{fig_highz_det} shows 
source index 229 in Tables \ref{tab_line_list1} and \ref{tab_line_list2} as a broadened, 
bright emission line detected in four fiber positions. The high flux 
($41.6_{-5.0}^{+4.2}\times10^{-17} $erg s$^{-1}$ cm$^{-2}$) and the lack of a 
spectral continuum detection are already sufficient to meet the classification cut as an 
LAE. However, classification from the spectrum alone is only possible 
for the brightest emission-line sources in this sample. 
There are no counterpart emission lines at any of the tested transitions (\S \ref{sec_spec_clas}), 
nor is there any associated X-ray detection. We next move to the deep, COSMOS $R$ image 
in Figure \ref{fig_highz_ima}. There, we find three plausible broadband counterparts 
with the brightest counterpart dominating the likelihood. There is no literature redshift for this 
object, but our Ly$\alpha$ line identification leads to EW$_0$=$51_{-8}^{+8}$\AA, so the 
object is classified as an LAE. We give compact detection images 
for five additional LAEs in Figure \ref{fig_high_mos}. 
\subsubsection{Low-z objects}
\label{sec_lowz}
\par
\begin{figure}
\centering
\includegraphics [scale=0.6,angle=-90]{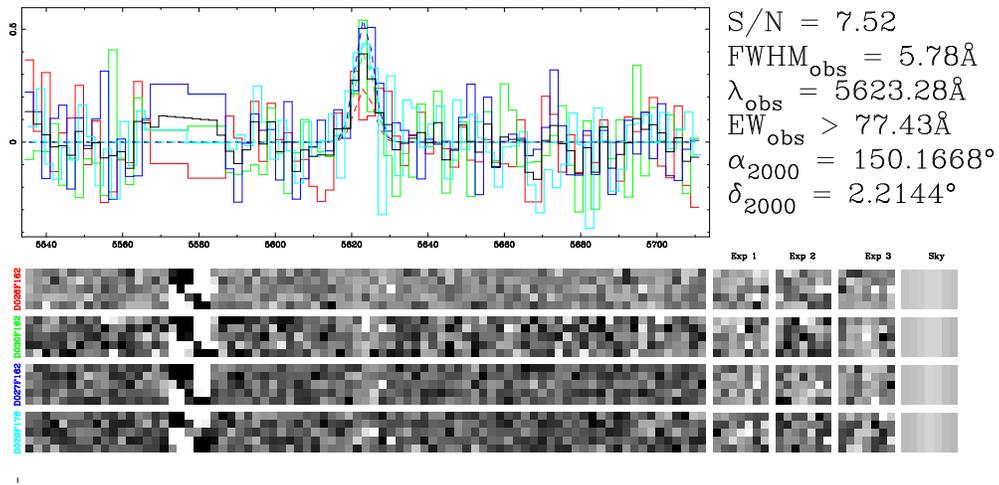}
\caption{The emission-line source detection for index 308 which lies in the COSMOS field. 
The format is the same as in Figure \ref{fig_highz_det}. The spectrum-based EW 
does not go deep enough to discriminate between the classifications. This source 
neither shows alternate emission lines nor has an X-ray counterpart. 
}
\label{fig_lowz_det}
\end{figure}

\begin{figure}
\centering
\includegraphics [scale=0.4,angle=-90]{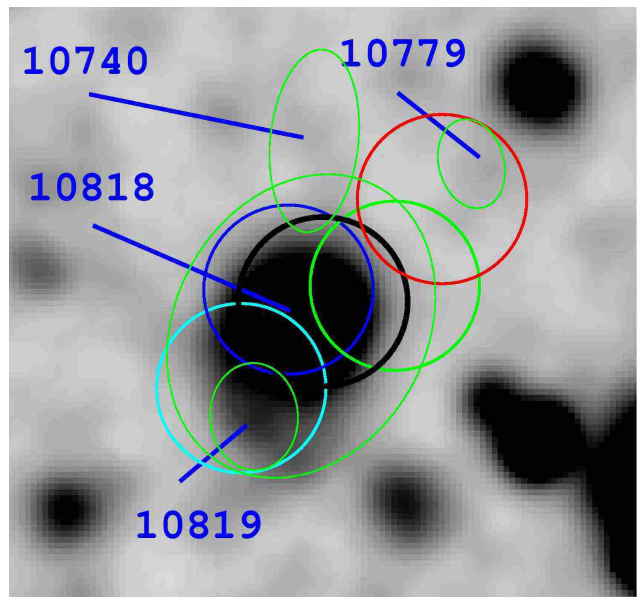}
\caption{The ground-based, $V_j$ imaging cutout for the Table 
\ref{tab_line_list1} index 308. The detection spectra are given 
in Figure \ref{fig_lowz_det}. The format is the same as in Figure 
\ref{fig_highz_ima}. The counterpart \#10818 at r$^+$=22.1 has a 
99.5\% likelihood of being associated. The observed EW of 
$30.9_{-9.5}^{+14.9}$\AA\ leads to a firm low-$z$ classification, presumably 
for [OII].}
\label{fig_lowz_ima}
\end{figure}

\begin{figure}
\centering
\subfigure{\includegraphics[bb=41 46 372 666,clip=true,scale=0.33,angle=-90] {quick_lowz2.eps}}
\subfigure{\includegraphics[scale=0.2,angle=-90] {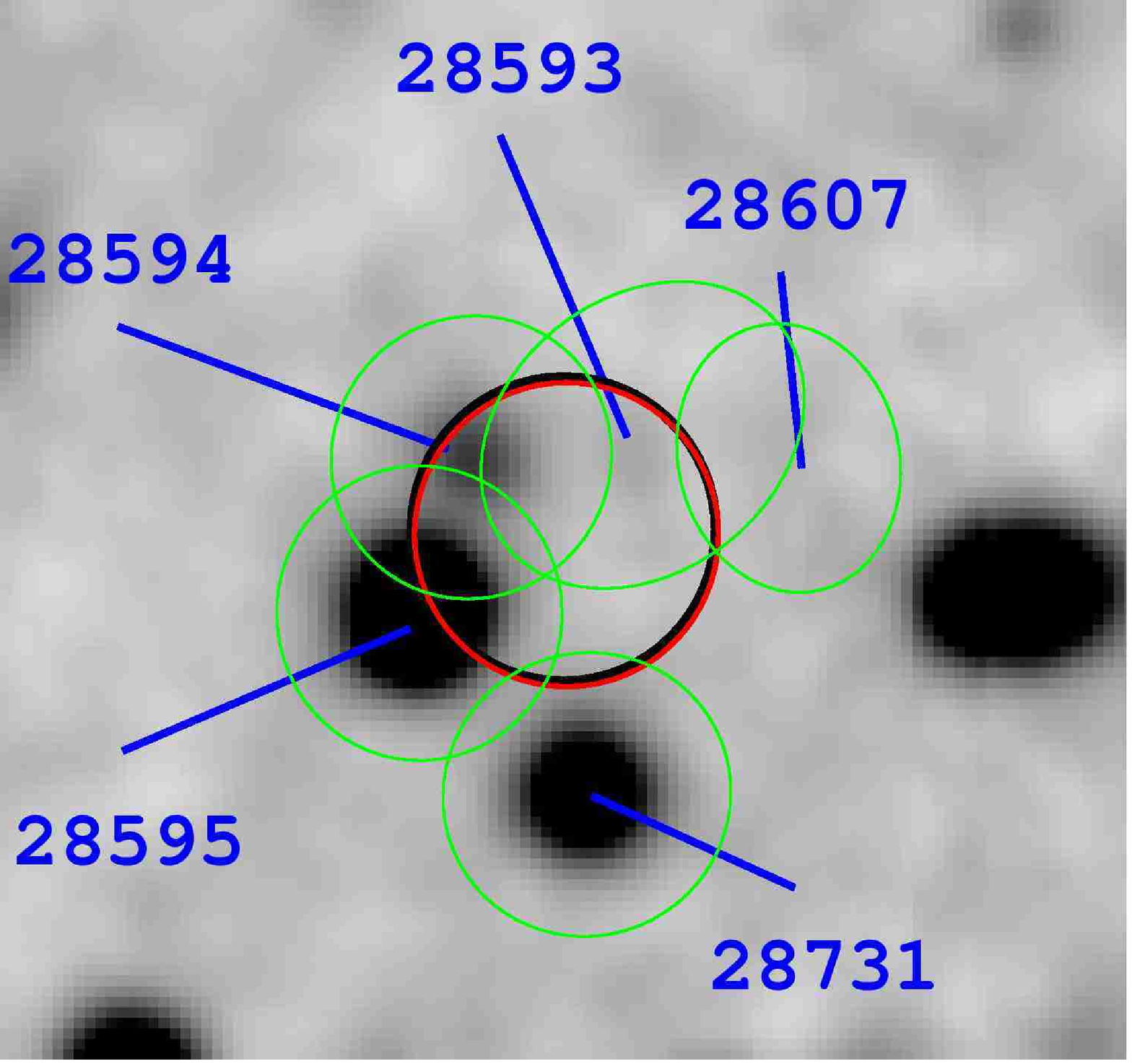}}
\vspace{-22pt}
\\
\subfigure{\includegraphics[bb=41 46 372 666,clip=true,scale=0.33,angle=-90] {quick_lowz3.eps}}
\subfigure{\includegraphics[scale=0.2,angle=-90] {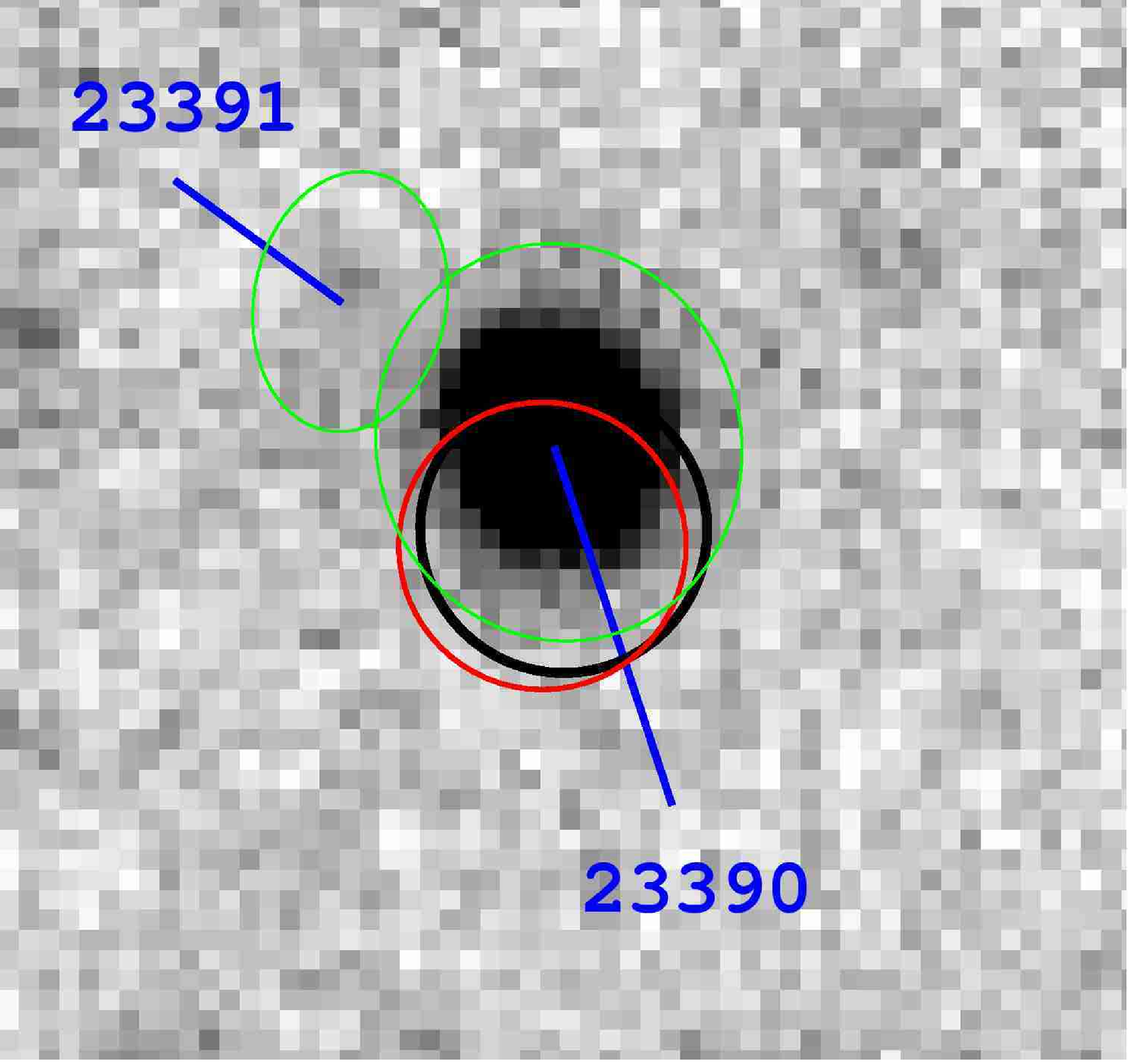}}
\vspace{-22pt}
\\
\subfigure{\includegraphics[bb=41 46 372 666,clip=true,scale=0.33,angle=-90] {quick_lowz4.eps}}
\subfigure{\includegraphics[scale=0.2,angle=-90] {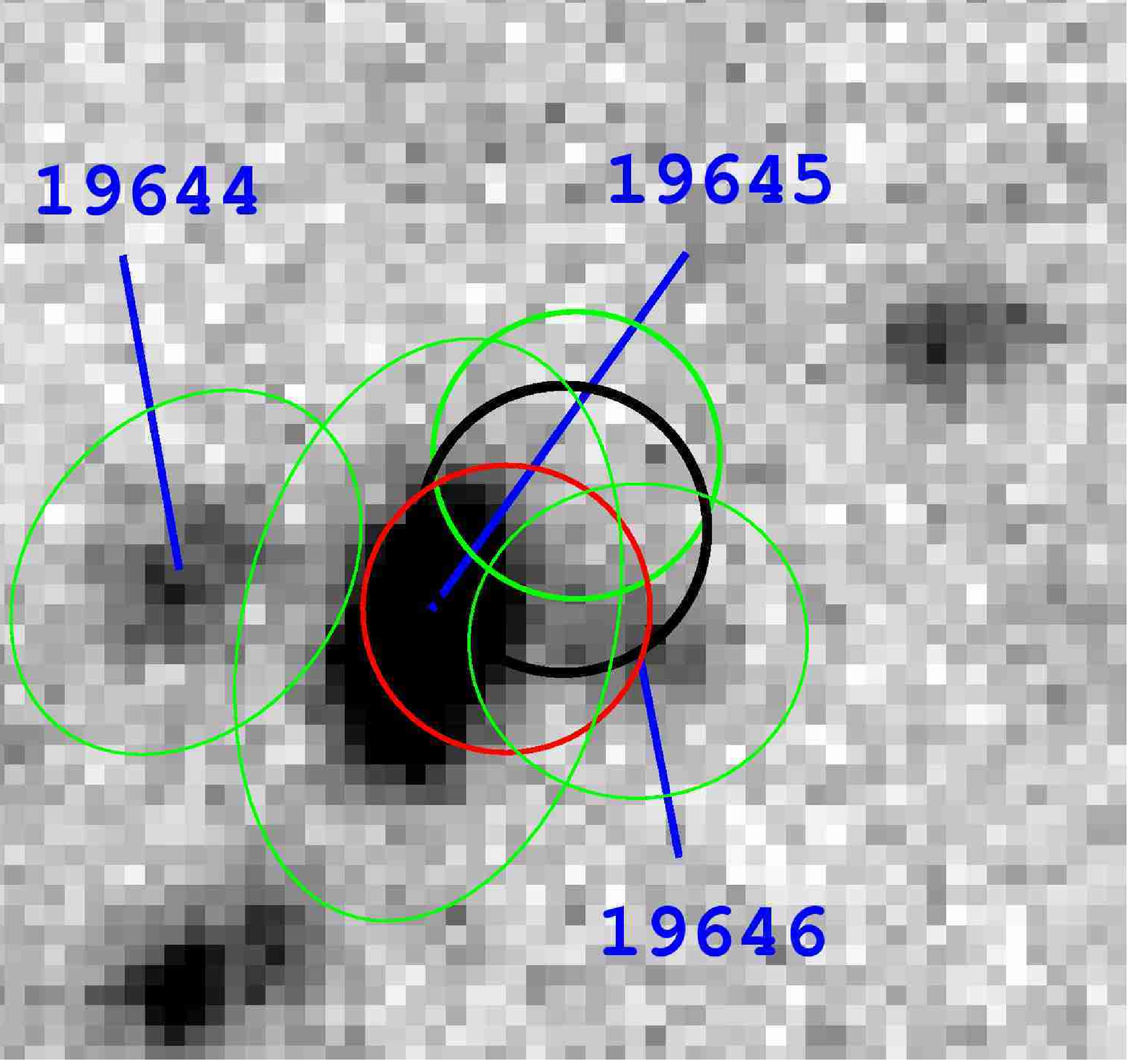}}
\vspace{-22pt}
\\
\subfigure{\includegraphics[bb=41 46 372 666,clip=true,scale=0.33,angle=-90] {quick_lowz5.eps}}
\subfigure{\includegraphics[scale=0.2,angle=-90] {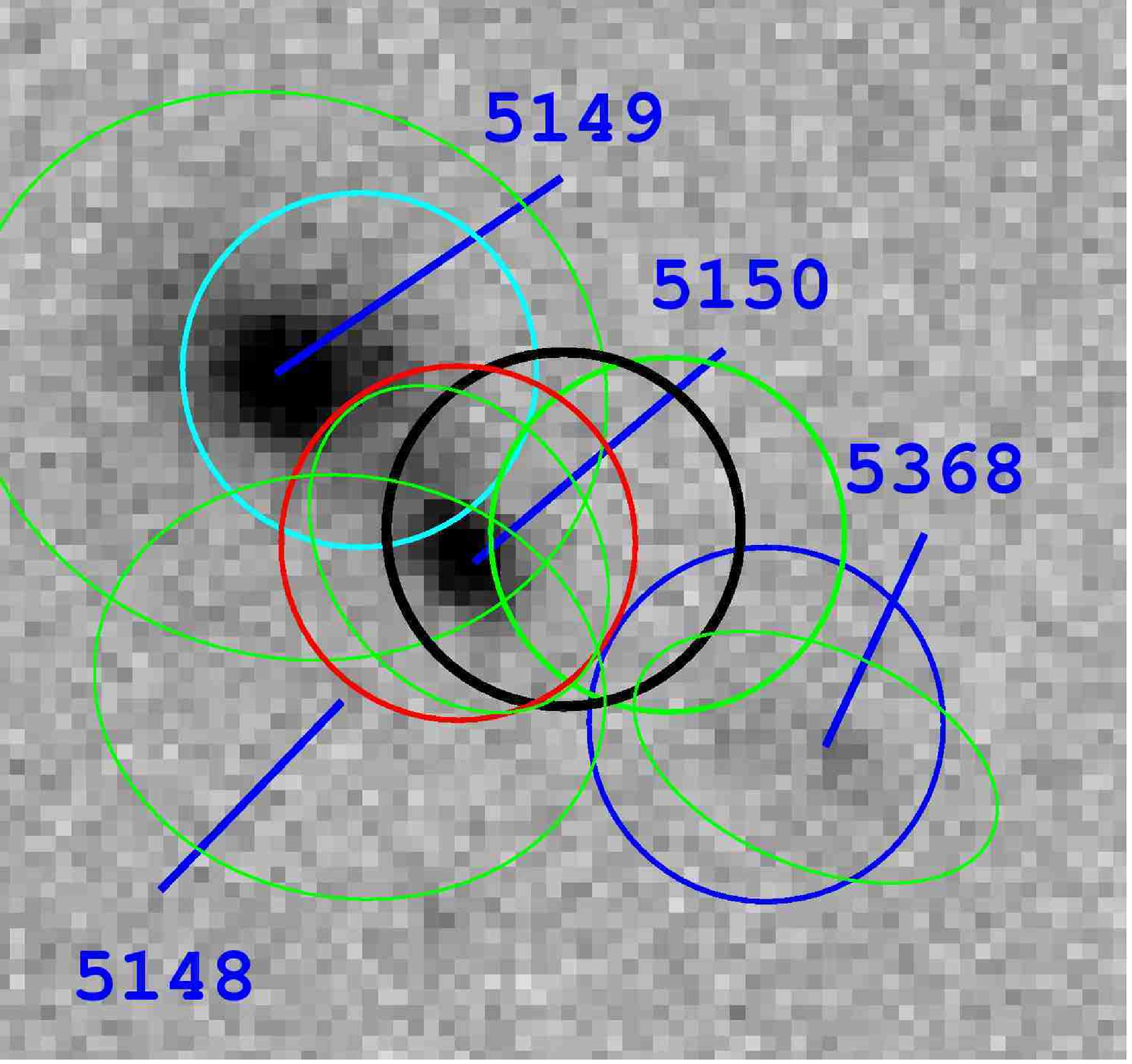}}
\vspace{-22pt}
\\
\subfigure{\includegraphics[bb=41 46 372 666,clip=true,scale=0.33,angle=-90] {quick_lowz1.eps}}
\subfigure{\includegraphics[scale=0.2,angle=-90] {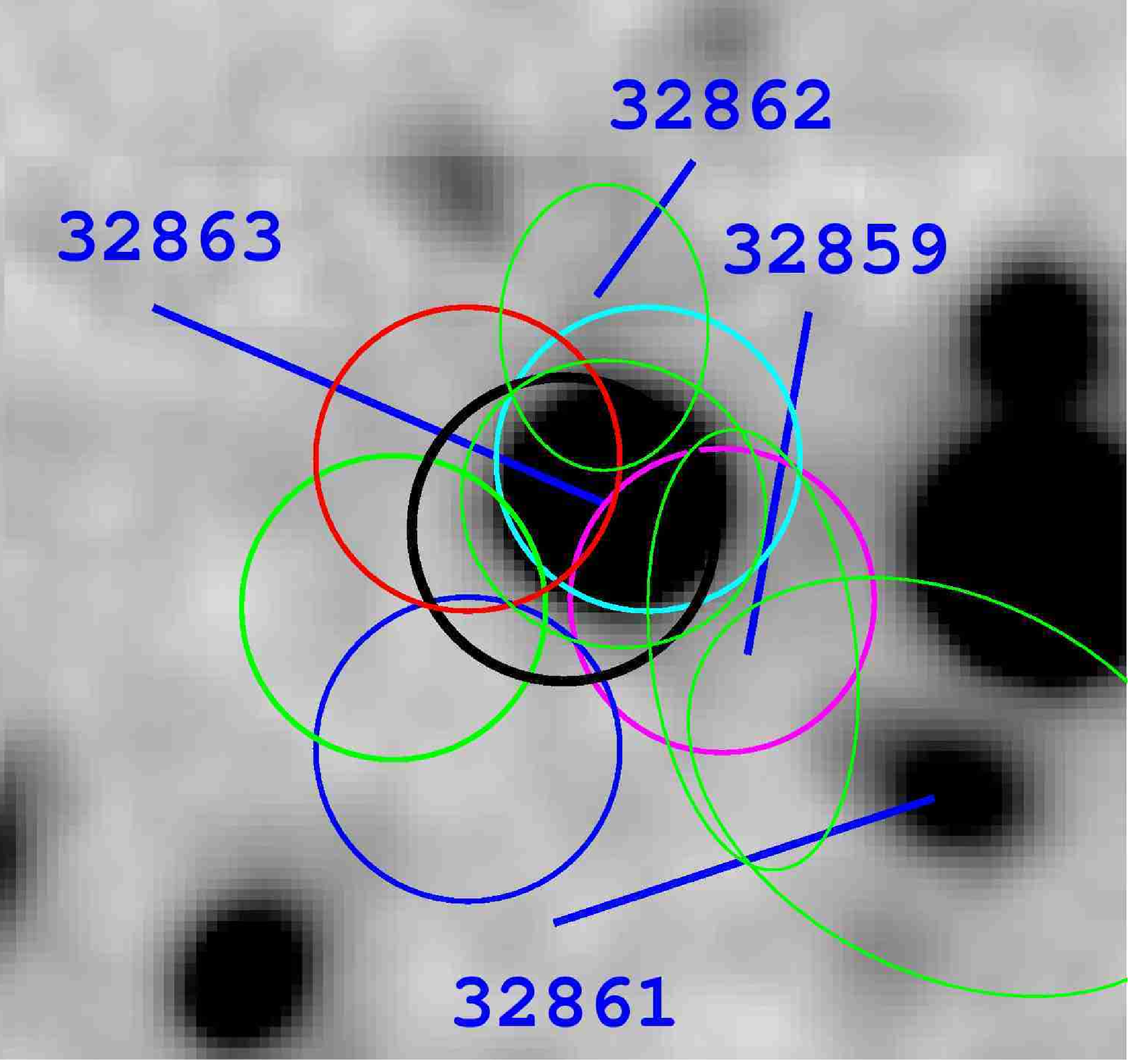}}
\vspace{0pt}
\caption{Five example detections of low-$z$ objects with the format of Figure 
\ref{fig_lowz_det}. The first and fifth lie in the COSMOS field, the 
second and third lie in the GOODS-N field both with previously measured 
redshifts, and the fourth lies in the XMM-LSS field. The entries 
from Table \ref{tab_line_list1} for these five are indicies 178, 351, 406, 33, and 192. The 
best continuum counterpart matches are to \#'s 28595, 23390, 19645, 5150, and 32863.}
\label{fig_low_mos}
\end{figure}

 The detection image in Figure \ref{fig_lowz_det} shows source index 308 as a high S/N, only 
slightly and not significantly broadened, emission-line source detected jointly 
in four fiber positions. The line flux is $18.4_{-4.2}^{+3.5}\times10^{-17} $erg s$^{-1}$ cm$^{-2}$, 
and the lack of a spectral continuum is insufficient for our EW-based classification scheme. 
There are no other emission lines detected in the object nor does the source have a 
X-ray counterpart. The COSMOS image (Figure 
\ref{fig_lowz_ima}) shows one bright continuum source barely offset from the 
emission-line centroid. A second, fainter object at larger separation is also 
analyzed at a much decreased likelihood, but would carry 
an LAE classification. Based on the most probable counterpart, we 
find a rest frame EW, assuming the line to be Ly$\alpha$, of EW$_0$=$8_{-2}^{+2}$\AA. 
This fails the EW cut, so we classify this as a low redshift object, 
presumably an [OII] emitter. The actual rest frame EW is then 
EW$_0$=$25_{-6}^{+6}$\AA. We note that association with the other 
possible counterpart would lead to the opposite conclusion. However, the 
likelihood of that association is quite low, P=0.1\%, so we confidently 
classify the source as an [OII] emitter. We give compact detection images
for five additional low redshift objects in Figure \ref{fig_low_mos}. 
\subsubsection{Objects with uncertain redshifts}
\label{sec_uncert}
\par
\begin{figure}
\centering
\includegraphics [scale=0.6,angle=-90]{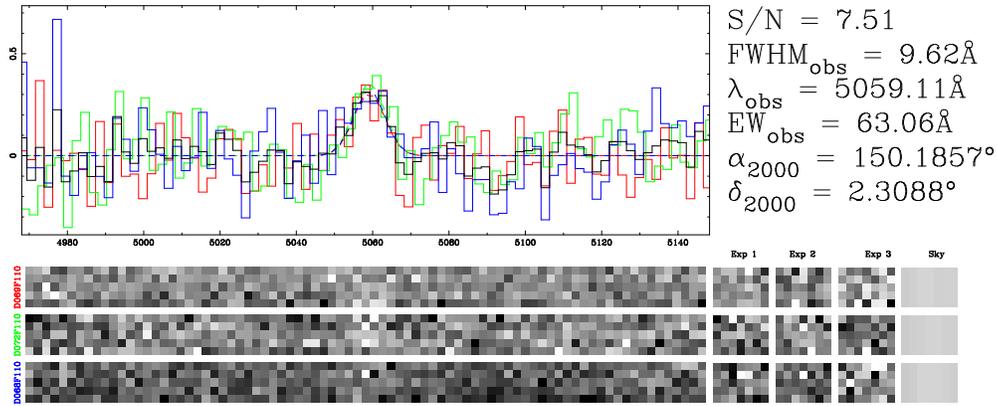}
\caption{The emission-line source detection for index 322 which 
lies in the COSMOS field. The 
spectrum-based EW suggests a low-$z$ classification. The format is the same as 
in Figure \ref{fig_highz_det}. Note that the emission line is broadened compared 
to the instrumental resolution.
}
\label{fig_unc_det}
\end{figure}

\begin{figure}
\centering
\includegraphics [scale=0.4,angle=-90]{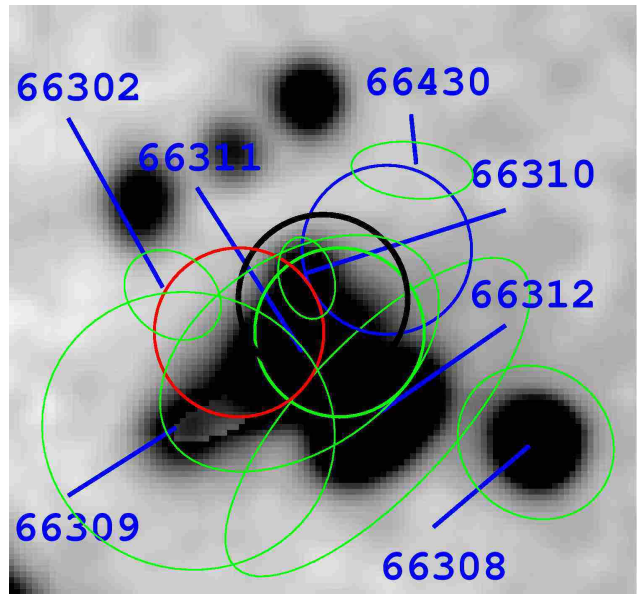}
\caption{The ground-based, $V_j$ imaging cutout for the Table
\ref{tab_line_list1} index 322. The detection spectra are given in
Figure \ref{fig_unc_det}. The format is the same as in Figure
\ref{fig_highz_ima}. The counterpart \#66311 at r$^+$=21.1 has a 84\% 
likelihood in association. The counterpart \#66312 is 
already assigned to the VIRUS-P detection of 
Table \ref{tab_line_list1} index 310 at $\lambda_{\mbox{obs}}=4948.2$\AA. 
The counterpart \#66310 at r$^+$=24.9 would be a LAE based on 
EW and looks like a reasonable candidate system resolved from \#66311 
in the HST image, but it only holds a 6\% chance of association. This 
source is associated to \#66311 as an [OII] emitter in the catalog.
}
\label{fig_unc_ima}
\end{figure}

 The detection image in Figure \ref{fig_unc_det} shows source index 322 as a high S/N, 
broadened line along with a continuum detection in three fiber positions. 
The emission line flux is $16.7_{-3.3}^{+4.4}\times10^{-17} $erg s$^{-1}$ cm$^{-2}$. 
The COSMOS image (Figure \ref{fig_unc_ima}) reveals three 
plausible counterparts. The most likely (84\%) counterpart implies an easy 
classification as an [OII] emitter with EW$_{obs}=37_{-9}^{+8}$\AA\. 
However, there is a non-trivial 
likelihood (8\%) that the counterpart not the bright galaxy, but instead the fainter object. 
In this case, the source would be classified as an LAE. Despite this uncertainty, 
we place this object in Tables \ref{tab_line_list2} with a [OII] classification.
\section{Emission line source catalog}
\label{sec_cat}
\subsection{GOODS-N comparisons}
\label{sec_compare}
\par Most of the detections and redshift classifications in this catalog are new to the 
literature. For instance in the COSMOS field, the magnitude limit of the spectroscopic cut 
(I$_{AB}<$22.5) to zCOSMOS \citep{Lil09} gives little overlap with this sample. 
Fortunately, the large number of deep spectroscopic observations in the GOODS-N field 
comprise a better test sample. We have made a detailed comparison of our measurements 
to those of \citet{Bar08}, which includes most previous GOODS-N measured redshifts. We further 
include one Ly$\alpha$ match from \citet{Low97} and 
one [OIII]5007 source from \citet{Wir04}. We note that the observations from the 
literature often have larger spectral coverage and higher resolution than our data, 
allowing alternate classification methods.
\par
\begin{figure}
\centering
\subfigure{\includegraphics[bb=41 46 372 666,clip=true,scale=0.25,angle=-90] {quick_hst1.eps}}
\subfigure{\includegraphics[scale=0.15,angle=-90] {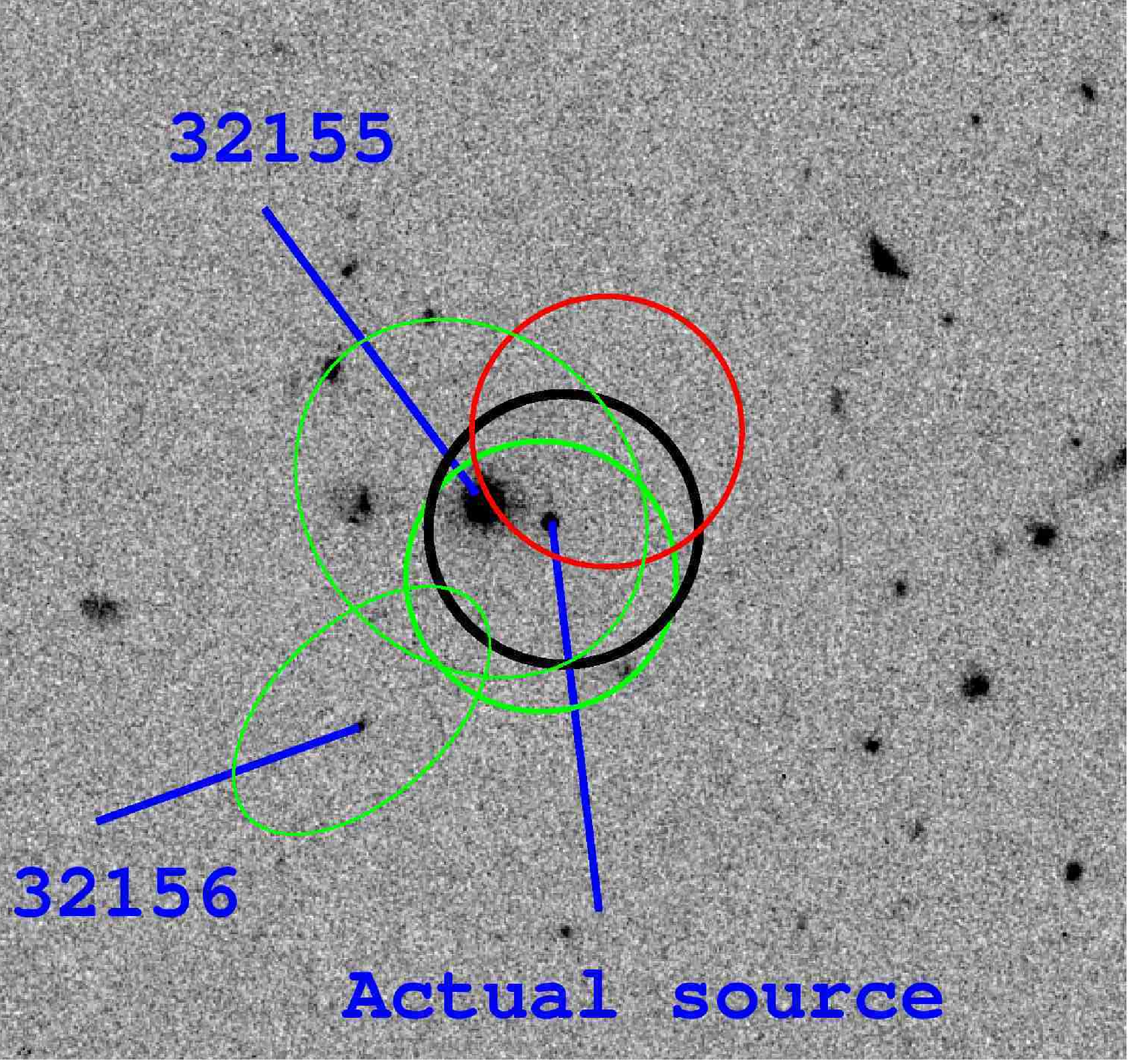}}
\vspace{-22pt}
\\
\subfigure{\includegraphics[bb=41 46 372 666,clip=true,scale=0.25,angle=-90] {quick_hst2.eps}}
\subfigure{\includegraphics[scale=0.15,angle=-90] {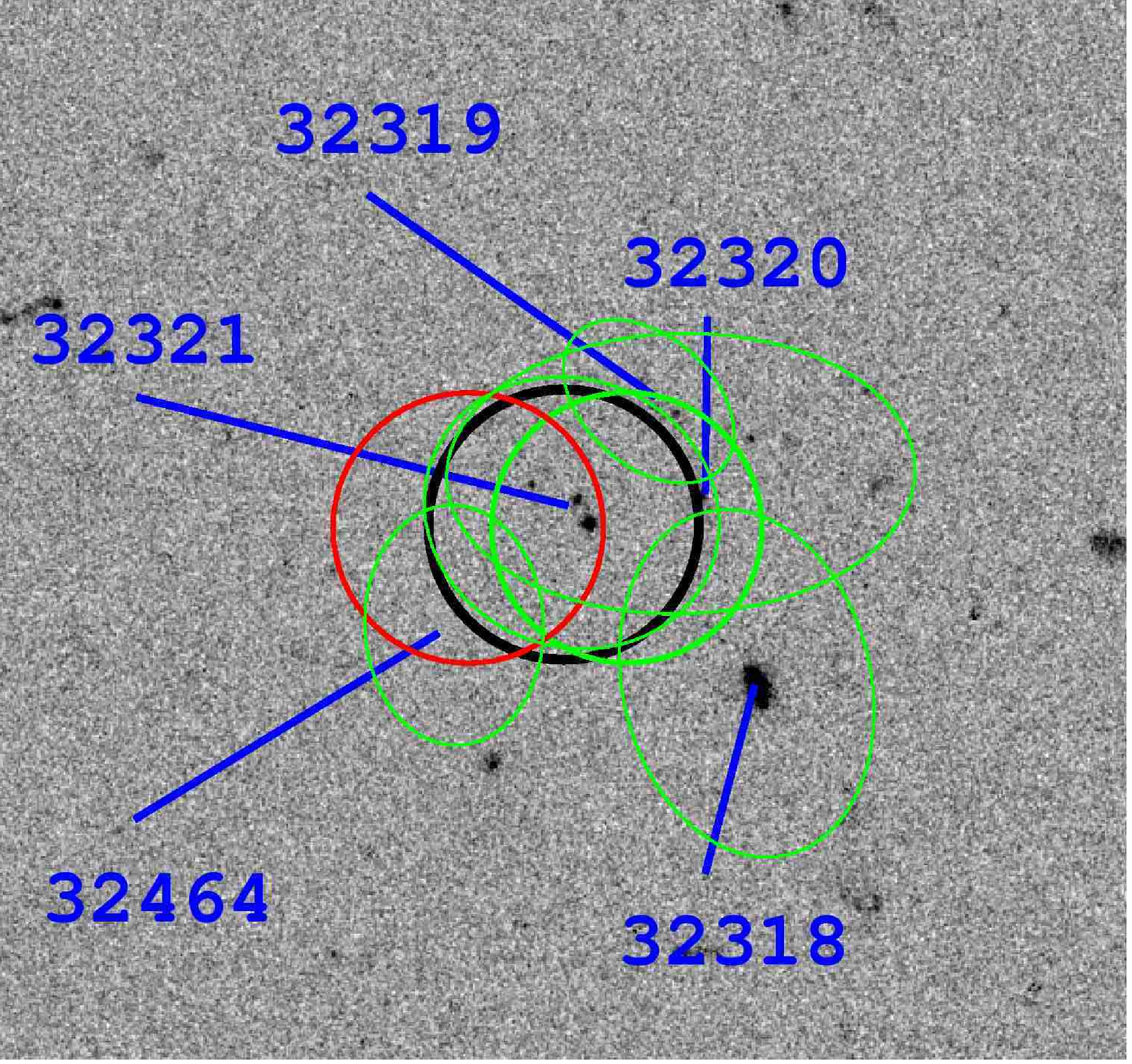}}
\vspace{-22pt}
\\
\subfigure{\includegraphics[bb=41 46 372 666,clip=true,scale=0.25,angle=-90] {quick_hst3.eps}}
\subfigure{\includegraphics[scale=0.15,angle=-90] {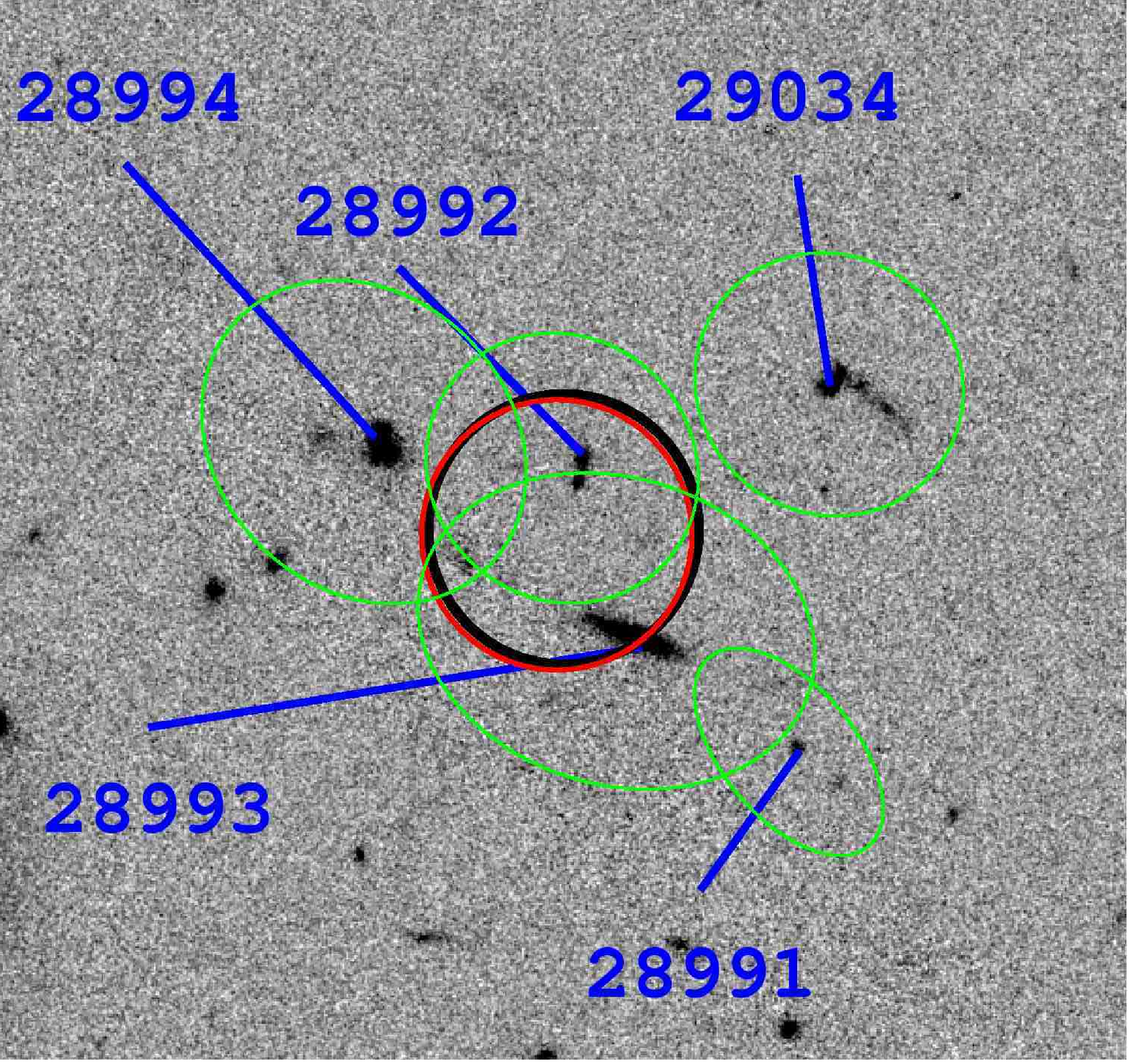}}
\vspace{-22pt}
\\
\subfigure{\includegraphics[bb=41 46 372 666,clip=true,scale=0.25,angle=-90] {quick_hst4.eps}}
\subfigure{\includegraphics[scale=0.15,angle=-90] {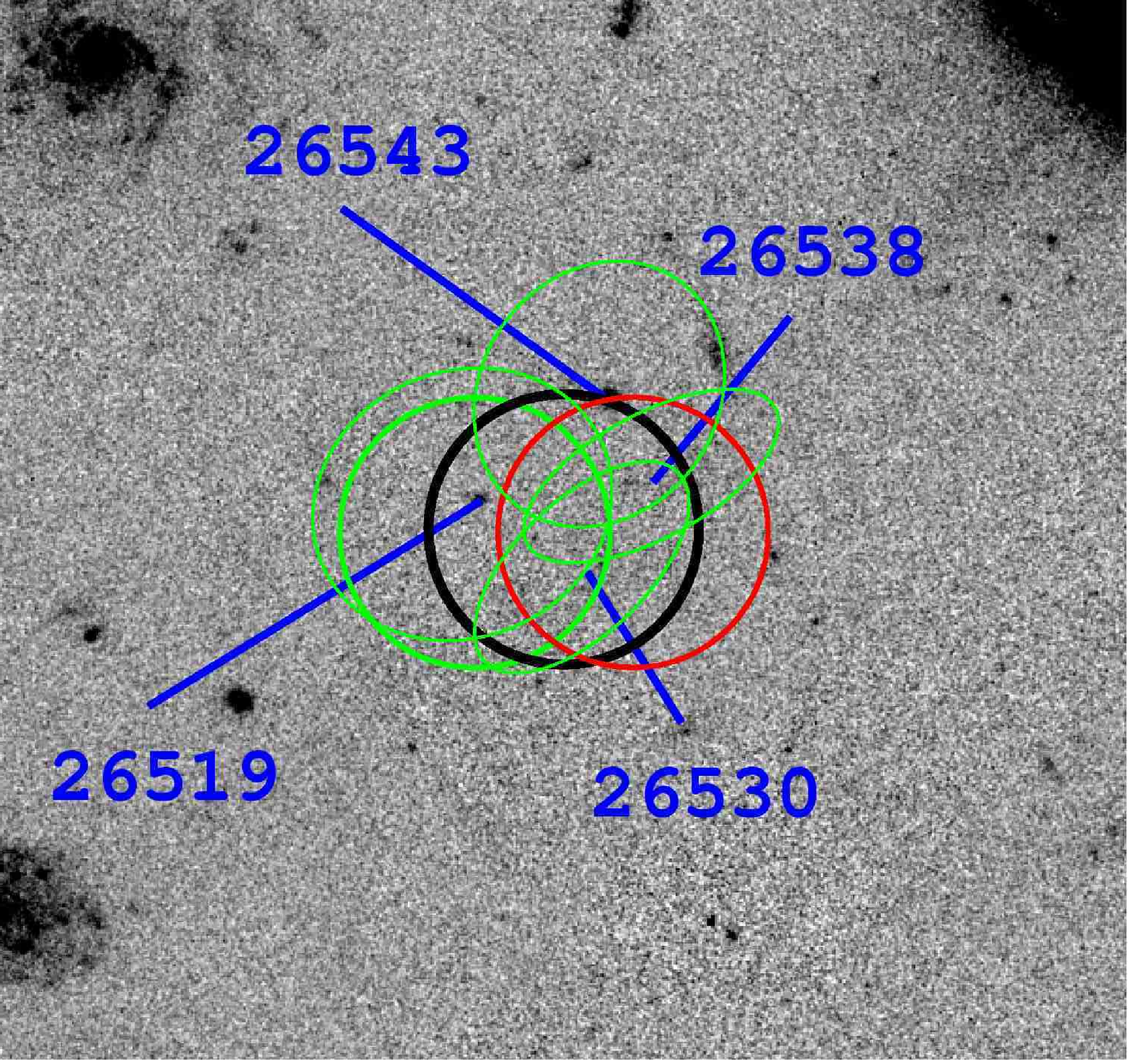}}
\vspace{-22pt}
\\
\subfigure{\includegraphics[bb=41 46 372 666,clip=true,scale=0.25,angle=-90] {quick_hst5.eps}}
\subfigure{\includegraphics[scale=0.15,angle=-90] {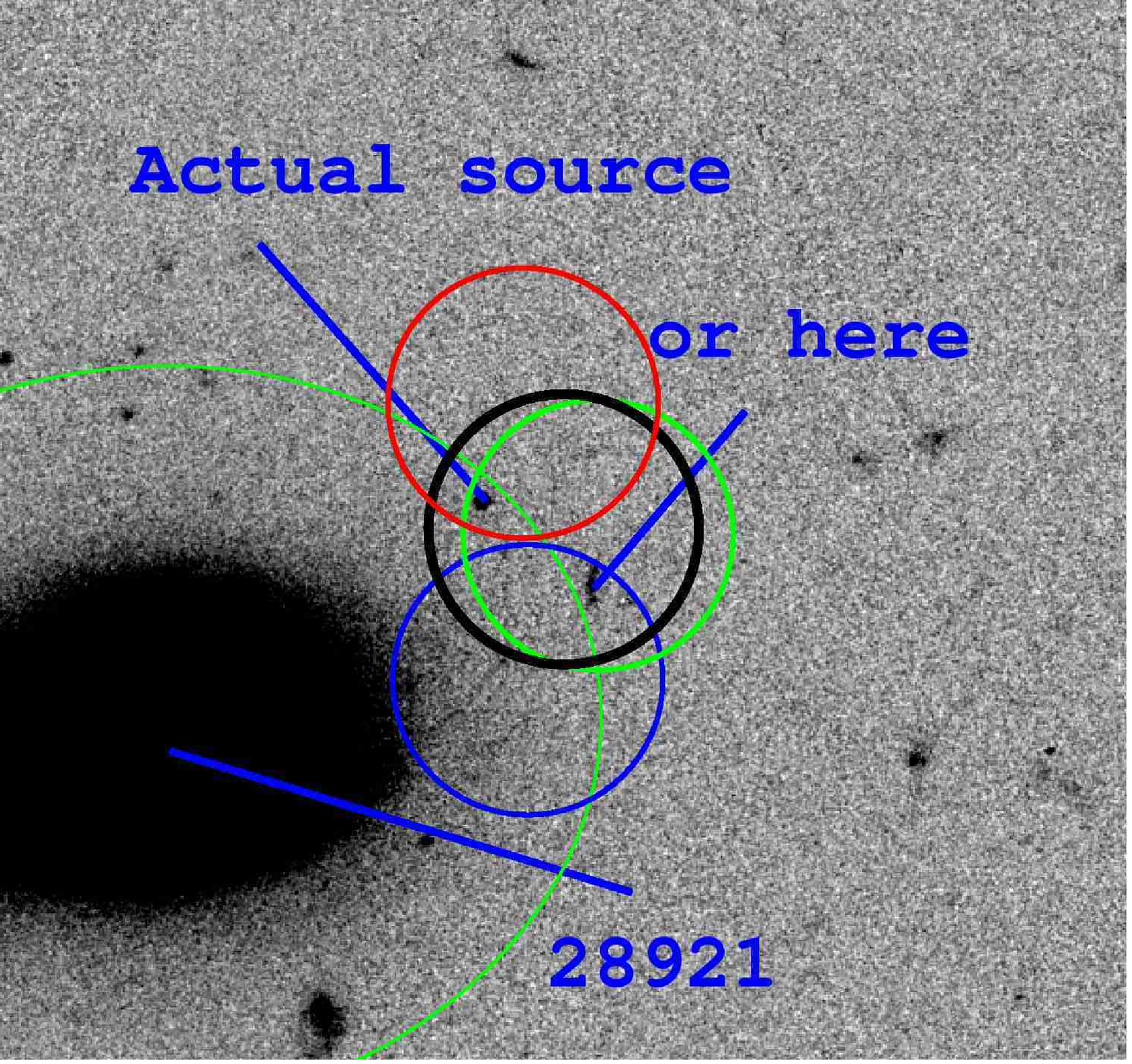}}
\vspace{-22pt}
\\
\subfigure{\includegraphics[bb=41 46 372 666,clip=true,scale=0.25,angle=-90] {quick_hst6.eps}}
\subfigure{\includegraphics[scale=0.15,angle=-90] {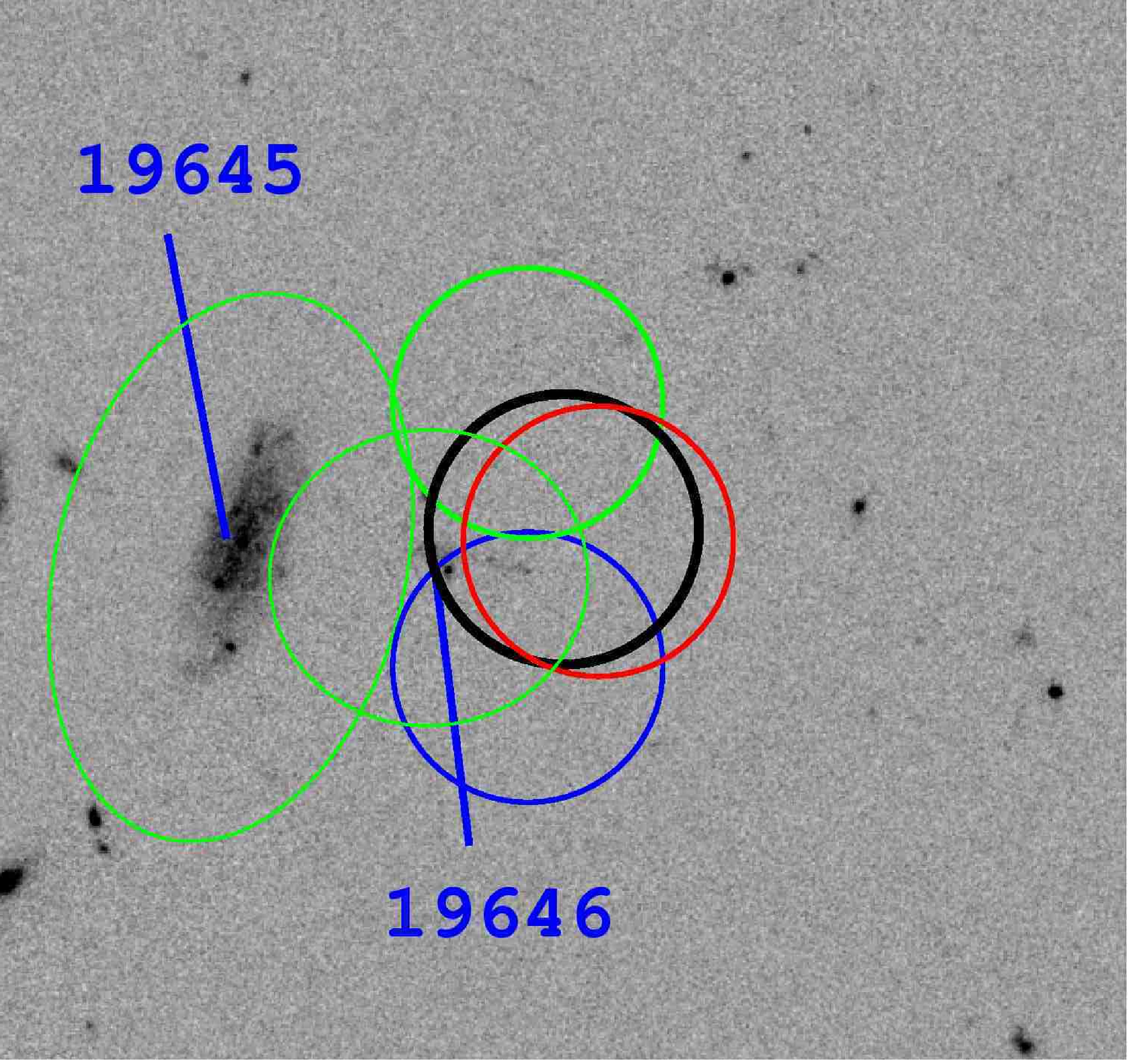}}
\vspace{0pt}
\caption{The spectral detection and HST ACS F606W \citep{Gia04} cutouts for six of the 
twelve new, high-$z$ redshift measurements in GOODS-N. From top to bottom, the 
objects are indicies 334, 338, 360, 372, 373, and 403 in Table \ref{tab_line_list2}. The 
first and fifth objects do not have identified counterparts in the ground-based 
images due to blending, although likely counterparts are identified in the HST data. 
The best counterparts for the second, third, fourth, and sixth objects are \#'s 32321, 
28992, 26519, and 19646.}
\label{fig_HDFN1}
\end{figure}

\begin{figure}
\centering
\subfigure{\includegraphics[bb=41 46 372 666,clip=true,scale=0.25,angle=-90] {quick_hst7.eps}}
\subfigure{\includegraphics[scale=0.15,angle=-90] {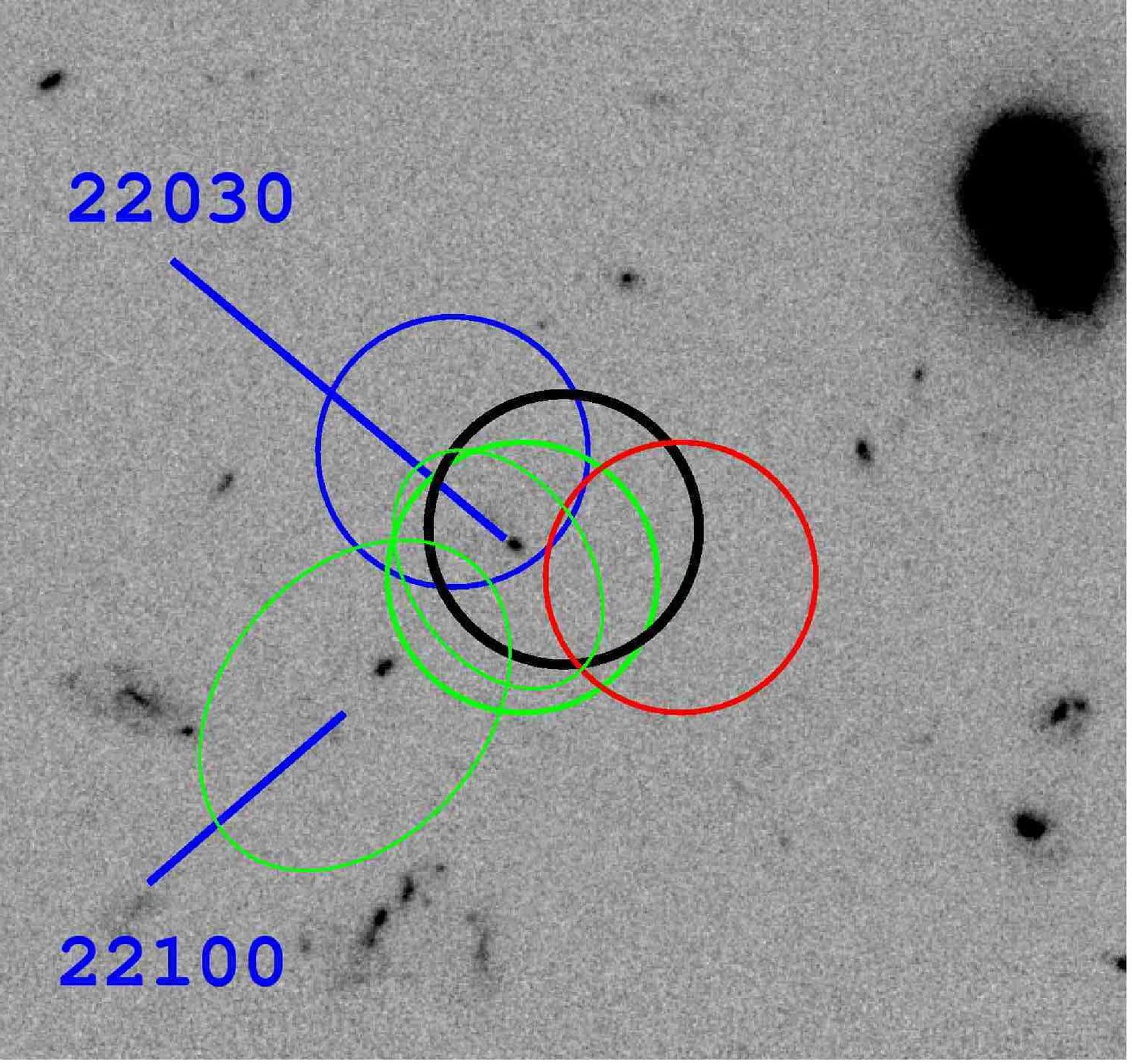}}
\vspace{-22pt}
\\
\subfigure{\includegraphics[bb=41 46 372 666,clip=true,scale=0.25,angle=-90] {quick_hst8.eps}}
\subfigure{\includegraphics[scale=0.15,angle=-90] {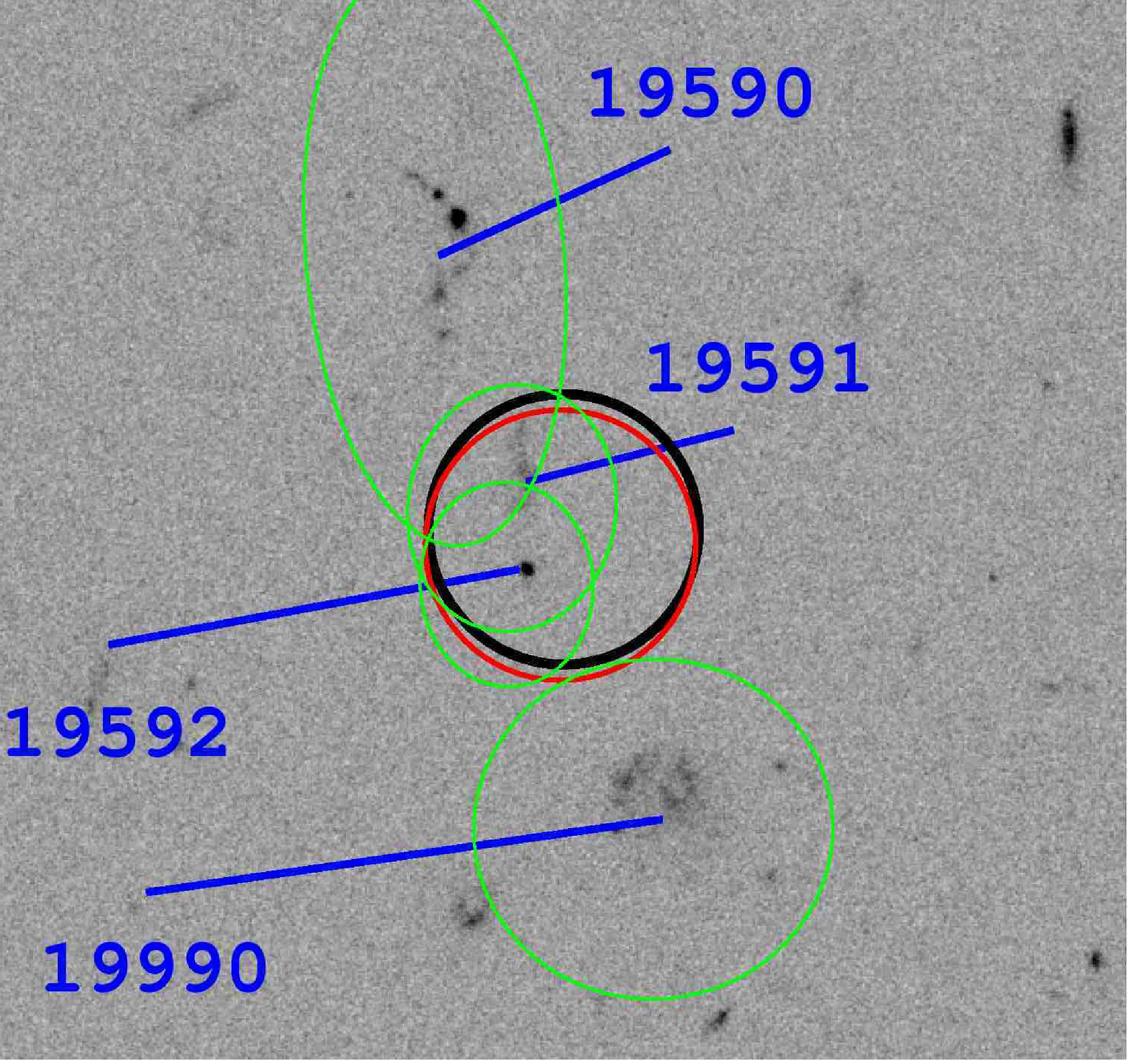}}
\vspace{-22pt}
\\
\subfigure{\includegraphics[bb=41 46 372 666,clip=true,scale=0.25,angle=-90] {quick_hst9.eps}}
\subfigure{\includegraphics[scale=0.15,angle=-90] {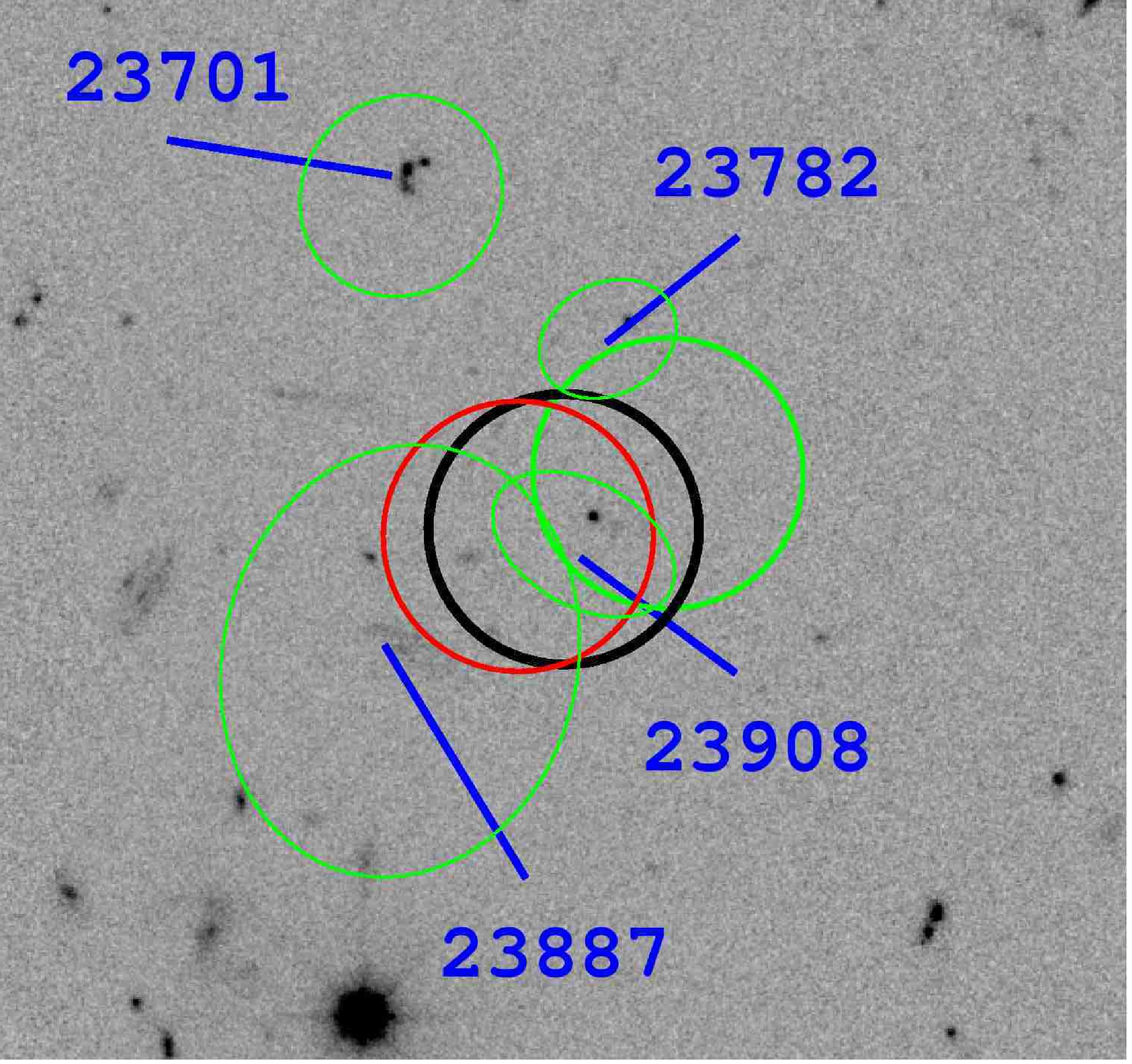}}
\vspace{-22pt}
\\
\subfigure{\includegraphics[bb=41 46 372 666,clip=true,scale=0.25,angle=-90] {quick_hst10.eps}}
\subfigure{\includegraphics[scale=0.15,angle=-90] {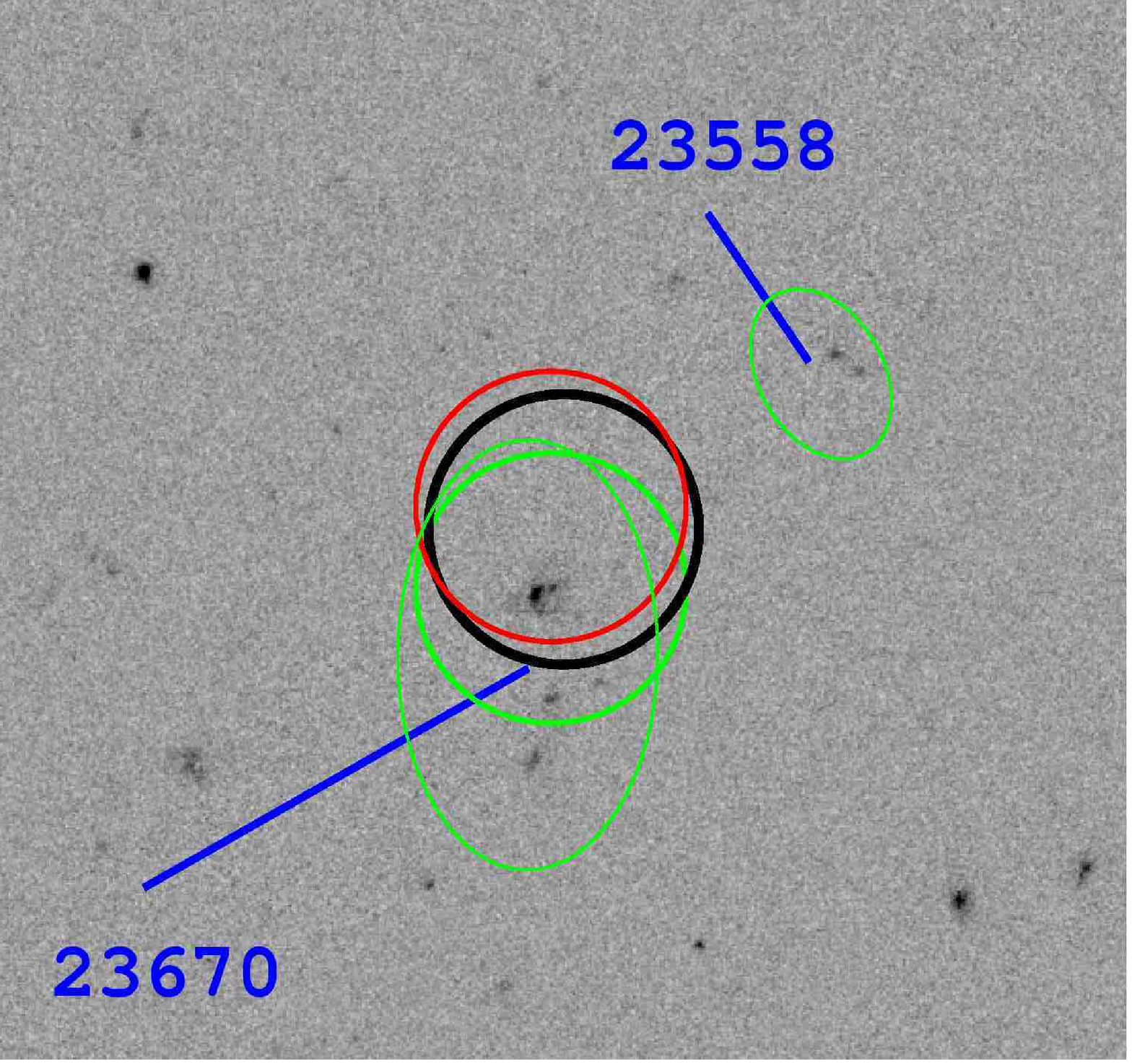}}
\vspace{-22pt}
\\
\subfigure{\includegraphics[bb=41 46 372 666,clip=true,scale=0.25,angle=-90] {quick_hst11.eps}}
\subfigure{\includegraphics[scale=0.15,angle=-90] {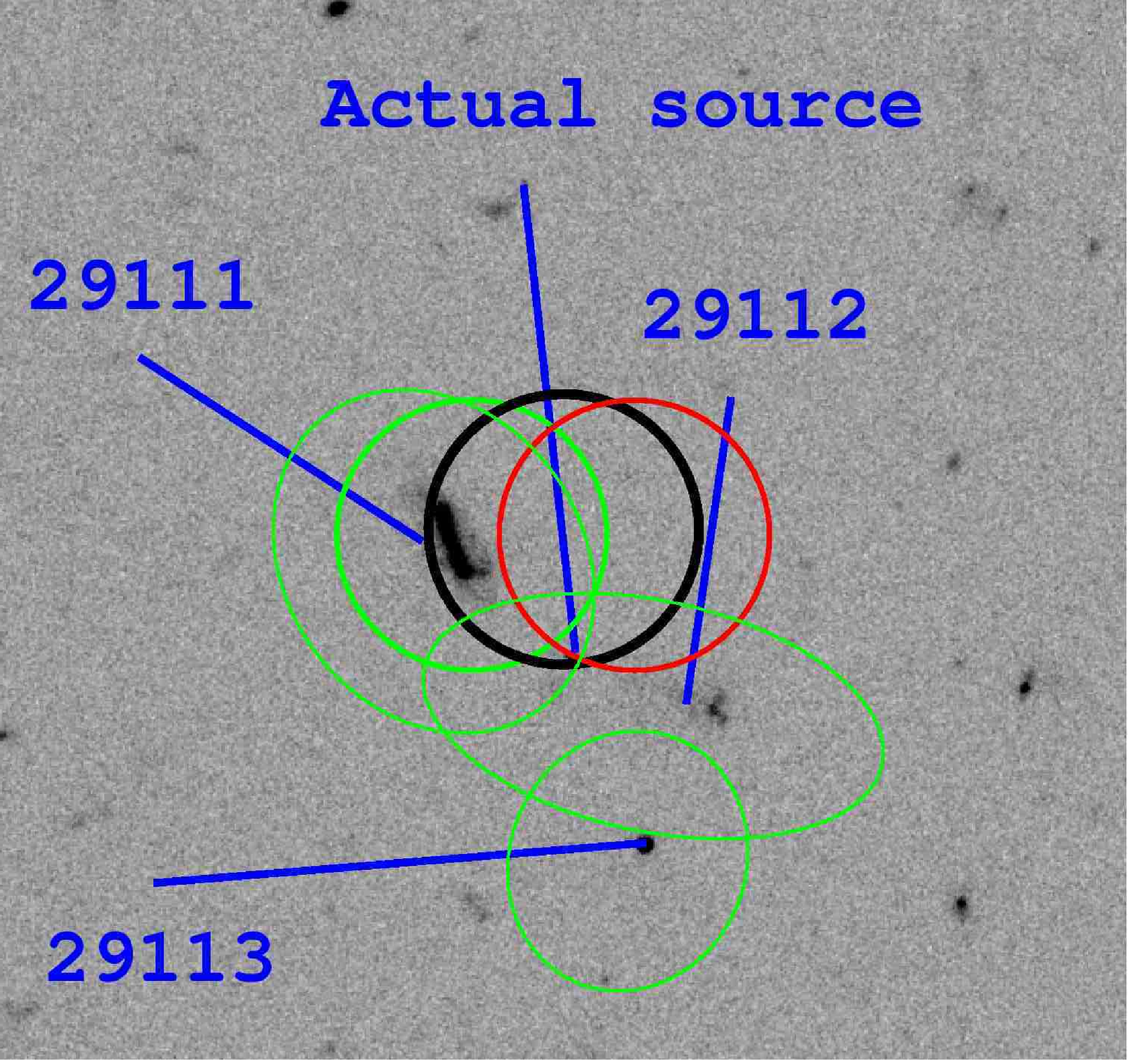}}
\vspace{-22pt}
\\
\subfigure{\includegraphics[bb=41 46 372 666,clip=true,scale=0.25,angle=-90] {quick_hst12.eps}}
\subfigure{\includegraphics[scale=0.15,angle=-90] {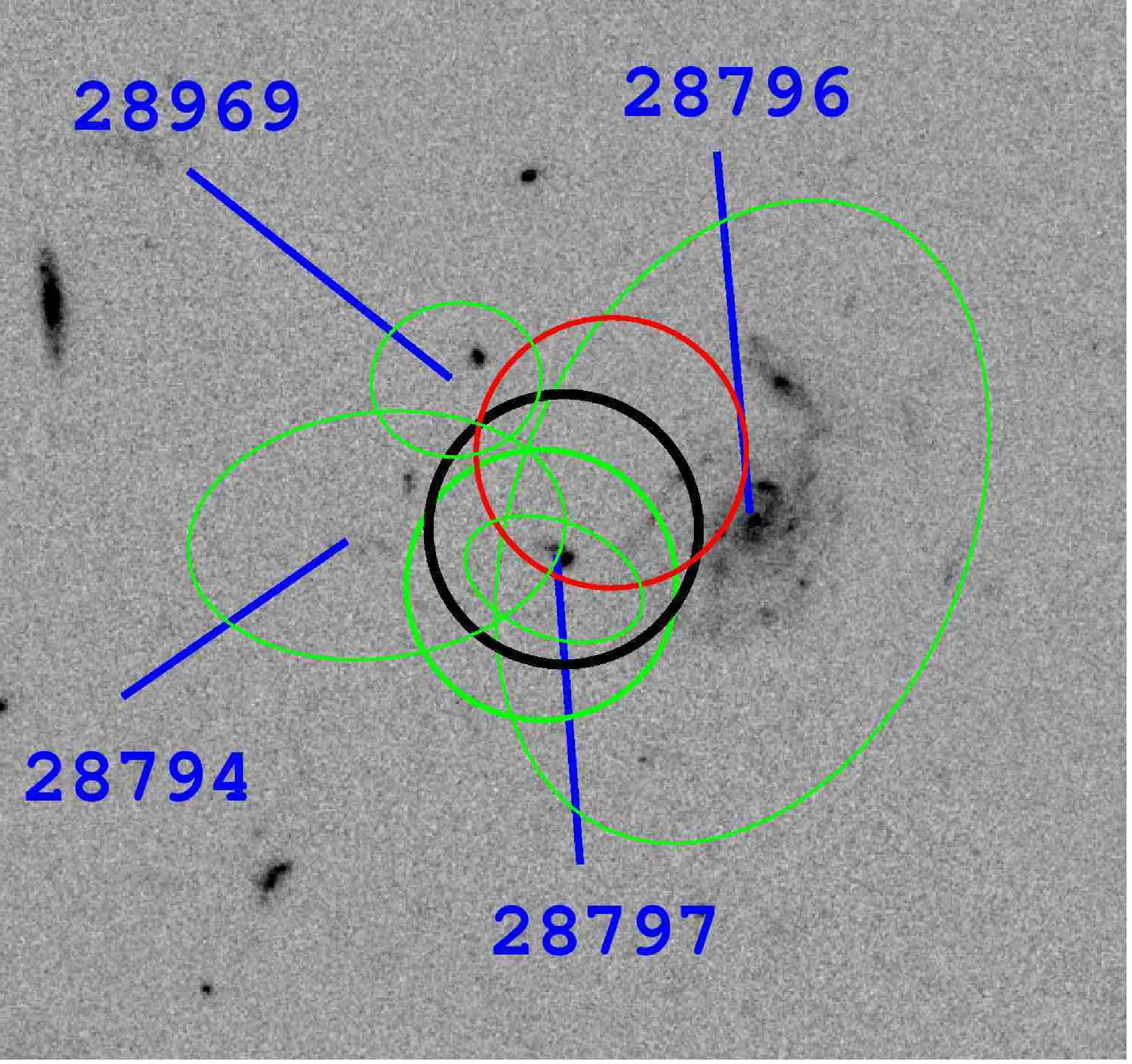}}
\vspace{0pt}
\caption{The spectral detection and HST ACS F606W \citep{Gia04} cutouts for the 
final six of twelve new, high-$z$ redshift measurements in GOODS-N. From top to bottom, the 
objects are indicies 415, 426, 434, 447, 467, and 474 in Table \ref{tab_line_list2}. The 
object with index 467 does not have an identified counterpart in the ground-based 
images due to blending, although a likely counterpart is identified in the HST data. 
The best counterparts for the remaining five, in order of listing, are 
22030, 19592, 23908, 23670, and 28797.}
\label{fig_HDFN2}
\end{figure}

 We find 119 unique emission-line sources in GOODS-N. 
Three of these do not have measured optical broadband counterparts from the 
ground-based imaging, appear to be blended with foreground objects 
when examined with the Hubble Space Telescope (HST) images of 
\citet{Gia04}, and are without published redshifts. We classify these as LAEs. In addition, 
there are nine other LAEs where we do measure a robust continuum counterpart but 
that are without published redshifts. We give the twelve new LAEs in 
GOODS-N in Figures \ref{fig_HDFN1} and \ref{fig_HDFN2}. In addition, we find 92 low-$z$ 
objects in common to the literature and 13 high-z objects (12 Ly$\alpha$ and 1 [CIV]). 
Finally, we find only two ojects in our catalog were mis-classified. 
Source index 371 was originally called an LAE, but the literature reveals it to 
be a CIV1549 emitter. Source index 400 was originally called an [OII] emitter, but the 
literature reveals it to be an [OIII]5007 emitter. 
We have rectified Table \ref{tab_line_list2} to reflect these two cases, and we then 
find an rms in $\frac{\Delta z}{1+z}$ of 0.001 and no offset compared 
to the literature, which is completely consistent with our 0.5\AA\ line center uncertainty. 
A weakness in the literature samples is the lack of emission-line flux 
calibration, so we cannot use the previous samples to quantitatively 
test this survey's completeness. We have qualitatively confirmed the 
completeness by searching for literature objects in our spectra and 
finding many dozen at $3<S/N<5$. 
\subsection{Catalog summary}
\label{sec_samp}
\par Table \ref{tab_line_list1} contains a segment of the detected emission-line catalog  
with the full version available electronically. 
The entry ".." is given where there is not an applicable value. Each emission-line 
is prefixed with the identifier ``HPS'' to stand for HETDEX Pilot Survey.
The column descriptions are: (1) the catalog number, 
(2) the emission-line right ascension in hrs:min:sec (J2000), 
(3) the emission-line declination \arcdeg:\arcmin:\arcsec\ (J2000), 
(4) the observed emission wavelength in vacuum air (\AA) (with 
an estimated 0.5\AA\ uncertainty based on simulations), 
(5) the intrinsic FWHM of the emission line (km s$^{-1}$) after removal of 
a 5\AA\ FWHM instrumental resolution and (with an estimated 
300 km s$^{-1}$ uncertainty), 
(6) the S/N of the emission-line flux detected within the 
aperture set of fibers, 
(7) the emission-line flux and error in 10$^{-17}$ erg s$^{-1}$ cm$^{-2}$ 
as measured with the curve of growth method (\S \ref{sec_fl_meas}), 
(8) the spatial FWHM of the emission line (\arcsec) 
as measured with the curve of growth method (\S \ref{sec_fl_meas}), and 
(9) any additional entries in the table 
that share a position and redshift with the emission line (i.e. as with detections 
of other emission lines from the same source). 
\par Table \ref{tab_line_list2} shows a segment of the counterpart and classification information for each emission-line detection 
with the full version available electronically. 
The entry ".." is given where there is not an applicable value. 
The column descriptions are: (1) the catalog number, 
(2) the best continuum-selected counterpart in the 
standard J2000 naming convention, 
(3) the $R$-band magnitude for this best counterpart (or 
the i' magnitude for MUNICS), 
(4) the probability of counterpart association (from Equation \ref{eq_spat_prob}), 
(5) the rest-frame EW and uncertainties for this 
counterpart and the selected transition based on the $R$-band 
photometry, or the i'-band in MUNICS where no $R$-band 
is available (\AA), 
(6) the rest-frame EW and uncertainties for this 
counterpart and the selected transition based on an 
interpolation between the two nearest filters (\AA), 
(7) the transition of the emission line based on the 
EW cut and/or the presence of multiple emission lines, 
(8) the estimated redshift, 
(9) the probability of the emission line being 
Ly$\alpha$ as calculated by marginalizing over all potential counterparts, and 
(10) the X-ray counterpart in the
standard J2000 naming convention. 
\subsection{Catalog properties}
\par
\begin{figure}
\centering
\subfigure{\includegraphics[scale=0.3,angle=-90] {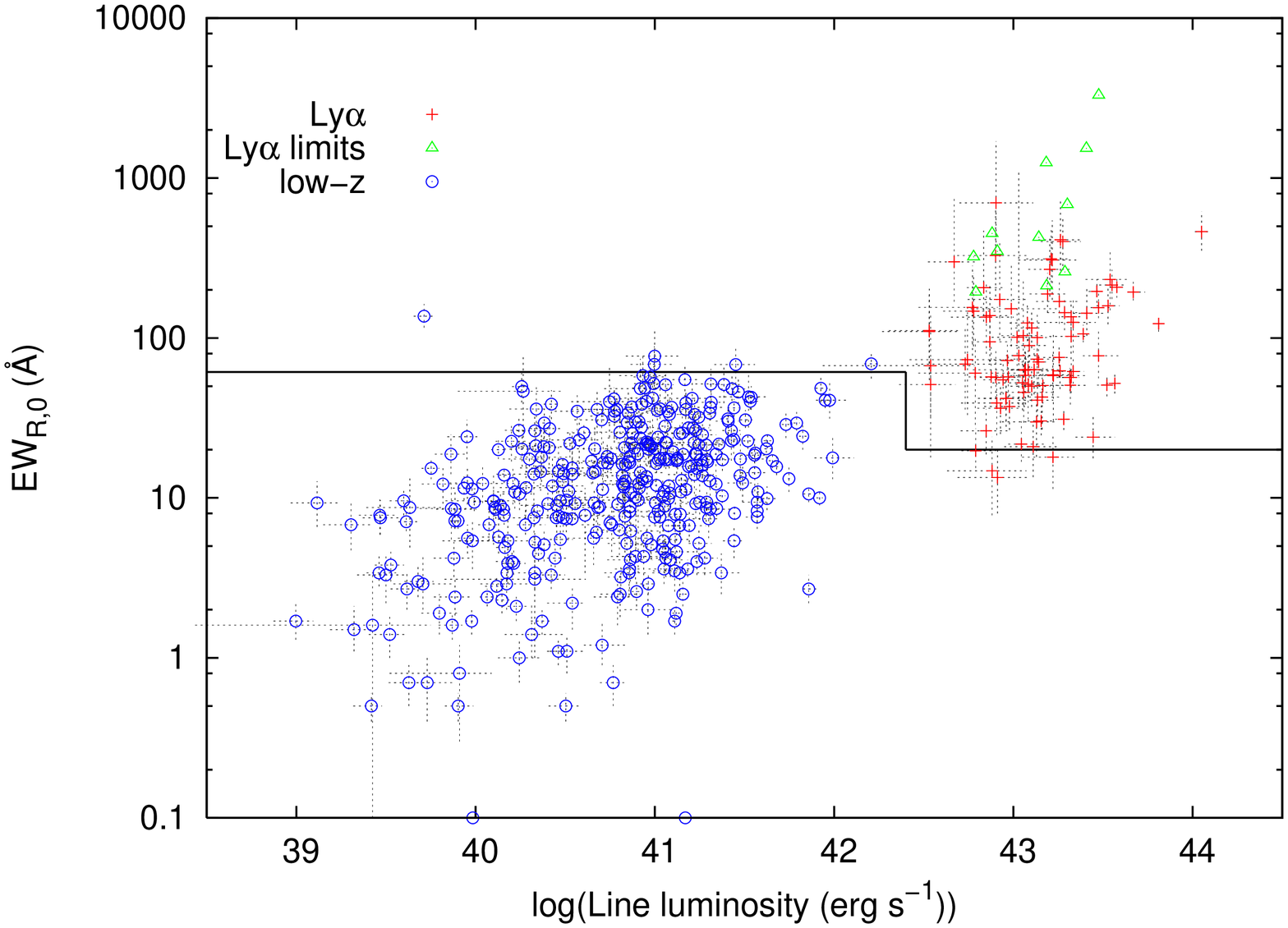}}
\subfigure{\includegraphics[scale=0.3,angle=-90] {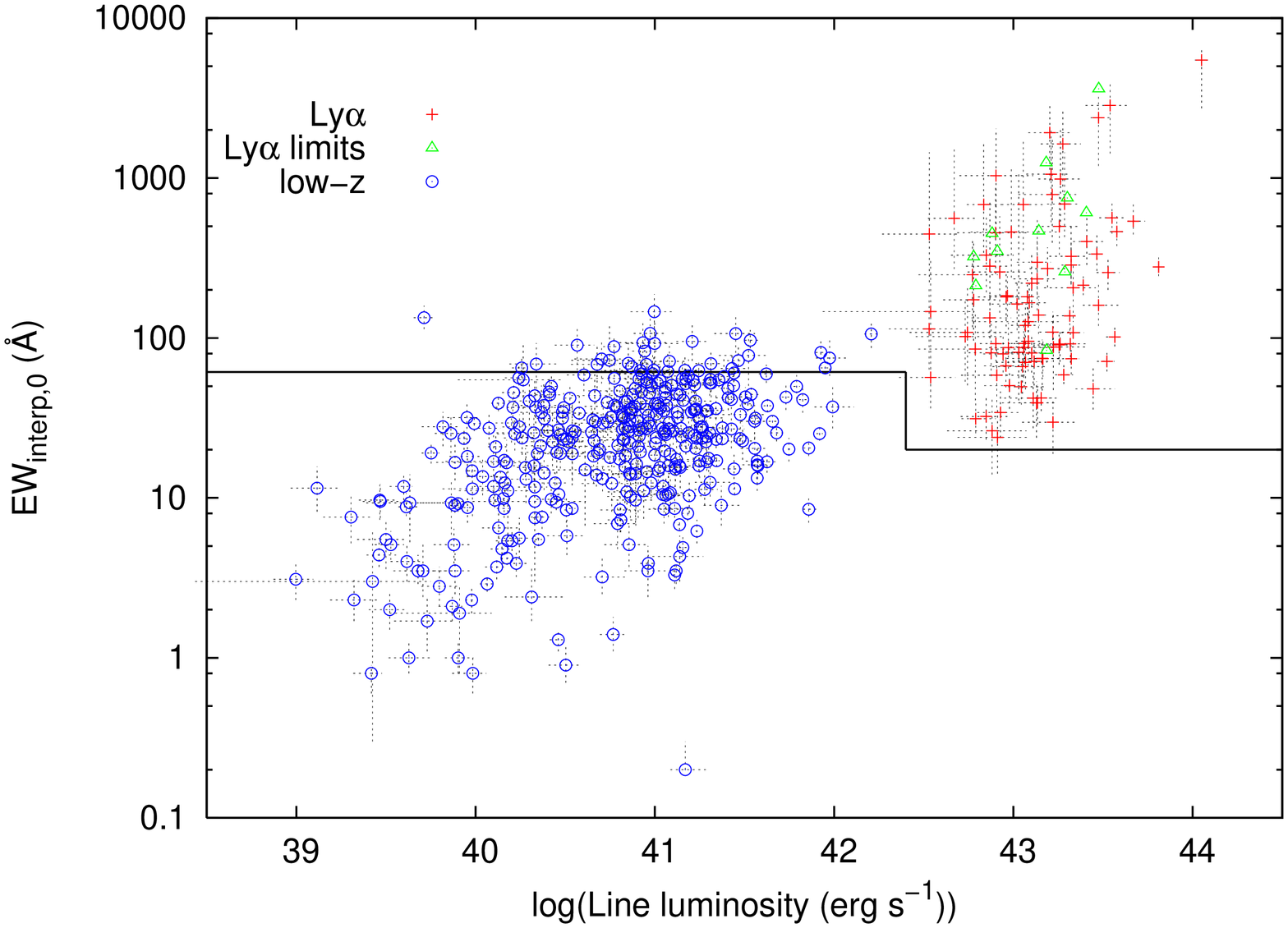}}
\caption{The distribution of rest-frame equivalent width 
and line luminosity for both the LAE and low-$z$ samples. The primary 
classification line based on EW is drawn. The jog in the 
EW cut line is simply due to the cut being defined as 
EW$_{\mbox{rest}}>20$\AA\ assuming the line to be Ly$\alpha$ so the 
equivalent threshold in the [OII] restframe is 61\AA. The 
drawn EW cut doesn't strictly apply to the low-$z$ objects with emission 
at transitions other than [OII]. A trend between higher 
EW and line luminosity in the LAEs is somewhat visible but 
noisy over this survey's dynamic range. 
The same trend is seen in surveys with lower flux limits and 
discussed in \citet{Cas10}.
\textit{Left} Continuum estimated only from the $R$-band photometry 
(or the i'-band in MUNICS). \textit{Right} Continuum estimated from 
interpolation with the two nearest filters bounding each emission line's 
wavelength.
}
\label{fig_EW}
\end{figure}

\begin{figure}
\centering
\includegraphics [scale=0.4,angle=-90]{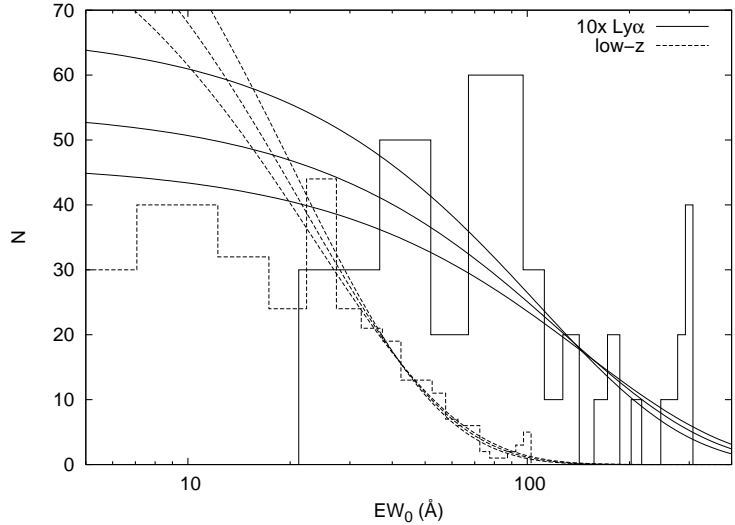}
\caption{The distribution of rest frame equivalent width 
for values with S/N$>$3 for both the LAE and low-$z$ sample by interpolating 
from the bounding broadband data. The LAE histogram has been scaled 
by 10$\times$ for visual clarity. A maximum likelihood fit 
was made by taking EW$_{\mbox{rest}}>20$\AA\ where the samples should be complete. An 
exponential scale length of 128$\pm$20\AA\ fits 
the LAE distribution and 22$\pm$1.6\AA\ fits
the [OII] distribution. The exponential fits and error ranges 
are also plotted. The largest plotted bins contain all 
values that lie higher than the histogram range.}
\label{fig_EWdist}
\end{figure}

\begin{figure}
\centering
\includegraphics [scale=0.4,angle=-90]{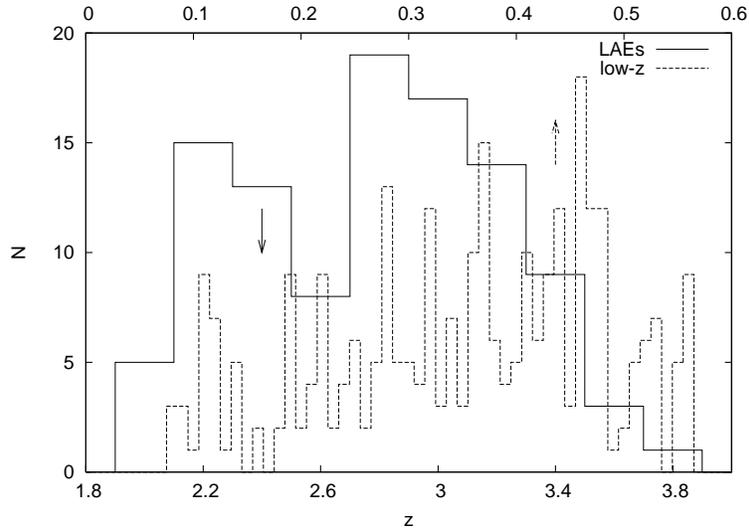}
\caption{The distribution in redshift for all the survey 
objects. The upper axis labels apply to the low-$z$ sample 
and the lower axis labels to the LAE sample. Each significantly 
overdense low-$z$ bin is primarily contained in one survey field. 
The overdensities may be early indicators of groups or clusters, but 
the low number statistics preclude firm classification. There are 
no clusters from the \citet{Koes07} catalog in any of this 
pilot survey's area.}
\label{fig_zdist}
\end{figure}

\begin{figure}
\centering
\subfigure{\includegraphics[scale=0.3,angle=-90] {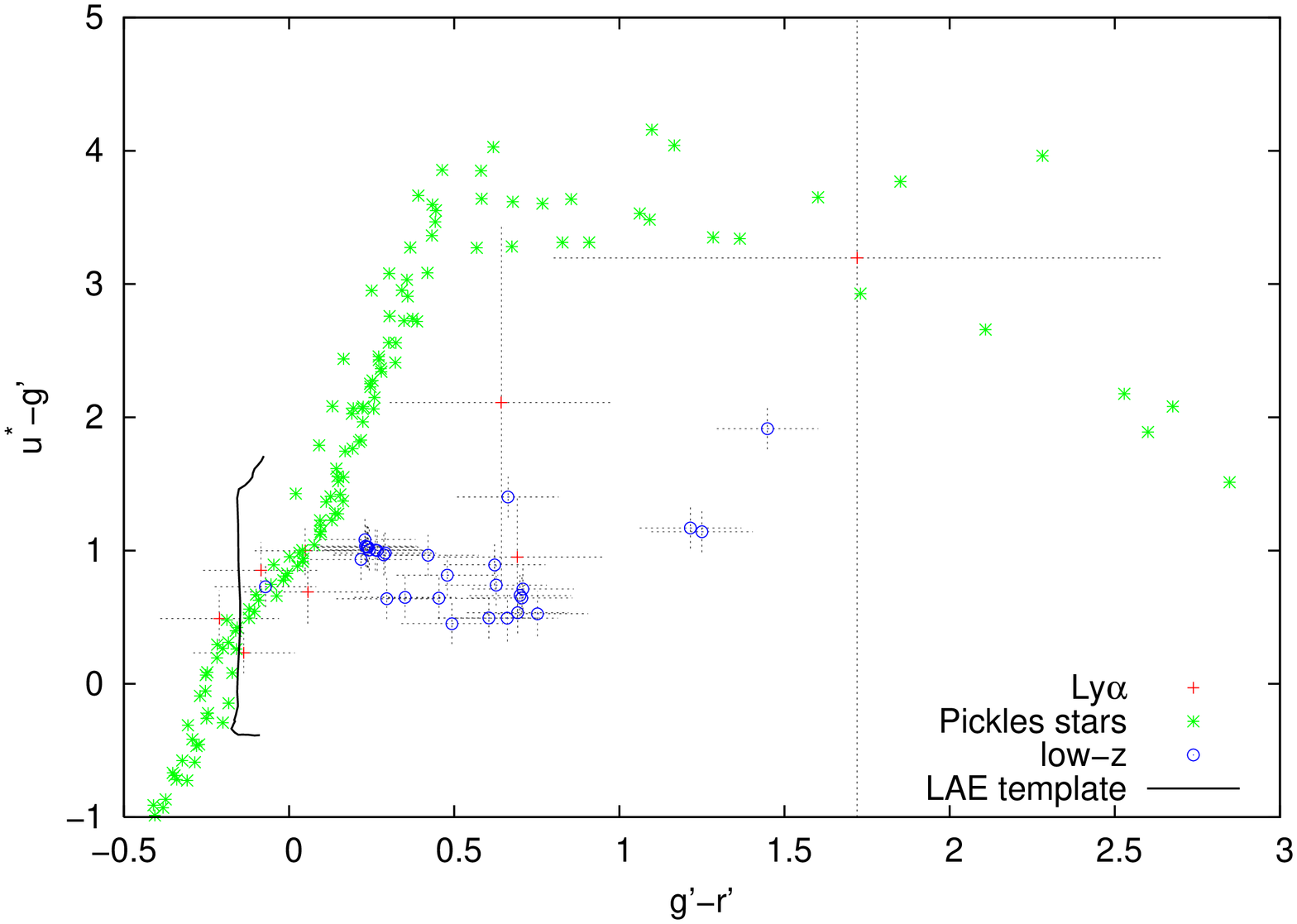}}\\
\subfigure{\includegraphics[scale=0.3,angle=-90] {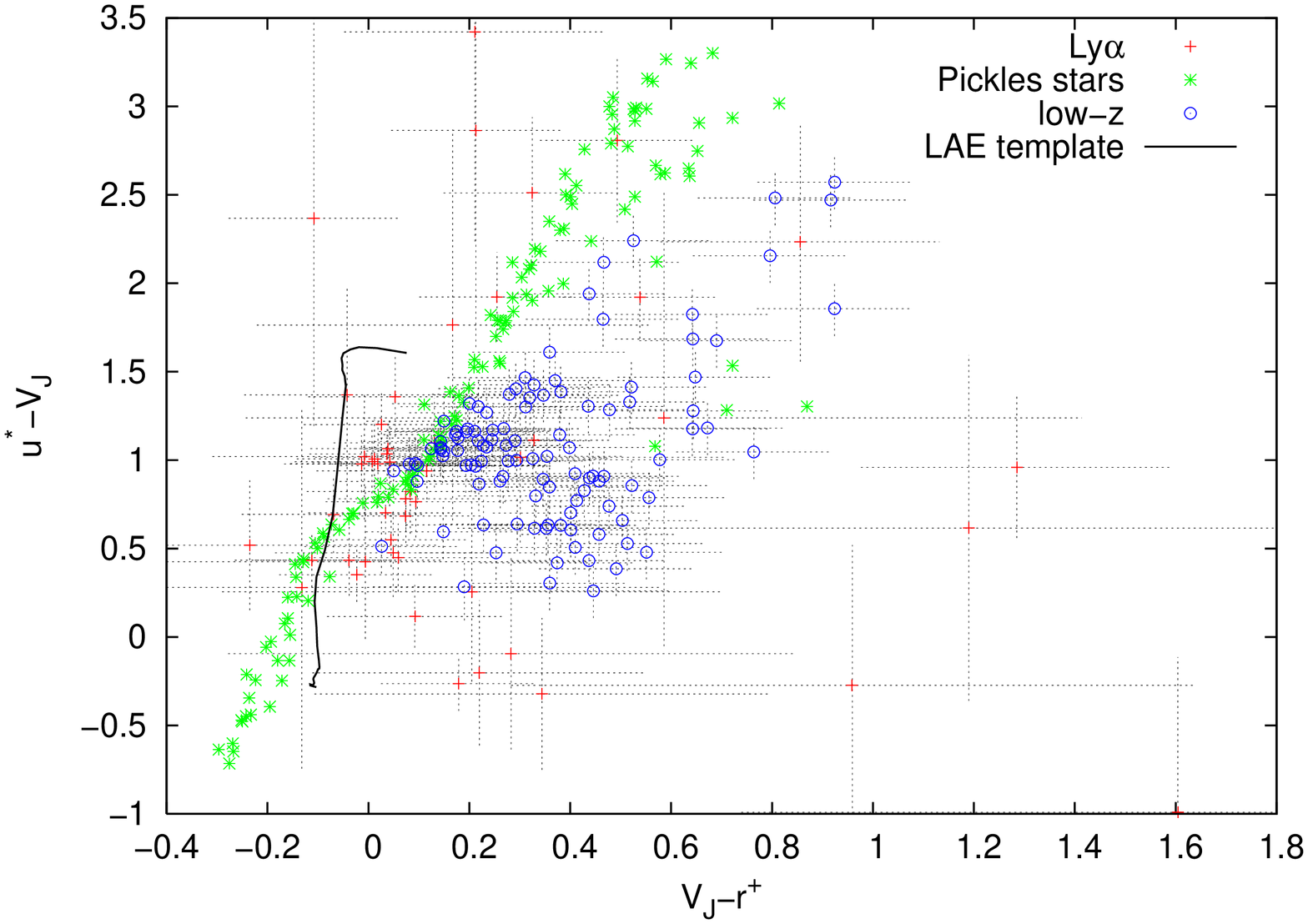}}\\
\subfigure{\includegraphics[scale=0.3,angle=-90] {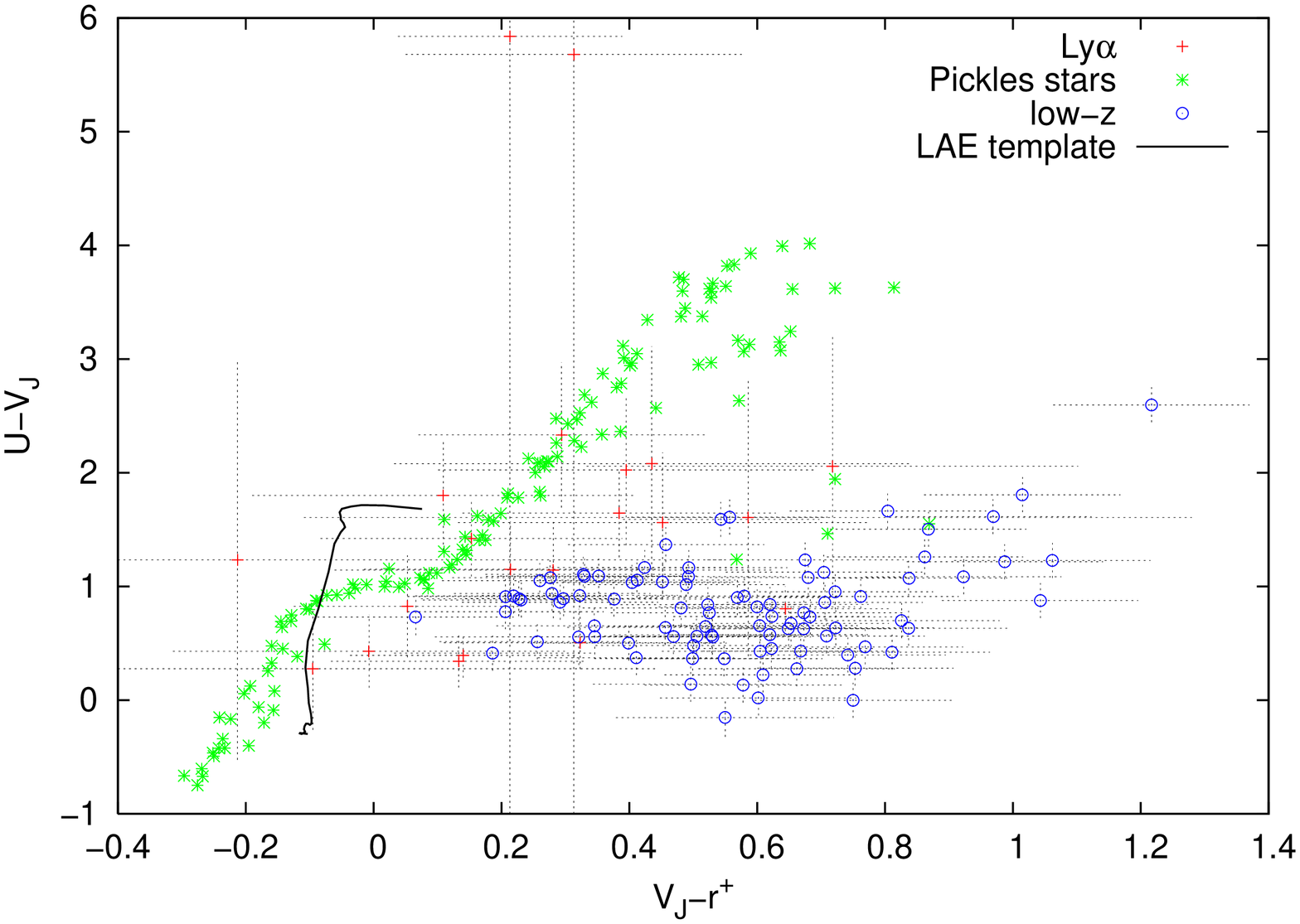}}
\caption{Color-color plots in the photometry bands that commonly define the 
Lyman Break Galaxy selection. Detections are not shown for the MUNICS 
field where we lack $U$-band data. The LBG selection rules 
are sensitive to the exact filter and telescope choice, so we do not 
transform these filter data into systems with published LBG rules. 
Instead, we synthesize colors of the \citet{Gaw07} LAE template 
as the solid, black curve for $1.3<z<4.5$ and stars \citep{Pic98}. Albeit with some 
exceptions and frequently large color errors, the 
LAE sample is segregated from the low-$z$ objects and 
lies where expected. \textit{Top} XMM-LSS objects, 
\textit{Middle} COSMOS objects, and \textit{Bottom} GOODS-N objects.}
\label{fig_color}
\end{figure}

 Figure \ref{fig_EW} compares the distribution of rest frame EW for LAEs and [OII] emitters 
to the emission-line luminosity. 
Histograms of the rest frame EW distributions of both low and high redshift sources 
are shown in Figure \ref{fig_EWdist}. A maximum likelihood fit 
was made by taking EW$_{\mbox{rest}}>20$\AA\ where the samples should be complete. An 
exponential scale length of 128$\pm$20\AA\ fits the LAE distribution and 22$\pm$1.6\AA\ fits 
the [OII] distribution. 
The redshift distribution of all sources is 
given in Figure \ref{fig_zdist}. No previously identified 
groups or clusters lie in our fields \citep[e.g.][]{Koek07}. 
In Figure \ref{fig_color}, we give the 
color-color diagram for the sample's LAEs. We do not try to transform the filter systems into 
filter sets which are usually applied to LBG and BX galaxy samples, but we do plot the 
location of the LAE spectral template from \citet{Gaw07}, made from \citet{Mara05} stellar 
population synthesis modeling, over the relevant redshifts. We 
also show the locus of stars from \citet{Pic98}. Many of the LAEs 
appear consistent with color space expectations based on continuum-selected samples.
\subsubsection{Spatially extended high-$z$ sources}
\par
\begin{figure}
\centering
\subfigure{\includegraphics[bb=41 46 372 666,clip=true,scale=0.33,angle=-90] {fig_extenda1.eps}}
\subfigure{\includegraphics[scale=0.2,angle=-90] {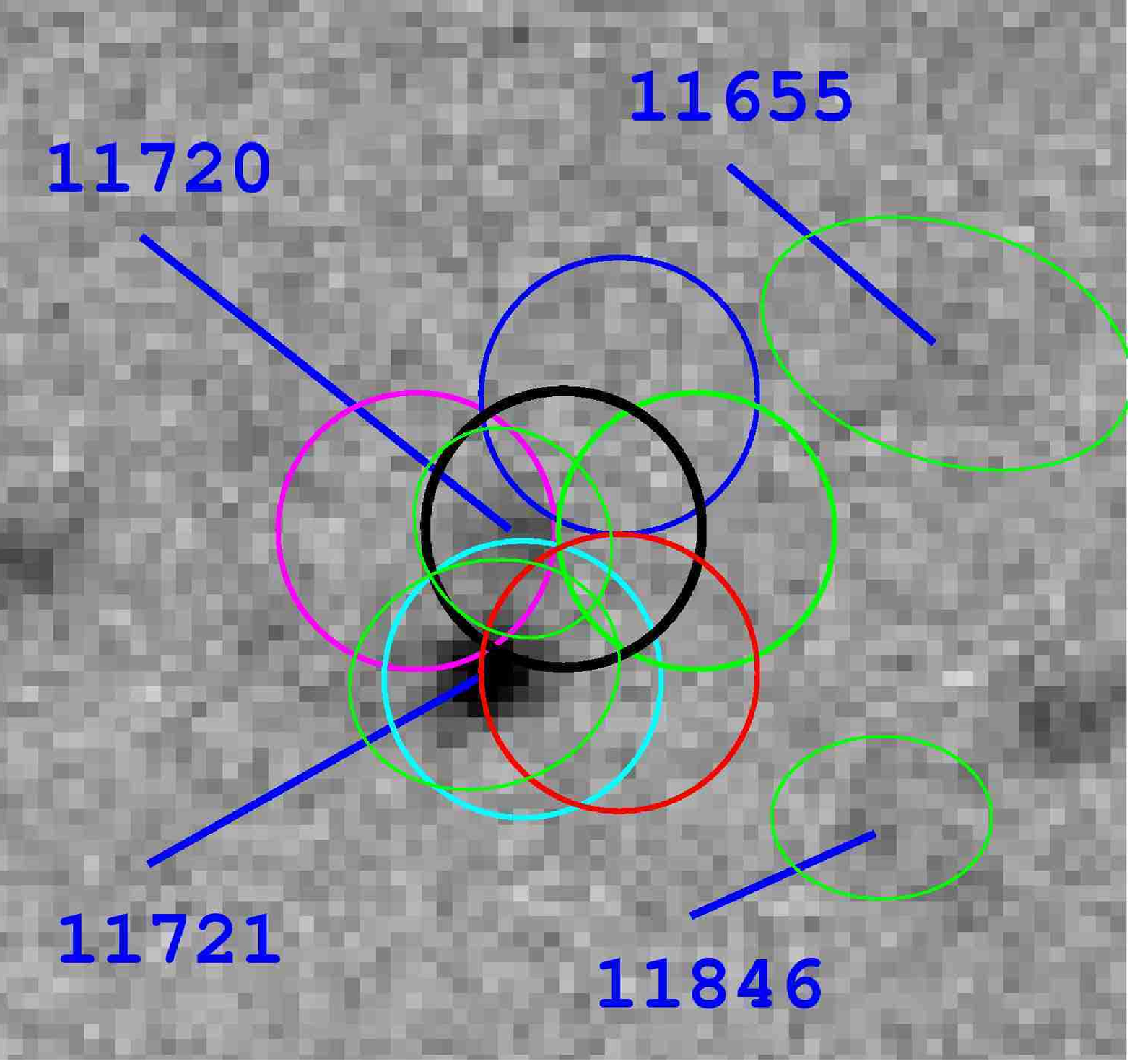}}
\vspace{-22pt}
\\
\subfigure{\includegraphics[bb=41 46 372 666,clip=true,scale=0.33,angle=-90] {fig_extenda2.eps}}
\subfigure{\includegraphics[scale=0.2,angle=-90] {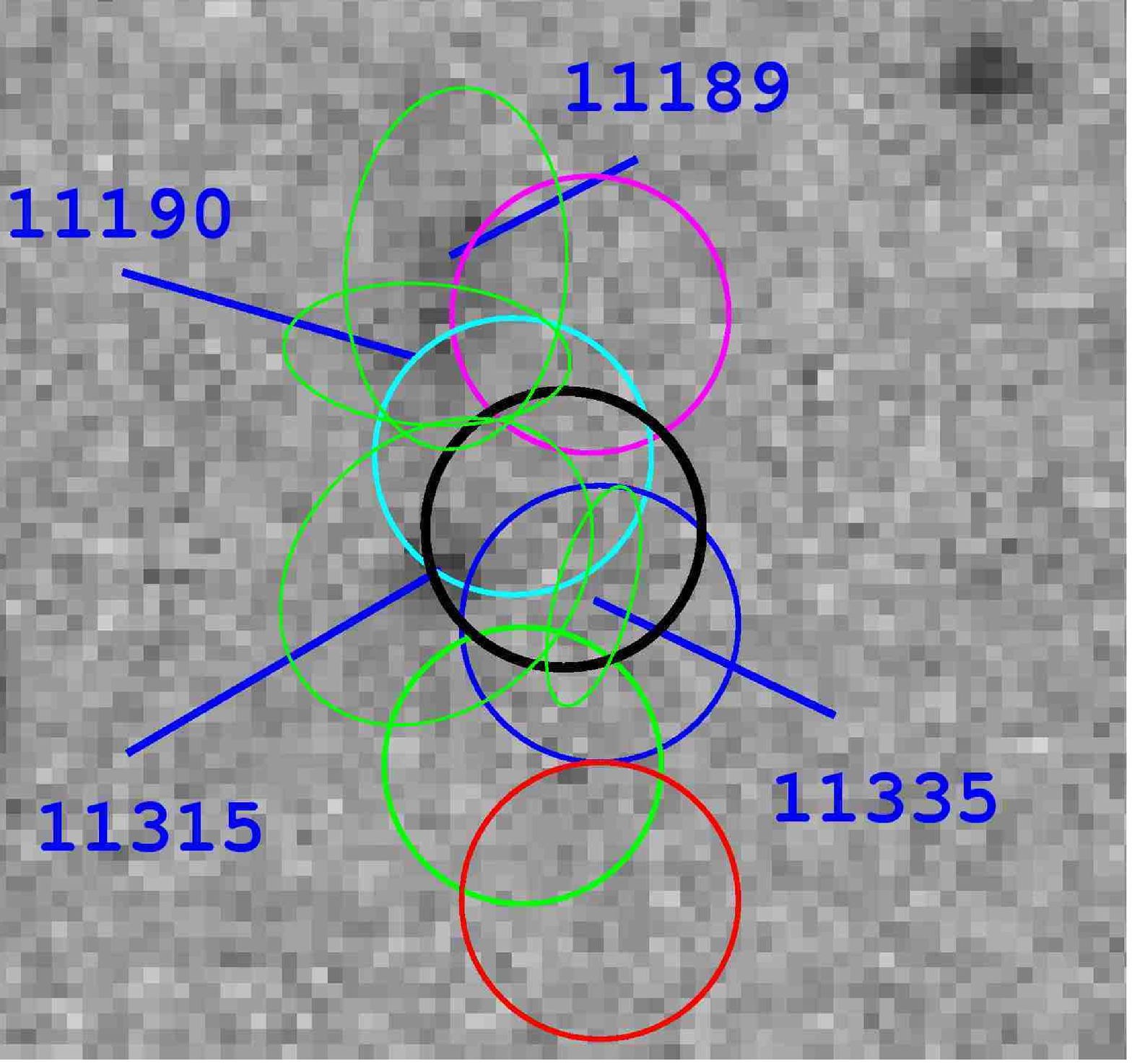}}
\vspace{-22pt}
\\
\subfigure{\includegraphics[bb=41 46 372 666,clip=true,scale=0.33,angle=-90] {fig_extenda3.eps}}
\subfigure{\includegraphics[scale=0.2,angle=-90] {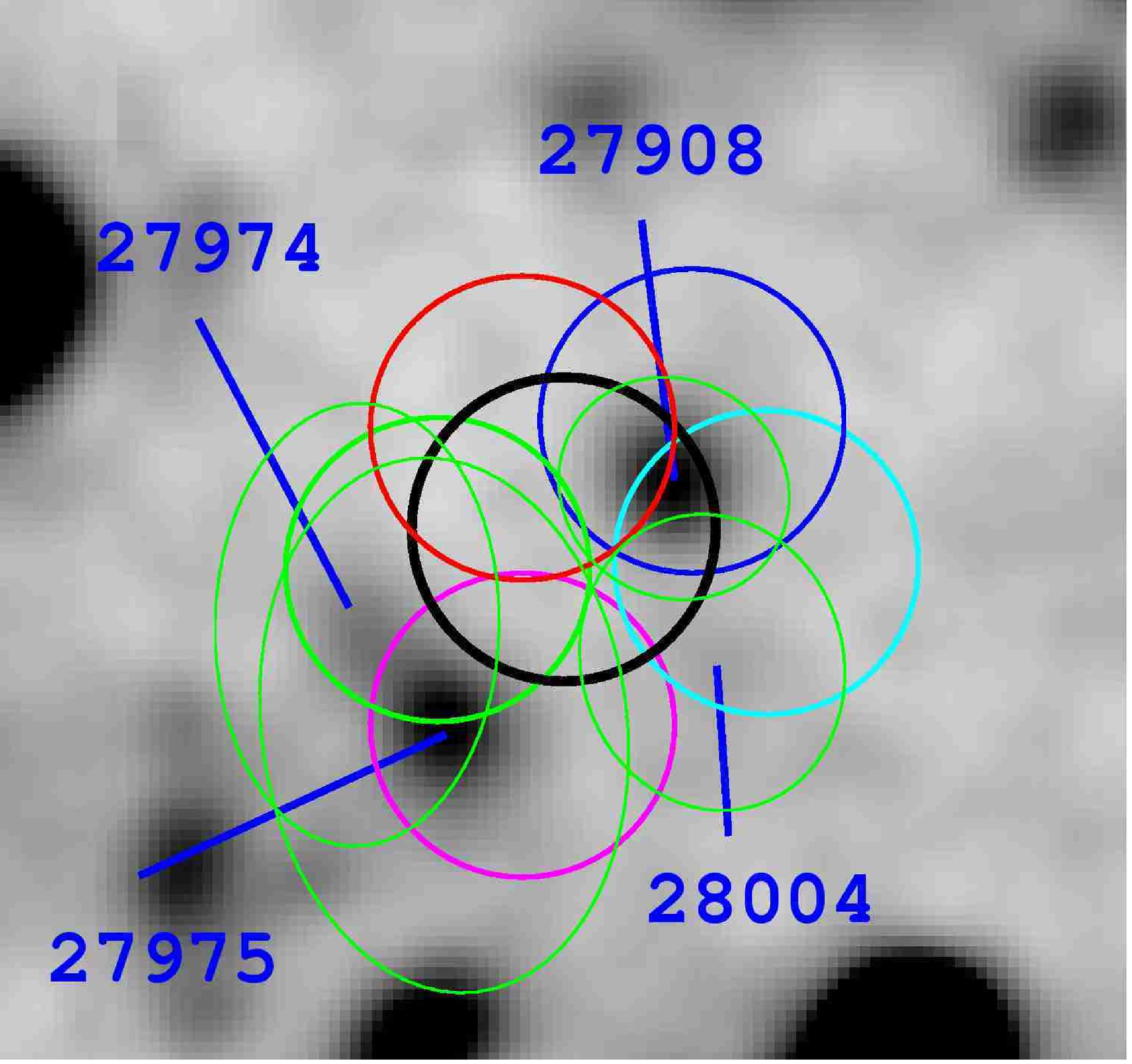}}
\vspace{-22pt}
\\
\subfigure{\includegraphics[bb=41 46 372 666,clip=true,scale=0.33,angle=-90] {fig_extenda4.eps}}
\subfigure{\includegraphics[scale=0.2,angle=-90] {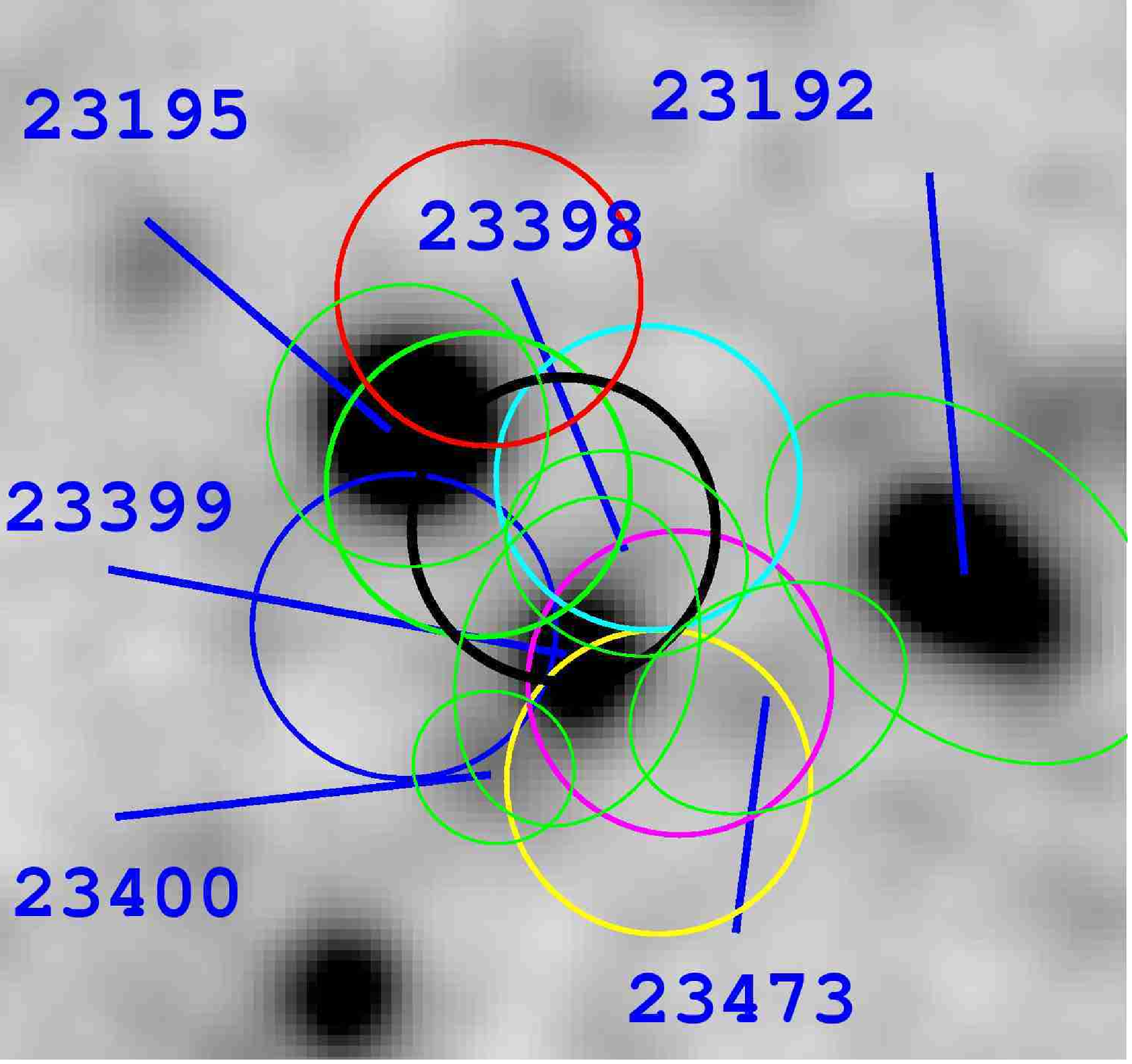}}
\vspace{-22pt}
\\
\subfigure{\includegraphics[bb=41 46 372 666,clip=true,scale=0.33,angle=-90] {fig_extenda5.eps}}
\subfigure{\includegraphics[scale=0.2,angle=-90] {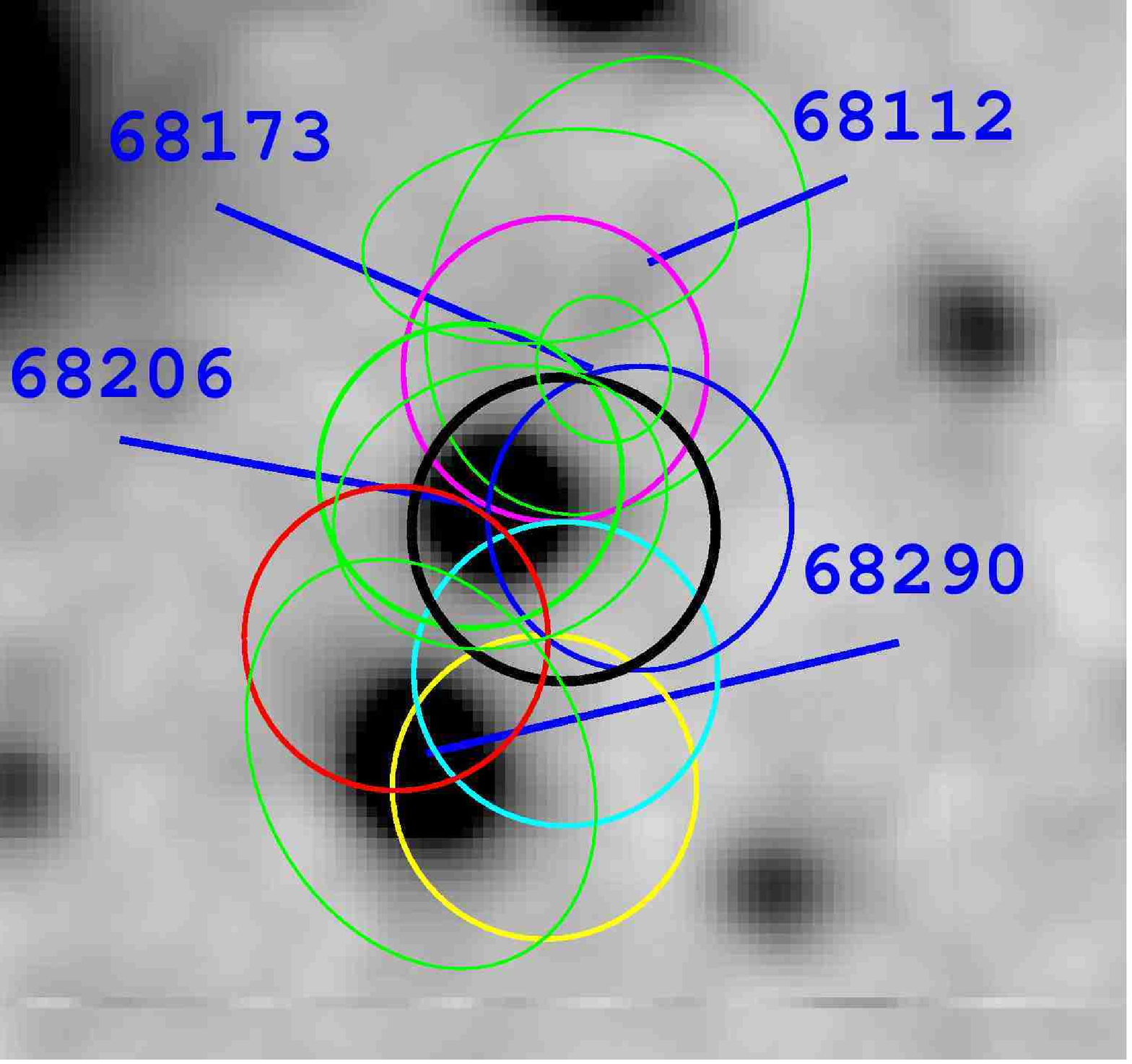}}
\vspace{0pt}
\caption{The five objects significantly extended in Ly$\alpha$. \textit{Left} The spectral 
detections. \textit{Right} The $R$-band images (i' for the first two MUNICS objects). 
\textit{First} Index 99 in MUNICS. The best counterpart is \#11720. 
\textit{Second} Index 126 in MUNICS. The best counterpart is \#11315. This source 
is also a high EW LAE.
\textit{Third} Index 162 in COSMOS. The best counterpart is \#27975. This source 
has an X-ray detection. 
\textit{Fourth} Index 164 in COSMOS. The best counterpart is \#23399.
\textit{Fifth} Index 261 in COSMOS. The best counterpart is \#68206. This source 
has an X-ray detection.
}
\label{fig_extend1}
\end{figure}

\begin{figure}
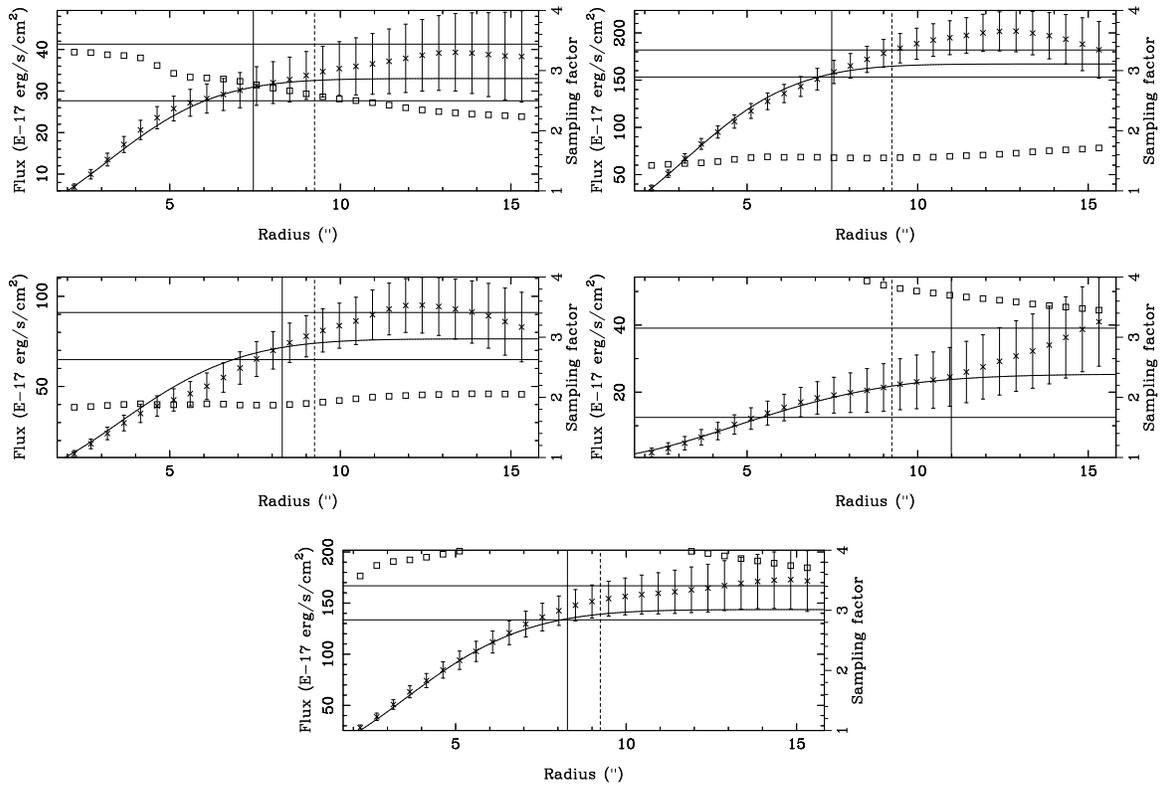

\centering
\subfigure{\includegraphics[scale=0.35,angle=-90] {fig_extendcov1.eps}}
\subfigure{\includegraphics[scale=0.35,angle=-90] {fig_extendcov2.eps}}
\\
\subfigure{\includegraphics[scale=0.35,angle=-90] {fig_extendcov3.eps}}
\subfigure{\includegraphics[scale=0.35,angle=-90] {fig_extendcov4.eps}}
\\
\subfigure{\includegraphics[scale=0.35,angle=-90] {fig_extendcov5.eps}}
\vspace{0pt}
\caption{The curve-of-growth plots for the five significant objects extended in Ly$\alpha$. 
The format is the same as in Figure \ref{fig_cog} and described therein. 
\textit{Top Left} Index 99 in MUNICS. The best counterpart is \#11720. 
\textit{Top Right} Index 126 in MUNICS. The best counterpart is \#11315. This source 
is also a high EW LAE.
\textit{Middle Left} Index 162 in COSMOS. The best counterpart is \#27975. This source 
has an X-ray detection. 
\textit{Middle Right} Index 164 in COSMOS. The best counterpart is \#23399.
\textit{Bottom} Index 261 in COSMOS. The best counterpart is \#68206. This source 
has an X-ray detection.
}
\label{fig_extend2}
\end{figure}

 Based on the detection threshold of \S \ref{sec_comp}, we find 
five objects whose Ly$\alpha$ emission appears significantly extended. 
Figure \ref{fig_extend1} gives the detection and broadband images for the objects, 
and Figure \ref{fig_extend2} shows
the curve-of-growth analysis which determines their sizes. 
These five objects are indices 99 and 126 in the MUNICS field and 
indices 162, 164, and 261 in the COSMOS field. Index 99 is also 
a high EW object. Indices 162 and 261 
are high EW if one use the $R$-band continua for the EW estimation, 
and they are both X-ray sources.
\subsubsection{High EW LAEs}
\par LAEs with EW$_{\mbox{rest}}>240$\AA\ are 
potential sites of exotic energy sources or unusual metallicity since 
stellar population modeling has shown that 
a normal initial mass function (IMF) cannot produce such high EWs \citep{Cha93}. 
If we consider our whole catalog and use the EW measurements derived from 
interpolating with the two nearest filters, we find 
11 LAEs without broadband counterparts and a further 
21 LAEs with counterparts that have EW$_{\mbox{rest}}>240$\AA\ at 
$>1\sigma$ significance. However, in order to 
make a conservative estimate, we instead use EW estimates based on the 
$R$-band photometry only and restrict the discussion to sources 
with emission-line detection S/N$>$6.5 to avoid false detections. 
This instead leaves only 
1 LAE without a counterpart and 2 LAEs with counterparts meeting the 
high EW criterion. We note that a number of the emission lines without 
broadband counterparts may have their origins obscured by ground-based seeing. 
For instance, three of the objects with new redshifts in GOODS-N are shown as part 
of Figures \ref{fig_HDFN1} and \ref{fig_HDFN2}. For 
homogeneity between all fields, we only measure continua 
from the ground-based images in this work. Some of the 
entries in Table \ref{tab_line_list2} as being without 
counterparts may be caused by blending and not image depth. In fact, the only emission-line 
detection with high confidence as being without a counterpart is index 314.
\par
\begin{figure}
\centering
\subfigure{\includegraphics[bb=41 46 372 666,clip=true,scale=0.33,angle=-90] {fig_highEWa1.eps}}
\subfigure{\includegraphics[scale=0.2,angle=-90] {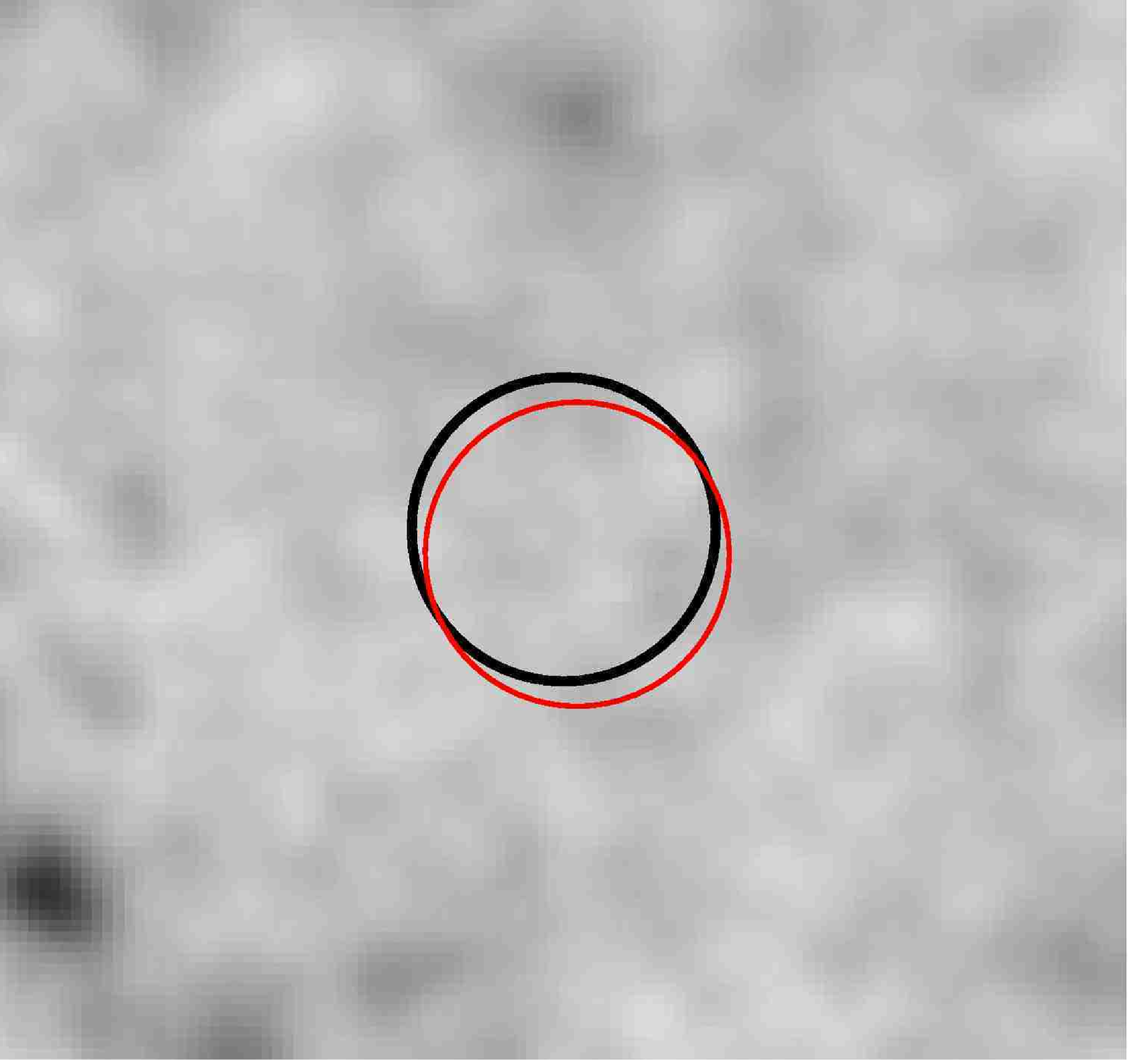}}
\vspace{-22pt}
\\
\subfigure{\includegraphics[bb=41 46 372 666,clip=true,scale=0.33,angle=-90] {fig_highEWa2.eps}}
\subfigure{\includegraphics[scale=0.2,angle=-90] {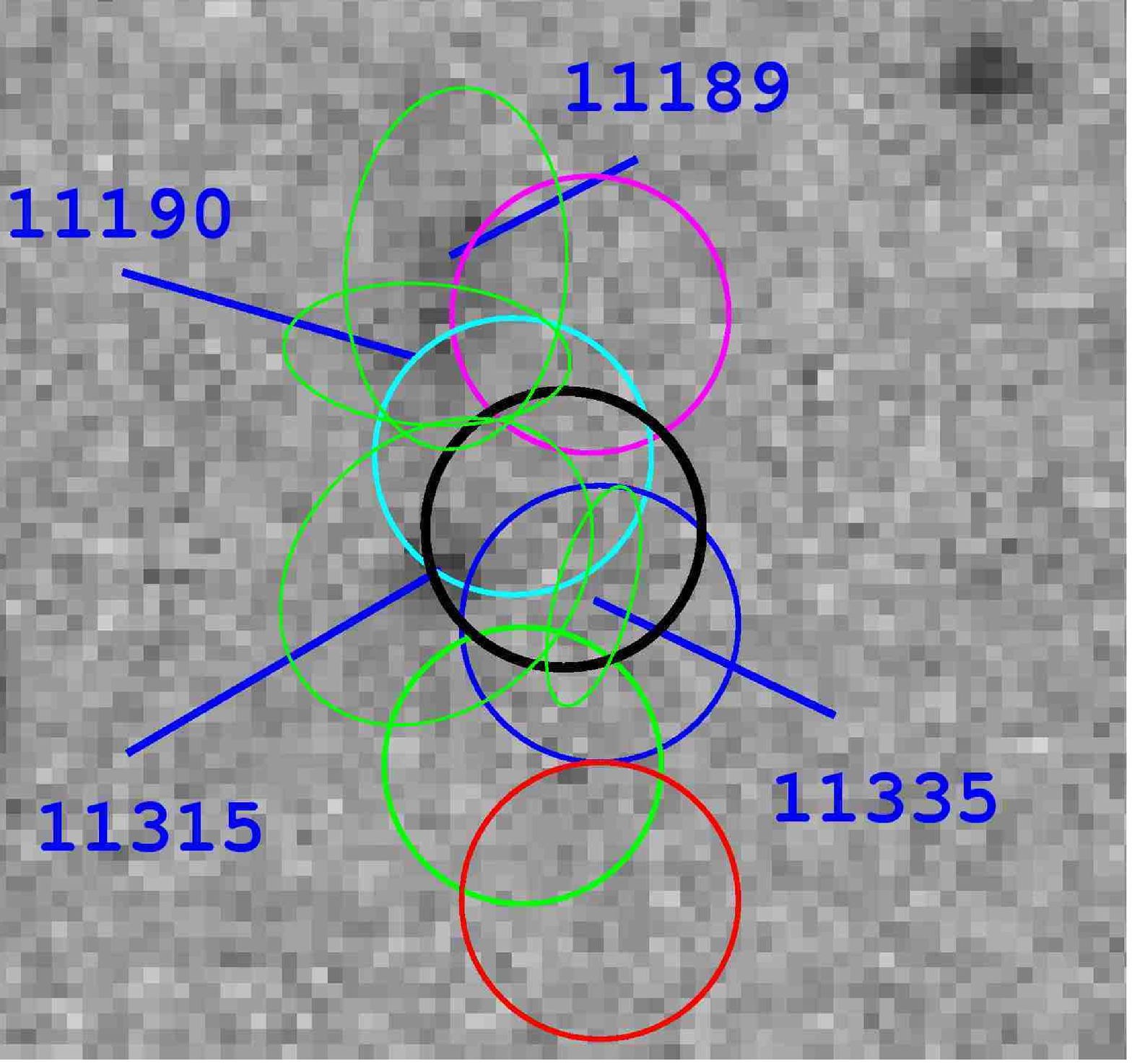}}
\vspace{-22pt}
\\
\subfigure{\includegraphics[bb=41 46 372 666,clip=true,scale=0.33,angle=-90] {fig_highEWa3.eps}}
\subfigure{\includegraphics[scale=0.2,angle=-90] {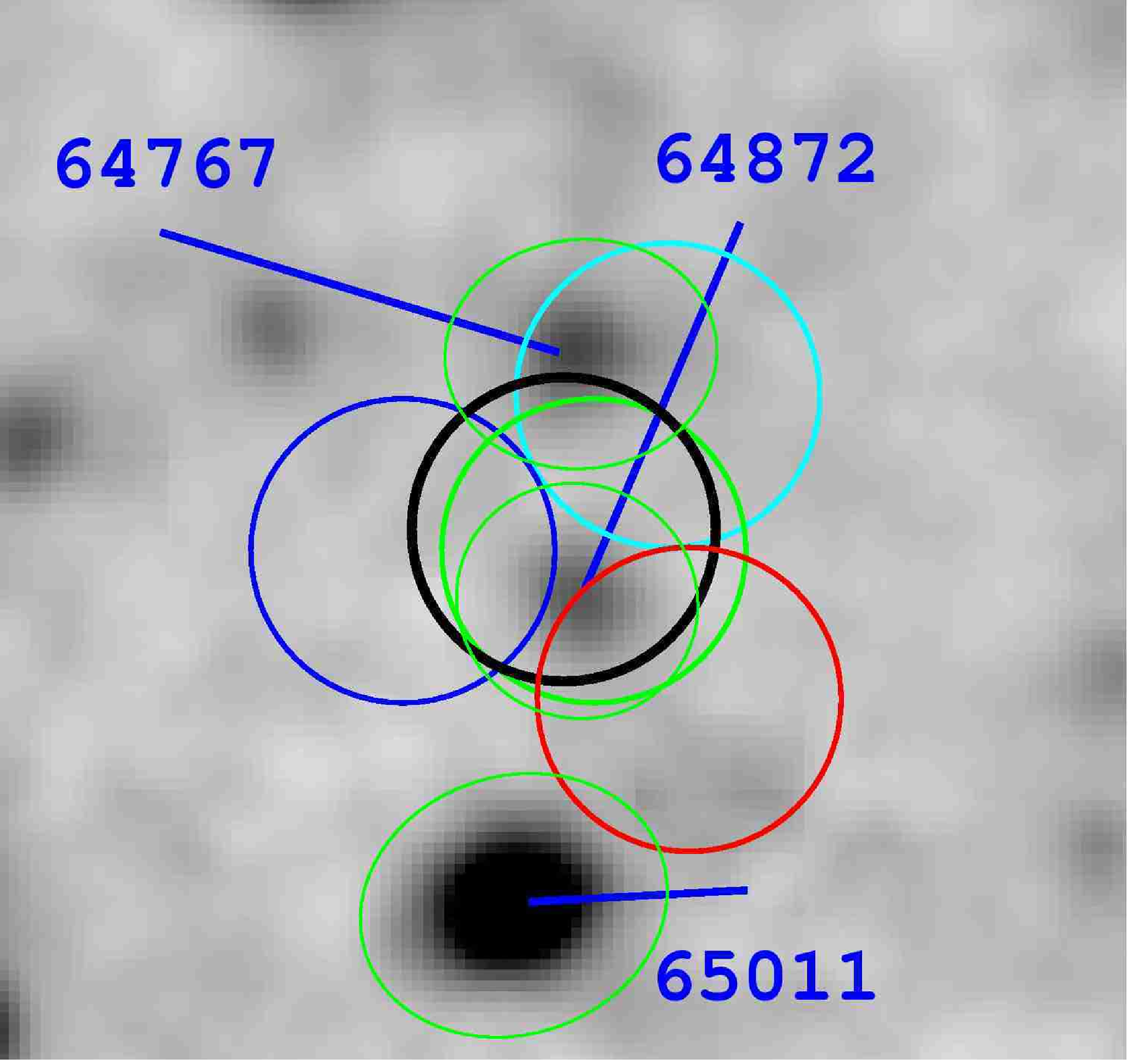}}
\vspace{0pt}
\caption{The three high-significance LAEs with EW$_{0}>$240\AA. \textit{Left} The spectral 
detection figures. \textit{Right} The $R$-band images. \textit{Top} Index 314 in the COSMOS field. 
No continuum counterpart is found. \textit{Middle} Index 126 in the MUNICS field. The 
best counterpart is listed as \#11815. The 
emission line is also significantly spatially extended. \textit{Bottom} Index 
231 in the COSMOS field. The best counterpart is listed as \#64872.
}
\label{fig_highEW1_det}
\end{figure}

 Figure \ref{fig_highEW1_det} shows the detection and image data 
for the three significant high EW LAEs. The top figures show the data for 
index 314, an LAE with z$=2.6312$ but with no counterpart in the COSMOS image 
and EW$_{\mbox{rest}}>348$\AA\ $(1\sigma)$. The middle figures show the 
data for index 126 in MUNICS. Although the counterpart is fairly bright at 
$m_{i'}=24.3$, the very bright emission line implies EW$_{\mbox{rest}}>352$\AA\ $(1\sigma)$. 
We note that this z$=2.8276$ object is also significantly extended in Ly$\alpha$ with FWHM=7\farcs5. 
Finally, the bottom figures show the data for index 231 in COSMOS with 
EW$_{\mbox{rest}}>282$\AA$(1\sigma)$ and z$=2.7215$. This object is 
marginally extended in Ly$\alpha$ (FWHM=6\farcs3), but also compatible with 
a point source and poor seeing. We find a 3\% high-EW fraction in the LAE sample 
by our best estimates. However, the fraction could be as high as 31\% by our 
most inclusive criteria.
\subsubsection{LAE number density expectation}
\par The spectral and spatial sensitivity limits along with the 
completeness simulation of \S \ref{sec_comp} completely define the survey's selection characteristics 
and are necessary inputs to the luminosity function calculations that will follow in future papers \citep{Bla10}. By considering 
all these effects, namely, the completeness distribution we are able to achieve with the 
detection routine (\S \ref{sec_cuts}) and simulated data (\S \ref{sec_comp}), and finally the LAE luminosity function of 
\citet{Gro07}, we predict this sample should contain 121 LAEs. The dominant uncertainty in this prediction is cosmic 
variance, which can be approximated by linear theory for a given redshift, volume, and number density 
from \citet{Som04} Figure 3. The effective Ly$\alpha$ survey volume in \citet{Gro07} is 
$1.1\times10^{5}$ h$_{70}^{-3}$ Mpc$^{3}$ which implies a relative cosmic variance of 
$\sigma_{v}\sim$35\% and $\sigma_{v}\sim$15\% for the volume of this survey. Within these factors, the 
LAE number statistics from this survey are low but not in serious conflict with earlier determinations. 
\section{Summary}
\label{sec_con}
\par We present untargeted integral field spectroscopic observations over 169 \sq\arcmin\ with the 
goal of characterizing emission-line galaxies at low ($z<0.56$) and high ($z\gtrapprox2$) 
redshifts. In this first of a series of papers, we describe the design, observations, 
calibrations, reductions, detections, measurements, and classification methods for the survey. The 
primary classification method we employ uses equivalent width cut computed by matching the 
emission-line objects to continuum counterparts in existing, deep images. We find that effective 
object classification can be made using EW$_{\mbox{rest}}>20$\AA\ 
where the continuum is defined using a single band of deep photometry, preferably 
in the $R$-band. We find 
397 unique emission-line galaxies: 168 over a 71.56 \sq\arcmin\ area 
in the COSMOS field, 118 over a 35.52 \sq\arcmin\ area
in the GOODS-N field, 79 over a 49.85 \sq\arcmin\ area
in the MUNICS field, and 32 over a 12.30 \sq\arcmin\ area
in the XMM-LSS field. The two transitions most frequently observed are 
[OII] (285 galaxies) and Ly$\alpha$ (104 galaxies). Based on a non-evolving 
\citet{Gro07} luminosity function, we should have detected 121 LAEs 
in this survey; the difference is within the range of cosmic variance. The field with 
the deepest X-ray data (GOODS-N) shows an AGN fraction in the LAE sample consistent with that of 
the shallower fields (6\%). We compare our data to the extensive GOODS-N targeted spectroscopy to 
verify our object classification and confirm 92 low-$z$ and 13 high-$z$ galaxies. Moreover, we 
derive new redshifts for a further 2 low-$z$ and 
12 high-$z$ galaxies in the GOODS-N field. Over all fields, 
eleven high-$z$ objects do not possess optical counterparts despite the imaging depth; these 
are either very high EW objects, contamination from noise, 
or objects whose counterparts have been blended by ground-based 
seeing. However, within the remaining LAEs we find 
a distribution of EW that can be described by an exponential scale length of 128$\pm$20\AA\, and with 
only three objects at EW$_{0}>240$\AA\ at $>$1$\sigma$ significance. 
Many of the newly discovered LAEs lie in the color ranges consistent with previous work. 
\par The main contaminant in our LAE sample is simply noise, which should be 10\% of the 
LAE sample based on simulations. A totally pure subsample of 68 LAEs 
can be defined using this catalog at $\mbox{S/N}>6$. 
We find five sources of Ly$\alpha$ emission that have a high significance 
as being spatially resolved, at least two of which are AGN. The pilot survey has validated that 
IFS searches for LAEs perform as expected. The 
forthcoming larger FOV of the full HETDEX survey will vastly improve the survey efficiency of this 
method. 

\acknowledgments
We thank the staff of McDonald Observatory and the engineers and machinists of UT-Austin 
and the Astrophysikalisches Institut Potsdam for their indispensable work. We thank Povilas Palunas for 
field planning and instrument commissioning work, Carlos Allende Prieto for Coud\'{e} 
reduction advice, Niel Brandt for advice on the field selection, and 
Hisanori Furusawa for generously providing his 
SXDS imaging number counts in electronic form. We thank the Cynthia and George Mitchell 
Foundation for funding the VIRUS-P instrument. This work required substantial time 
investment from the McDonald Observatory and was only possible with the 
support and patience of the director, Dr. David Lambert. Salary and 
travel support for students was provided by 
a Texas Norman Hackerman Advanced Research Program under grants ARP 
003658-0005-2006 and 003658-0295-2007. 
JJA acknowledges the support of a National Science Foundation
Graduate Research Fellowship and the University of Texas at Austin's 
William S. Livingston Award, David Bruton Jr. Fellowship, and Donald P. Harrington Fellowship. 
This research has made use of the NASA/IPAC 
Extragalactic Database (NED) 
which is operated by the Jet Propulsion Laboratory, California Institute of Technology, 
under contract with the National Aeronautics and Space Administration. Some of the 
data presented in this paper were obtained from the Multimission Archive at the Space 
Telescope Science Institute (MAST). STScI is operated by the Association of Universities for 
Research in Astronomy, Inc., under NASA contract NAS5-26555. Support for MAST for non-HST data is 
provided by the NASA Office of Space Science via grant NNX09AF08G and by other grants and contracts.
{\it Facilities:} \facility{Smith (VIRUS-P)}.
\appendix
\section{Atmospheric differential refraction astrometry correction}
\label{ap_adr}
\par Atmospheric differential refraction (ADR) effects over this dataset's wavelength and 
airmass ranges are of the order of the astrometric solution errors, so we have made 
an astrometry correction to the guider-based positions considering the emission-line 
source wavelength. There are two ADR effects on the observed fiber positions at any 
given wavelength: 
the atmosphere's wavelength-dependent index of refraction at a fixed airmass and the 
different airmasses between the science and guider FOVs. As we have stated in 
\S \ref{sec_setup}, the guider's effective wavelength is 5000\AA\, and we 
ignore color corrections for different guide stars. In order to retain the ability to stack 
exposures taken at the same dither position, we average the positional 
differences over the $N$ exposures and apply Equation \ref{eq_ADR} 
from \citet{Sma77} as the average positional corrections for an emission-line source 
at wavelength $\lambda$ due to ADR where $\phi$ is 
the site latitude, $\delta$ is the declination, $\theta_g$ is the 
distance angle between the guider and IFU centers, H is the hour angle at the 
middle of the frame's exposure, and k is the constant of mean refraction 
calculated \citep{Fil82} for average 2 km altitude conditions \citep{All73} 
and related to the atmosphere's index of refraction. Common 
corrections derived this way are 0-2\arcsec\ with a median of 0\farcs3 
for our sample. 
\begin{eqnarray}
\label{eq_ADR}
\Delta\alpha=\sum_{i=1}^N \frac{k(\lambda)\sec^2\delta\sin H}{N\times(\tan\delta\tan\phi+\cos H)}-\frac{k(5000\AA)\sec^2(\delta+\theta_g)\sin H}{N\times(\tan(\delta+\theta_g)\tan\phi+\cos H)} \nonumber \\
\Delta\delta=\sum_{i=1}^N \frac{k(\lambda)\times(\tan\phi-\tan\delta\cos H)}{N\times(\tan\delta\tan\phi+\cos H)}-\frac{k(5000\AA)\times(\tan\phi-\tan(\delta+\theta_g)\cos H)}{N\times(\tan(\delta+\theta_g)\tan\phi+\cos H)}
\end{eqnarray}

\bibliographystyle{apj}   
\bibliography{VP_Survey}   
\newpage

\clearpage

\clearpage

\begin{deluxetable}{ccccrc}
\tabletypesize{\scriptsize}
\tablecaption{Summary of VIRUS-P Observations\label{tab_obs_run}}
\tablewidth{0pt}
\tablehead{
\colhead{Date} & \colhead{Number of} & \colhead{Fields} & \colhead{Median} 
& \colhead{Range of} & \colhead{Number of}\\
& \colhead{pointings} && \colhead{A$_{\mbox{V}}$ used} & \colhead{A$_{\mbox{V}}$ used} & \colhead{emission line}\\
&&&&& \colhead{detections}\\
}
\startdata
2007 Oct 04-09&3&MUNICS S2&0.29&0.18-0.44&21\\
2007 Nov 05-10&5&MUNICS S2;COSMOS&0.33&0.00-1.48&26\\
2007 Dec 04-09&6&MUNICS S2;COSMOS&0.24&0.23-0.42&36\\
2008 Jan 03-10&8&MUNICS S2;COSMOS&0.25&0.00-2.84&71\\
2008 Feb 01-12&6&MUNICS S2;COSMOS;GOODS-N&0.28&0.00-1.63&27\\
2008 Apr 01-07&3&COSMOS;GOODS-N&0.30&0.00-0.97&30\\
2008 Apr 28-May 03&3&GOODS-N&0.39&0.00-2.77&43\\
2008 Jun 03-09&2&GOODS-N&0.31&0.00-1.69&0\\
2008 Sep 24-29&2&XMM-LSS&0.36&0.19-0.51&28\\
2008 Nov 24-30&2&COSMOS&0.13&0.07-0.46&8\\
2008 Dec 22-27&4&COSMOS;XMM-LSS&0.23&0.13-0.47&40\\
2009 Jan 21-27&3&COSMOS;XMM-LSS&0.23&0.13-0.67&46\\
2009 Feb 19-23&4&COSMOS;GOODS-N&0.19&0.00-0.53&27\\
2009 Mar 20-25&4&COSMOS;GOODS-N&0.34&0.00-0.92&29\\
2009 Apr 20-25&2&GOODS-N&0.20&0.07-0.48&28\\
2009 May 20-25&1&GOODS-N&0.21&0.00-1.55&5\\
2010 Feb 09-11&1&COSMOS&0.30&0.10-0.50&0\\
\enddata
\end{deluxetable}

\begin{deluxetable}{cccccccccc}
\tabletypesize{\scriptsize}
\tablecaption{Ancillary broadband imaging properties\label{tab_Schlegel}}
\tablewidth{0pt}
\tablehead{
\colhead{Field} & \colhead{Central $\alpha$} & \colhead{Central $\delta$} & 
\colhead{E(B-V)} & \colhead{Filters} & \colhead{FWHM} & 
\colhead{Stack} & \colhead{Band$_K$} & 
\colhead{Depth} & \colhead{VIRUS-P}\\
&\colhead{(J2000)}&\colhead{(J2000)}&&&\colhead{\tablenotemark{$\ast$}}
&\colhead{\tablenotemark{$\dagger$}}&\colhead{\tablenotemark{$\ddagger$}}&
\colhead{\tablenotemark{$\ast\ast$}}&\colhead{area (\sq\arcmin)}\\
}
\startdata
COSMOS&10:00:30&+02:15:04&0.018&u*,B$_J$,V$_J$,r',i',z'&1.33&B$_J$r'i'&V$_J$&26.5&71.6\\
GOODS-N&12:36:51&+62:12:51&0.012&U$_J$,B$_J$,V$_J$,R$_J$,I$_J$,z'&1.26&B$_J$R$_J$I&V$_J$&26.6&35.5\\
MUNICS-S2&03:06:41&+00:01:15&0.083&B$_J$,g',i',z'&0.99&B$_J$g'i'&g'&25.8&49.9\\
XMM-LSS&02:21:20&-04:30:00&0.027&u*,g',r',i',z'&0.97&g'r'i'&g'&25.8&12.3\\
\enddata
\tablenotetext{$\ast$}{The worst seeing FWHM in \arcsec\ to which all bands are matched.}
\tablenotetext{$\dagger$}{Filters combined to form the detection image.}
\tablenotetext{$\ddagger$}{The band chosen for Kron aperture measurement.}
\tablenotetext{$\ast\ast$}{The 5$\sigma$ limit in AB magnitudes for a point source in a 2\arcsec\ aperture for the 
band with the Kron aperture measurement.}
\end{deluxetable}

\begin{centering}
\begin{deluxetable}{lllrrrrrc} \tabletypesize{\tiny}
\tablecaption{HETDEX Pilot Survey Emission Line Catalog (Abridged)\label{tab_line_list1}} \tablewidth{0pt}
\tablehead{	\colhead{HPS Index}	&
		\colhead{$\alpha$}	&
		\colhead{$\delta$}	&
		\colhead{$\lambda_{\mbox{det}}$}	&
		\colhead{FWHM}	&
		\colhead{S/N$_{\mbox{det}}$}	&
		\colhead{Flux}	&
		\colhead{Spatial}	&
		\colhead{Matching}	\\
		\colhead{}	&
		\colhead{(J2000)}	&
		\colhead{(J2000)}	&
		\colhead{(\AA)}	&
		\colhead{(km s$^{-1}$)}	&
		\colhead{}	&
		\colhead{(10$^{-17}$ cgs)}	&
		\colhead{FWHM (\arcsec)}	&
		\colhead{indices}	\\
		\colhead{(1)}	&
		\colhead{(2)}	&
		\colhead{(3)}	&
		\colhead{(4)}	&
		\colhead{(5)}	&
		\colhead{(6)}	&
		\colhead{(7)}	&
		\colhead{(8)}	&
		\colhead{(9)}	}

\startdata

001&02:21:11.16&-04:31:25.0&5219.16&229&8.1&17.4$_{-3.9}^{3.5}$&4.7$_{-0.6}^{0.8}$&..\\
002&02:21:12.21&-04:32:25.3&5448.72&307&5.6&12.2$_{-4.3}^{3.6}$&4.3$_{-1.0}^{0.8}$&..\\
003&02:21:14.28&-04:31:38.2&4973.93&422&7.5&19.9$_{-3.1}^{4.7}$&4.4$_{-0.5}^{0.8}$&..\\
004&02:21:14.86&-04:31:56.6&5261.37&1285&6.3&42.6$_{-12.4}^{11.2}$&5.1$_{-0.9}^{0.7}$&5\\
005&02:21:15.14&-04:31:54.0&4270.67&1841&33.1&342.1$_{-14.3}^{16.5}$&4.8$_{-0.1}^{0.2}$&4\\
006&02:21:16.26&-04:29:32.8&4591.58&399&14.8&32.7$_{-3.6}^{3.5}$&4.6$_{-0.2}^{0.4}$&..\\
007&02:21:16.35&-04:31:14.6&5161.72&293&19.5&49.4$_{-4.4}^{2.6}$&4.7$_{-0.2}^{0.2}$&..\\
008&02:21:17.25&-04:27:55.7&5820.13&118&6.7&19.1$_{-3.2}^{5.0}$&6.6$_{-0.6}^{0.9}$&..\\
009&02:21:17.25&-04:30:10.4&5464.33&78&12.1&14.1$_{-2.5}^{1.5}$&3.6$_{-0.4}^{0.5}$&..\\
010&02:21:17.47&-04:27:30.6&4808.33&357&15.9&38.9$_{-3.6}^{3.4}$&4.6$_{-0.3}^{0.2}$&..\\
011&02:21:18.48&-04:27:32.2&4590.82&441&6.6&21.2$_{-4.6}^{4.7}$&5.9$_{-0.7}^{0.7}$&..\\
012&02:21:19.20&-04:30:28.9&4398.59&344&9.9&26.4$_{-3.9}^{4.0}$&5.7$_{-0.5}^{0.6}$&..\\
013&02:21:19.22&-04:30:39.9&5250.96&378&7.8&11.7$_{-1.9}^{2.6}$&3.6$_{-0.5}^{0.7}$&..\\
014&02:21:19.47&-04:28:21.9&4863.62&328&21.8&49.6$_{-3.8}^{2.4}$&5.5$_{-0.2}^{0.2}$&..\\
015&02:21:19.48&-04:29:34.0&4919.31&243&21.7&36.2$_{-2.8}^{2.8}$&4.9$_{-0.3}^{0.2}$&..\\
016&02:21:19.83&-04:28:08.8&4901.07&397&5.9&13.7$_{-3.1}^{5.2}$&5.3$_{-1.2}^{1.2}$&..\\
017&02:21:19.85&-04:27:43.4&4588.13&93&5.9&13.2$_{-3.4}^{3.9}$&5.3$_{-0.9}^{1.1}$&..\\
018&02:21:20.77&-04:30:56.6&5655.44&972&8.2&45.4$_{-7.5}^{8.4}$&5.6$_{-0.6}^{0.4}$&19\\
019&02:21:20.78&-04:30:56.4&5452.07&438&26.8&83.6$_{-4.0}^{4.2}$&5.6$_{-0.1}^{0.1}$&18\\
020&02:21:21.27&-04:27:44.6&5788.83&303&8.9&21.8$_{-3.9}^{3.9}$&5.1$_{-0.6}^{0.8}$&..\\
021&02:21:22.13&-04:30:35.0&5448.33&277&8.6&8.8$_{-1.9}^{2.4}$&3.3$_{-0.8}^{0.7}$&..\\
022&02:21:22.84&-04:29:25.2&4586.30&603&7.4&17.7$_{-4.3}^{5.5}$&5.2$_{-0.9}^{0.8}$&..\\
023&02:21:22.88&-04:29:44.3&4540.36&323&9.4&18.5$_{-4.1}^{3.2}$&4.6$_{-0.6}^{0.7}$&..\\
024&02:21:23.28&-04:29:22.3&4592.57&204&8.8&14.3$_{-2.5}^{3.7}$&4.0$_{-0.5}^{0.6}$&..\\
...&&&&&&&&\\
033&02:21:26.83&-04:30:05.8&5460.20&343&9.4&25.4$_{-4.6}^{3.6}$&5.0$_{-0.6}^{0.5}$&..\\
051&03:06:34.61&-00:00:49.6&4980.74&428&5.2&9.1$_{-4.0}^{5.0}$&3.5$_{-1.7}^{1.4}$&..\\
062&03:06:38.12&+00:00:40.0&3741.83&392&5.5&67.7$_{-21.0}^{22.9}$&3.4$_{-1.0}^{1.3}$&..\\
085&03:06:41.48&+00:01:10.8&3699.08&2323&7.5&288.3$_{-59.8}^{90.6}$&3.5$_{-0.7}^{0.7}$&86\\
092&03:06:44.44&+00:01:46.7&5682.19&363&6.0&13.2$_{-4.3}^{4.2}$&4.4$_{-0.7}^{1.0}$&..\\
094&03:06:46.23&+00:02:18.3&5767.11&370&22.7&134.7$_{-11.4}^{9.6}$&6.0$_{-0.4}^{0.2}$&..\\
099&03:06:48.36&+00:00:17.0&4868.35&505&8.3&33.0$_{-5.3}^{8.3}$&7.4$_{-0.6}^{0.7}$&..\\
126&03:06:55.26&+00:00:33.8&4653.11&815&14.2&167.1$_{-13.8}^{14.8}$&7.5$_{-0.3}^{0.5}$&..\\
160&10:00:08.61&+02:17:38.6&4175.32&663&6.0&17.1$_{-6.4}^{10.5}$&5.2$_{-1.6}^{1.5}$&..\\
162&10:00:08.73&+02:15:32.0&4167.83&1063&7.3&76.4$_{-11.5}^{14.6}$&8.3$_{-0.9}^{1.3}$&..\\
164&10:00:08.95&+02:17:23.2&4196.31&482&6.6&25.4$_{-12.9}^{13.7}$&11.0$_{-3.3}^{3.3}$&..\\
178&10:00:11.39&+02:15:14.1&5484.87&283&5.8&10.1$_{-2.6}^{3.5}$&4.8$_{-0.7}^{0.8}$&..\\
192&10:00:13.57&+02:13:16.6&4893.46&406&9.0&23.6$_{-4.9}^{3.9}$&5.4$_{-0.6}^{0.6}$&..\\
223&10:00:18.56&+02:14:59.8&4018.61&1102&5.7&39.0$_{-9.4}^{11.5}$&6.4$_{-0.9}^{1.0}$&..\\
229&10:00:19.39&+02:13:12.6&4910.43&625&14.2&41.6$_{-5.0}^{4.2}$&5.4$_{-0.3}^{0.5}$&..\\
231&10:00:20.80&+02:19:18.8&4524.17&551&7.0&30.8$_{-7.6}^{8.0}$&6.3$_{-0.7}^{1.1}$&..\\
234&10:00:21.49&+02:13:51.5&5465.91&78&14.1&24.4$_{-3.0}^{3.6}$&4.5$_{-0.4}^{0.4}$&..\\
261&10:00:28.57&+02:17:48.4&3763.70&886&9.8&143.7$_{-10.1}^{23.2}$&8.3$_{-0.6}^{0.9}$&..\\
289&10:00:35.24&+02:18:07.3&5235.12&428&6.4&17.1$_{-4.5}^{4.5}$&6.5$_{-1.3}^{1.2}$&..\\
308&10:00:40.04&+02:12:51.7&5622.45&157&7.5&18.4$_{-4.2}^{3.5}$&5.4$_{-0.9}^{0.8}$&..\\
310&10:00:40.70&+02:17:41.5&4948.20&532&7.7&10.7$_{-2.4}^{2.9}$&4.0$_{-0.6}^{0.7}$&..\\
313&10:00:40.78&+02:18:23.6&3765.58&249&5.9&25.1$_{-10.1}^{12.4}$&5.0$_{-1.3}^{1.8}$&..\\
314&10:00:41.09&+02:17:03.6&4414.31&352&6.8&14.4$_{-2.9}^{5.7}$&3.7$_{-0.7}^{1.1}$&..\\
322&10:00:44.57&+02:18:31.6&5057.65&486&7.5&16.7$_{-3.3}^{4.4}$&4.7$_{-0.8}^{0.9}$&..\\
323&10:00:45.12&+02:18:22.0&5617.75&213&15.3&36.7$_{-2.8}^{3.9}$&5.1$_{-0.3}^{0.4}$&324,325,326\\
341&12:36:17.52&+62:13:10.0&4778.04&298&9.3&12.8$_{-2.8}^{4.1}$&3.8$_{-0.6}^{0.8}$&..\\
351&12:36:23.05&+62:13:45.0&5530.63&352&6.0&33.7$_{-12.1}^{13.8}$&7.1$_{-1.4}^{1.5}$&..\\
356&12:36:29.24&+62:11:53.4&5698.97&315&9.0&25.6$_{-6.1}^{7.2}$&7.1$_{-0.7}^{1.0}$&..\\
371&12:36:35.34&+62:14:23.4&4660.93&957&6.0&13.9$_{-5.5}^{4.1}$&3.7$_{-1.4}^{1.1}$&..\\
400&12:36:46.37&+62:14:08.4&5651.77&76&7.2&5.9$_{-2.0}^{1.7}$&2.5$_{-1.4}^{1.0}$&..\\
402&12:36:46.86&+62:12:27.4&4821.05&630&8.3&16.8$_{-3.2}^{1.7}$&3.9$_{-0.5}^{0.5}$&..\\
406&12:36:47.53&+62:15:14.3&5715.47&209&7.4&10.6$_{-2.3}^{1.2}$&4.1$_{-0.7}^{0.8}$&..\\
430&12:36:52.03&+62:11:25.9&5452.19&79&5.9&6.6$_{-1.9}^{2.3}$&4.3$_{-1.2}^{1.1}$&..\\
439&12:36:56.87&+62:11:52.0&5760.95&232&6.5&8.4$_{-2.6}^{3.0}$&4.2$_{-0.9}^{1.2}$&..\\
447&12:36:59.37&+62:13:42.6&5016.05&318&5.8&7.2$_{-1.2}^{2.1}$&4.4$_{-0.9}^{0.9}$&..\\

\enddata
\tablecomments{Table \ref{tab_line_list1} is published in its entirety in the electronic 
edition of the \apjs. A portion is shown here to display its form and content.}
\end{deluxetable}
\end{centering}

\begin{centering}
\begin{deluxetable}{llccrrcccc} \tabletypesize{\tiny}
\tablecaption{HETDEX Pilot Survey Emission Line Classifications (Abridged)\label{tab_line_list2}}
\tablewidth{0pt}
\tablehead{	\colhead{HPS Index}	&
		\colhead{Counter-}	&
		\colhead{Counter-}	&
		\colhead{Counter-}	&
		\colhead{EW$_{\mbox{R,rest}}$\tablenotemark{$\ast$}}	&
		\colhead{EW$_{\mbox{interp,rest}}$}	&
		\colhead{Trans-}	&
		\colhead{$z_{\mbox{est}}$}	&
		\colhead{Ly$\alpha$}	&
		\colhead{X-ray}	\\
		\colhead{}	&
		\colhead{part}	&
		\colhead{part m$_{\mbox{R}}$\tablenotemark{$\ast$}}	&
		\colhead{part P}	&
		\colhead{(\AA)}	&
		\colhead{(\AA)}	&
		\colhead{ition}	&
		\colhead{}	&
		\colhead{P}	&
		\colhead{counterpart}	\\
		\colhead{(1)}	&
		\colhead{(2)}	&
		\colhead{(3)}	&
		\colhead{(4)}	&
		\colhead{(5)}	&
		\colhead{(6)}	&
		\colhead{(7)}	&
		\colhead{(8)}	&
		\colhead{(9)}	&
		\colhead{(10)}	}

\startdata

001&J0221112-043126&23.05&0.93&51.9$_{-14.6}^{14.7}$&62.9$_{-17.4}^{17.0}$&[OII]&0.4004&0.07&..\\
002&J0221122-043225&23.17&0.96&42.0$_{-16.3}^{14.9}$&59.8$_{-23.0}^{20.5}$&[OII]&0.4620&0.04&..\\
003&J0221143-043138&24.31&0.98&58.8$_{-15.1}^{22.2}$&109.0$_{-26.5}^{36.8}$&Ly$\alpha$&3.0915&1.00&..\\
004&J0221150-043156&21.05&0.98&10.4$_{-3.5}^{3.4}$&7.1$_{-2.4}^{2.3}$&CIII]1909&1.7561&0.02&J0221151-043156\\
005&J0221150-043156&21.05&1.00&54.9$_{-9.7}^{10.9}$&55.1$_{-9.2}^{9.9}$&CIV1549&1.7570&0.00&J0221151-043156\\
006&J0221164-042933&23.82&0.89&56.7$_{-11.5}^{12.6}$&74.1$_{-14.8}^{15.8}$&Ly$\alpha$&2.7770&1.00&..\\
007&J0221163-043116&21.38&0.98&31.2$_{-6.0}^{6.2}$&48.3$_{-8.9}^{8.8}$&[OII]&0.3850&0.02&..\\
008&J0221171-042757&22.82&0.67&51.2$_{-12.1}^{16.8}$&57.4$_{-13.3}^{18.3}$&[OII]&0.5616&0.01&..\\
009&J0221174-043001&23.21&0.98&51.0$_{-12.4}^{11.2}$&49.0$_{-11.6}^{9.9}$&[OII]&0.4661&0.02&..\\
010&J0221174-042729&21.43&0.99&24.0$_{-4.6}^{5.1}$&39.1$_{-7.5}^{8.2}$&[OII]&0.2901&0.01&..\\
011&J0221185-042734&23.93&0.56&40.8$_{-11.1}^{12.2}$&39.3$_{-10.6}^{11.3}$&Ly$\alpha$&2.7764&1.00&..\\
012&J0221194-043029&20.27&1.00&5.1$_{-1.1}^{1.3}$&14.3$_{-3.1}^{3.4}$&[OII]&0.1802&0.00&..\\
013&J0221193-043039&24.91&0.70&62.9$_{-17.5}^{26.5}$&92.3$_{-23.3}^{31.4}$&Ly$\alpha$&3.3194&1.00&..\\
014&J0221194-042822&21.69&1.00&39.3$_{-7.3}^{7.8}$&55.5$_{-10.3}^{10.9}$&[OII]&0.3050&0.00&..\\
015&J0221196-042934&20.62&1.00&10.7$_{-2.0}^{2.2}$&18.8$_{-3.5}^{3.8}$&[OII]&0.3199&0.00&..\\
016&J0221198-042801&22.17&0.84&17.0$_{-4.8}^{7.3}$&25.8$_{-7.2}^{11.1}$&[OII]&0.3150&0.00&..\\
017&J0221199-042743&24.33&0.96&36.4$_{-11.3}^{13.7}$&34.2$_{-10.4}^{12.0}$&Ly$\alpha$&2.7742&1.00&..\\
018&J0221208-043057&20.49&0.74&14.4$_{-3.4}^{3.9}$&20.4$_{-4.7}^{5.3}$&[NeIII]3869&0.4617&0.00&..\\
019&J0221208-043057&20.48&0.75&24.3$_{-4.3}^{4.8}$&41.1$_{-7.0}^{7.5}$&[OII]&0.4629&0.00&..\\
020&J0221211-042744&21.80&0.99&22.7$_{-5.6}^{6.0}$&27.0$_{-6.6}^{6.9}$&[OII]&0.5532&0.01&..\\
021&J0221223-043034&23.13&0.60&29.5$_{-7.9}^{10.1}$&40.2$_{-10.6}^{13.2}$&[OII]&0.4619&0.03&..\\
022&J0221230-042925&25.14&0.81&103.4$_{-42.9}^{122.1}$&682.6$_{-328.6}^{1000.0}$&Ly$\alpha$&2.7727&1.00&..\\
023&J0221229-042944&22.44&1.00&27.1$_{-7.5}^{7.1}$&44.1$_{-12.1}^{11.1}$&[OII]&0.2182&0.00&..\\
024&J0221233-042923&22.03&1.00&14.6$_{-3.5}^{4.8}$&21.2$_{-5.0}^{6.8}$&[OII]&0.2322&0.00&..\\
...&&&&&&&&&\\
033&J0221269-043006&22.59&0.54&51.6$_{-12.7}^{12.4}$&69.2$_{-16.6}^{15.5}$&[OII]&0.4650&0.01&..\\
051&J0306348-000051&23.67&0.47&14.8$_{-7.0}^{9.3}$&26.3$_{-12.3}^{15.5}$&Ly$\alpha$&3.0971&0.52&..\\
062&J0306382+000039&24.11&0.95&125.3$_{-44.5}^{52.1}$&205.9$_{-73.0}^{84.9}$&Ly$\alpha$&2.0780&1.00&..\\
085&J0306417+000108&19.53&0.99&17.8$_{-4.7}^{6.6}$&37.1$_{-9.9}^{13.9}$&MgII2798&0.3220&0.01&J0306417+000109\\
092&J0306444+000146&23.33&0.98&18.0$_{-6.6}^{7.2}$&29.7$_{-10.8}^{11.1}$&Ly$\alpha$&3.6741&1.00&..\\
094&J0306463+000219&21.03&0.99&68.9$_{-13.1}^{14.1}$&105.9$_{-19.3}^{19.7}$&[OII]&0.5474&0.01&..\\
099&J0306484+000017&24.75&0.55&142.8$_{-39.7}^{63.9}$&401.3$_{-114.5}^{190.1}$&Ly$\alpha$&3.0047&1.00&..\\
126&J0306554+000033&24.31&0.77&461.7$_{-109.9}^{156.3}$&5461.0$_{-2730.5}^{1000.0}$&Ly$\alpha$&2.8276&1.00&..\\
160&J1000086+021739&27.35&0.61&698.3$_{-333.5}^{1000.0}$&1034.3$_{-559.0}^{1000.0}$&Ly$\alpha$&2.4346&1.00&..\\
162&J1000088+021529&24.45&0.20&214.4$_{-39.0}^{52.8}$&564.3$_{-114.8}^{165.4}$&Ly$\alpha$&2.4284&1.00&J100008.8+021528\\
164&J1000089+021721&24.32&0.31&63.3$_{-32.3}^{35.6}$&126.4$_{-64.5}^{70.5}$&Ly$\alpha$&2.4518&1.00&..\\
178&J1000115+021513&23.27&0.64&38.5$_{-10.4}^{14.4}$&59.7$_{-16.1}^{22.3}$&[OII]&0.4717&0.36&..\\
192&J1000135+021317&22.02&0.99&25.3$_{-5.6}^{5.2}$&44.1$_{-9.4}^{8.2}$&[OII]&0.3130&0.01&..\\
223&J1000187+021460&25.46&0.31&268.2$_{-86.7}^{157.1}$&1919.9$_{-959.9}^{1000.0}$&Ly$\alpha$&2.3057&1.00&..\\
229&J1000194+021312&23.37&0.87&50.7$_{-7.5}^{7.8}$&71.3$_{-9.6}^{9.2}$&Ly$\alpha$&3.0393&1.00&..\\
231&J1000208+021918&26.02&0.74&400.3$_{-118.3}^{168.7}$&1633.7$_{-816.9}^{1000.0}$&Ly$\alpha$&2.7215&1.00&..\\
234&J1000215+021350&23.37&0.63&136.8$_{-20.6}^{26.0}$&133.8$_{-20.1}^{25.4}$&[OIII]5007&0.0917&0.37&..\\
261&J1000286+021749&23.76&0.87&193.4$_{-21.9}^{38.9}$&536.7$_{-92.4}^{157.8}$&Ly$\alpha$&2.0960&1.00&J100028.6+021745\\
289&J1000351+021806&23.37&0.63&68.4$_{-18.7}^{20.2}$&146.0$_{-39.6}^{41.9}$&[OII]&0.4046&0.37&..\\
308&J1000401+021251&22.11&1.00&24.8$_{-6.0}^{5.6}$&26.6$_{-6.4}^{5.7}$&[OII]&0.5086&0.00&..\\
310&J1000407+021741&24.92&0.72&54.7$_{-13.7}^{18.7}$&79.5$_{-18.6}^{24.0}$&Ly$\alpha$&3.0703&0.98&..\\
313&J1000408+021823&22.75&0.98&13.4$_{-5.4}^{6.9}$&23.9$_{-9.7}^{12.3}$&Ly$\alpha$&2.0975&1.00&..\\
314&..&..&0.97&435.4$_{-87.7}^{173.4}$&435.4$_{-87.7}^{173.4}$&Ly$\alpha$&2.6312&1.00&..\\
322&J1000446+021830&21.14&0.84&8.2$_{-1.8}^{2.4}$&27.9$_{-5.8}^{8.0}$&[OII]&0.3570&0.08&..\\
323&J1000453+021822&19.88&0.89&8.5$_{-1.0}^{1.3}$&9.9$_{-1.0}^{1.4}$&[OIII]5007&0.1220&0.11&..\\
341&J1236175+621301&24.81&0.97&57.1$_{-14.6}^{23.4}$&87.1$_{-22.8}^{37.3}$&Ly$\alpha$&2.9304&1.00&..\\
351&J1236231+621346&21.10&1.00&17.5$_{-6.4}^{7.6}$&33.2$_{-12.0}^{14.2}$&[OII]&0.4839&0.00&J1236230+621347\\
356&J1236292+621153&22.84&0.98&68.2$_{-16.9}^{21.1}$&106.6$_{-25.9}^{31.4}$&[OII]&0.5291&0.02&..\\
371&J1236356+621424&23.45&0.36&22.2$_{-9.0}^{7.1}$&30.5$_{-12.3}^{9.7}$&CIV1549&2.0090&0.27&J1236356+621424\\
400&J1236465+621408&23.75&0.59&48.1$_{-16.9}^{15.3}$&50.2$_{-17.6}^{15.4}$&[OIII]&0.1290&0.01&J1236463+621405\\
402&J1236470+621226&25.27&0.43&115.5$_{-29.2}^{35.8}$&220.0$_{-65.2}^{108.5}$&Ly$\alpha$&2.9658&1.00&..\\
406&J1236478+621513&21.69&0.90&9.9$_{-2.3}^{1.6}$&15.9$_{-3.6}^{2.2}$&[OII]&0.5335&0.10&..\\
430&J1236519+621125&21.88&1.00&9.3$_{-2.7}^{3.4}$&11.5$_{-3.3}^{4.2}$&[OIII]5007&0.0889&0.00&..\\
439&J1236573+621153&24.17&0.54&77.0$_{-25.3}^{33.4}$&92.5$_{-29.0}^{34.8}$&[OII]&0.5457&0.46&..\\
447&J1236595+621341&24.22&0.93&19.7$_{-4.3}^{7.5}$&31.3$_{-5.9}^{10.1}$&Ly$\alpha$&3.1262&1.00&..\\

\enddata
\tablecomments{Table \ref{tab_line_list2} is published in its entirety in the electronic 
edition of the \apjs. A portion is shown here to display its form and content.}
\tablenotetext{$\ast$}{The Johnson or SDSS R band filters used are listed in Table \ref{tab_Schlegel}. 
The MUNICS field, at $\alpha\approx$3 hours, instead uses an SDSS $i$ filter.}
\end{deluxetable}
\end{centering}

\begin{deluxetable}{crrr}
\tablecaption{Emission line/X-ray counterpart statistics\label{tab_AGN}}
\tablewidth{0pt}
\tablehead{
\colhead{Field} & \colhead{Low-z counterparts} & \colhead{High-z counterparts} & \colhead{Depth}\\
& \colhead{\tablenotemark{$\ast$}} & \colhead{\tablenotemark{$\ast$}} & \colhead{\tablenotemark{$\dagger$}}\\
}
\startdata
COSMOS&2/112&4/55&0.73\\
GOODS-N&27/94&2/25&0.14\\
XMM-LSS&1/24&0/8&$\sim$27\\
MUNICS&4/63&0/16&$\sim$20
\enddata
\tablenotetext{$\ast$}{(Counterparts/total emission lines)}
\tablenotetext{$\dagger$}{Assuming a point source, 10$^{-15}$ erg s$^{-1}$ cm$^{-2}$ in 2-10 keV, 
if available, or 2-8 keV. The depth for XMM-LSS varies over the observed regions, and 
the sensitivity map for the X-ray coverage in MUNICS is not published.}
\end{deluxetable}

\end{document}